\documentclass[aps, prd, english, superscriptaddress, longbibliography, 11pt, notitlepage, nofootinbib]{revtex4}

\immediate\write16{<<WARNING: LINEDRAW macros work with emTeX-dvivers and other drivers supporting emTeX \special's (dviscr, dvihplj, dvidot, dvips, dviwin, etc.)>>}

\usepackage[colorlinks=true, pdfstartview=FitV, linkcolor=blue, citecolor=blue, urlcolor=blue]{hyperref}

\usepackage[english]{babel}
\usepackage{graphicx,float,caption,subcaption}
\graphicspath{{figures/}}
\usepackage{amsmath,amsthm,amssymb,mathtools}
\usepackage{physics,slashed,tensor,cases}
\usepackage{mwe}
\usepackage{hhline}
\usepackage{textcomp}
\usepackage{natbib}
\usepackage{xcolor}
\usepackage{bm}

\usepackage{stackengine}
\stackMath

\makeatletter

\DeclareMathOperator{\diag}{diag}
\DeclareMathOperator{\SU}{SU}
\DeclareMathOperator{\e}{e}
\DeclareMathOperator{\Ei}{Ei}

\begin{document}

\title{Generalized BPS magnetic monopoles in inhomogeneous Yang-Mills-Higgs models}
\date{\today}

\author{F. R. Silva}
\email{filipe.rodrigues@ufpe.br}
\affiliation{Departamento de Física, Universidade Federal de Pernambuco, Av. Prof. Moraes Rego, 1235, Recife - PE - 50670-901, Brazil}

\author{A. Mohammadi}
\email{azadeh.mohammadi@ufpe.br}
\affiliation{Departamento de Física, Universidade Federal de Pernambuco, Av. Prof. Moraes Rego, 1235, Recife - PE - 50670-901, Brazil}

\begin{abstract}
    We present a non-Abelian model for magnetic monopoles in inhomogeneous media, based on a generalization of the standard 't~Hooft-Polyakov model. The medium is described by spatially dependent couplings in the gauge and scalar sectors, constrained by $P(|\Phi|,r)M(|\Phi|,r)=1$ so that the Bogomol'nyi-Prasad-Sommerfield (BPS) bound is preserved. For static spherically symmetric configurations, we study the first-order monopole equations for the class of generalized permeabilities $M(H,r)=f(r)/H^\alpha$. For the power-law profile $f(r)=r^\beta$, we determine the domain in the $(\alpha,\beta)$ plane where regular BPS solutions exist. On the line $\alpha=1$, the system becomes exactly integrable, with closed-form monopole solutions in an inhomogeneous background. Away from this analytical sector, the solutions are constructed numerically. The model supports a rich spectrum of configurations, including effectively point-like monopoles, compact-core monopoles, hollow monopoles, shell-like structures, and multi-shell monopoles characterized by multiple concentric peaks in the energy density.
\end{abstract}

\maketitle

\section{Introduction}\label{sec:intro}
The introduction of non-canonical kinetic terms in field theories provides a simple way to modify the standard dynamics by altering the kinetic structure of the model. These generalized theories are commonly known as k-field models. The idea originated in cosmology, where it was shown in~\cite{Picon-1999-kInflation} that cosmic inflation can be driven purely by higher-order kinetic terms, without the need for a potential. This mechanism is known as k-inflation.

Extending this idea to topological defects leads to the class of so-called k-defects~\cite{Babichev-2006-Global}. In models with spontaneous symmetry breaking, these defects can have sizes and profiles that differ significantly from those of their canonical counterparts, even when the potential is the same.

The combination of a modified kinetic term and a potential can alter the asymptotic behavior of the fields~\cite{Adam-2007-kdefects,Adam-2009-Corrigendum}. Depending on the form of the kinetic term, the fields may approach the vacuum with a power-law behavior, leading to the appearance of compactons. This flexibility makes k-field models particularly useful for describing defects in confined environments or in the presence of local inhomogeneities.

Within this framework, we focus on magnetic monopoles, which arise as static, spherically symmetric solutions of a $(3+1)$-dimensional non-Abelian gauge theory coupled to an adjoint scalar field with a quartic Higgs potential~\cite{Hooft-1974-Magnetic,Polyakov-1974-Particle}. These configurations are regular and have finite energy under suitable boundary conditions. In the BPS limit, where the scalar self-interaction vanishes, monopoles minimize the energy and satisfy first-order equations that saturate the Bogomol'nyi bound~\cite{Bogomolny-1975-Stability,Prasad-1975-Exact}.

Extensions of magnetic monopoles beyond the canonical ’t Hooft-Polyakov model are motivated by the need to describe these objects in more complex environments, such as effective dielectric or chromoelectric media. This has led to generalized Yang-Mills-Higgs models in which the dynamics are governed by scalar-dependent coupling functions~\cite{Casana-2012-Generalized}. These models preserve the BPS limit and admit first-order equations, while allowing for significant modifications of the energy density~\cite{Atmaja-2018-BPS,Atmaja-2022-Are,Mulyanto-2026-On}. As a result, monopoles can develop a wide variety of internal structures~\cite{Bazeia-2013-Deformed,Bazeia-2018-Magnetic,Bazeia-2018-Small,Bazeia-2018-Bimagnetic,Bazeia-2021-Multimagnetic,Benes-2023-Shapes,Benes-2026-Magnetic}.

The inclusion of spatial inhomogeneities, or impurities, in topological defect dynamics has traditionally led to the loss of integrability and of the Bogomol'nyi bound. However, this picture changed with the introduction of impurity models that preserve BPS properties. In particular, it was shown in $(1+1)$ dimensions~\cite{Adam-2019-BPS,Adam-2019-phi4,Adam-2019-Solvable} that translation invariance can be broken without affecting the minimum-energy configurations, provided the impurity coupling preserves half of the original BPS structure.

The $(1+1)$-dimensional formalism was soon extended to more general interaction and coupling scenarios. For example, it has been shown that impurities preserving half of the BPS structure in the $\phi^4$ model can provide a controlled mechanism for fermion transfer during defect scattering, as well as for the radiation emitted in the process~\cite{Campos-2020-Fermion}. More recently, it was shown that singular impurities can completely screen the static repulsion between solitons, allowing multi-defect BPS configurations (kink-kink states) to remain stable at arbitrary distances~\cite{Slawinska-2025-Kink}.

It is known from~\cite{Tong-2014-Vortices} that the BPS impurity mechanism extends to gauge theories and more complex topologies in $(2+1)$ dimensions. In this context, the vortex moduli space remains well defined in the presence of both electric and magnetic impurities, although its geometry is affected differently in each case. Since then, magnetic impurities in two-dimensional systems have been widely studied~\cite{Cockburn-2017-Dynamics,Almeida-2022-Selfdual,Bazeia-2024-Abelian,Gonzalez-2026-Selfdual}.

A BPS impurity framework for generalized scalar fields in arbitrary dimensions was introduced in~\cite{Bazeia-2025-Generalized}, but its application to magnetic monopoles remained largely unexplored until recently. In~\cite{Bazeia-2025-Magnetic}, this idea was extended to Yang-Mills-Higgs theory in $(3+1)$ dimensions. It has been shown that background impurities, whether scalar or non-Abelian vector fields, can preserve the BPS limit while significantly modifying the monopole structure. In particular, they affect the asymptotic behavior of the fields and the range of magnetic interactions.

In this paper, we introduce an explicit radial dependence $r$ in the coupling functions, breaking translational invariance. We assume that this dependence is determined by the medium's effective permeability. In this way, spatial inhomogeneity directly affects the monopole solutions, linking their internal structure to the medium's properties through the first-order equations. The paper is organized as follows. In Sec.~\ref{sec:model}, we present the model and derive the corresponding BPS equations. In Sec.~\ref{sec:pfunc}, we analyze the solutions for a power-law inhomogeneous coupling, identifying the parameter region that supports regular monopoles and discussing their properties. In Sec.~\ref{sec:conclusion}, we summarize our results and outline possible directions for future work. Technical details on the asymptotic analysis and the numerical implementation are presented in the appendices~\ref{apx:asymp} and~\ref{apx:axl_func}.

\section{The model}\label{sec:model}

We consider an $\SU(2)$ Yang-Mills-Higgs theory in $(3+1)$-dimensional Minkowski spacetime, with coordinates $x^{\mu}$ and metric $\eta_{\mu\nu}=\diag(-1,1,1,1)$. The gauge field $A_{\mu}(x)$ and the Higgs field $\Phi(x)$ take values in the Lie algebra $\mathfrak{su}(2)$ and can be written as
\[
A_{\mu}(x)=A^a_{\mu}(x)T^a,
\qquad 
\Phi(x)=\phi^a(x)T^a,
\]
where $T^a$ are the generators of $\SU(2)$ satisfying $[T^a,T^b]=f^{abc}T^c$. We adopt the convention $T^a=-\frac{i}{2}\sigma^a$, with $\sigma^a$ the Pauli matrices, so that the structure constants are $f^{abc}=\varepsilon^{abc}$.

The components $\phi^a(x)$ are Lorentz scalars transforming in the adjoint representation of $\SU(2)$, while $A^a_{\mu}(x)$ are Lorentz vectors and adjoint gauge fields. The adjoint scalar triplet ${\phi^a}$ and the adjoint gauge field ${A^a_{\mu}}$ together account for fifteen real field components at each point in spacetime.

\medskip

We study the generalized Yang-Mills-Higgs Lagrangian density
\begin{equation}\label{eq:lagrangian}
    \mathcal{L} = \frac{P(|\Phi|,r)}{2}\Tr(F_{\mu\nu}F^{\mu\nu}) + M(|\Phi|,r)\Tr(D_{\mu}\Phi D^{\mu}\Phi) - V(\abs{\Phi}),
\end{equation}
where $D_{\mu}\Phi=\partial_{\mu}\Phi+e[A_{\mu},\Phi]$ is the covariant derivative, and the field strength tensor is given by
\begin{equation}
    F_{\mu\nu} = \partial_{\mu}A_{\nu}-\partial_{\nu}A_{\mu}+e[A_{\mu},A_{\nu}].
\end{equation}
It satisfies the Bianchi identity $D_{[\rho}F_{\mu\nu]}=0$, or equivalently $D_{\mu}\widetilde{F}^{\mu\nu}=0$, where the dual tensor is defined as $\widetilde{F}^{\mu\nu}=\frac12\varepsilon^{\mu\nu\alpha\beta}F_{\alpha\beta}$.

\medskip

The Higgs potential is chosen in the following symmetry-breaking form
\begin{equation}
    V(|\Phi|) = \frac{\lambda}{4}\big(|\Phi|^2-\nu^2\big)^2,
\end{equation}
with $\lambda>0$. In natural units $\hbar=c=1$, the canonical mass dimensions are $[\phi^a]=[A^a_{\mu}]=1$, $[e]=0$, $[\lambda]=0$, and $[\nu]=1$.

\medskip

The functions $P(|\Phi|,r)$ and $M(|\Phi|,r)$, with $\abs{\Phi}=(\phi^a\phi^a)^{1/2}$ and $r=(x^ix^i)^{1/2}$, are assumed to be positive and dimensionless. They can be interpreted as generalized electric permittivity and magnetic permeability, respectively, describing an effective inhomogeneous medium in which the gauge and scalar fields propagate. Because these functions depend explicitly on the radial coordinate $r$, the model breaks spacetime translational invariance and hence Poincaré symmetry, while preserving rotational symmetry, time translations, and gauge invariance. In particular, it explicitly localizes the monopole configurations at the origin.

Throughout this work, we impose the constraint
\begin{equation}\label{eq:PM_constraint}
    P(|\Phi|,r)\,M(|\Phi|,r)=1,
\end{equation}
which guarantees that the Bogomol'nyi saturation of the energy density is possible and that the total energy of BPS configurations remains independent of the detailed form of the inhomogeneous couplings.

\medskip

Varying the action with respect to $\phi^a$ and $A^a_{\mu}$ yields the field equations
\begin{subequations}
    \begin{align}
        D_{\mu}(MD^{\mu}\phi^a) &= \frac{P_{\phi^a}}{4} F^b_{\mu\nu}F^{b\mu\nu} + \frac{M_{\phi^a}}{2}D_{\mu}\phi^b D^{\mu}\phi^b + V_{\phi^a},
        \\
        D_{\mu}(PF^{a\mu\nu}) &= Me\varepsilon^{abc}\phi^bD^{\nu}\phi^c,
    \end{align}
\end{subequations}
where subscripts $\phi^a$ denote derivatives with respect to $\phi^a$, and $V_{\phi^a}=\lambda(\abs{\Phi}^2-\nu^2)\phi^a$. Since we are interested in magnetic monopole solutions, we focus on static, purely magnetic configurations and adopt the temporal gauge $A_0^a=0$. The static field equations then reduce to
\begin{subequations}\label{eq:static}
    \begin{align}
        D_i(MD_i\phi^a) &= \frac{P_{\phi^a}}{4}(F^b_{ij})^2 + \frac{M_{\phi^a}}{2}(D_{i}\phi^b)^2 + V_{\phi^a},
        \\
        D_i(PF^a_{ij}) &= Me\varepsilon^{abc}\phi^bD_j\phi^c.
    \end{align}
\end{subequations}

The energy-momentum tensor is given by $T_{\mu\nu}=2\fdv*{\mathcal{L}}{\eta^{\mu\nu}}$. For static fields, the energy density reads
\begin{equation}
    \mathcal{E} = \frac{1}{2M}(B^a_i)^2 + \frac{M}{2}(D_i\phi^a)^2 + V,
\end{equation}
where $B_i^a=\frac12\varepsilon_{ijk}F^a_{jk}$ is the non-Abelian magnetic field. We now perform the Bogomol'nyi trick by completing the square in $\mathcal{E}$, which leaves us with
\begin{equation}\label{eq:squared}
    \mathcal{E} = \frac{1}{2M}(B^a_i \mp MD_i\phi^a)^2 \pm B^a_iD_i\phi^a + V.
\end{equation}

\noindent Since the squared term is non-negative, the energy density satisfies the condition $\mathcal{E}\geq\pm B^a_iD_i\phi^a + V$. The bound is saturated when the fields satisfy the first-order Bogomol'nyi (BPS) equations
\begin{equation}\label{eq:bogo}
    B^a_i \mp M D_i\phi^a = 0.
\end{equation}

\medskip

To construct monopole solutions, we employ the standard spherically symmetric ansatz
\begin{subequations}\label{eq:ansatz}
    \begin{align}
        \phi^a(\vec{r}) &= \hat{r}^a\big(\nu H(r)\big),
        \\
        A^a_i(\vec{r}) &= \varepsilon_{aij}\hat{r}^j\left(\frac{1-K(r)}{er}\right),
    \end{align}
\end{subequations}
where $\hat{r}^a=x^a/r$ represents the $a$-th component of the position unit vector $\hat{r}$. We impose the following boundary conditions
\begin{subequations}\label{eq:bc}
    \begin{align}
        H(0)=0 &\quad\mathrm{and}\quad K(0)=1,
        \\
        \lim\limits_{r\to\infty}H(r)=\pm1 &\quad\mathrm{and}\quad \lim\limits_{r\to\infty}K(r)=0.
    \end{align}
\end{subequations}
which ensure regularity at the origin and finite total energy.

Inserting the ansatz into the Bogomol’nyi equations yields the first-order BPS system
\begin{subequations}\label{eq:1st-order}
    \begin{align}
        H' &= \pm\frac{(1-K^2)}{e\nu r^2M},
        \\
        K' &= \mp e\nu MHK,
    \end{align}
\end{subequations}
where the prime denotes a derivative with respect to $r$. These equations automatically satisfy the second-order Euler-Lagrange equations in the BPS limit.

In the BPS sector, the energy density simplifies considerably. Using the first-order equations, one finds
\begin{align}
    \mathcal{E} &= \pm\frac{\nu[H(1-K^2)]'}{er^2},
    \label{eq:eps1}
    \\[4pt]
    &= \frac{1}{er^2}\left[\frac{(1-K^2)^2}{er^2M} + 2e\nu^2MH^2K^2\right].
    \label{eq:eps2}
\end{align}
Integrating (\ref{eq:eps1}) and using the boundary conditions (\ref{eq:bc}), the total energy is obtained as
\[
E=4\pi\int_0^{\infty}\dd r\,r^2\mathcal{E}=\frac{4\pi\nu}{e},
\]
which coincides with the standard monopole mass and is independent of the specific form of $P$ and $M$ functions.

\medskip

From this point on, we set $e=\nu=1$ and choose the upper sign in (\ref{eq:1st-order}). The sign choice reflects the symmetry $\Phi\to-\Phi$ of the Lagrangian (\ref{eq:lagrangian}): the upper sign corresponds to monopoles, while the lower sign corresponds to anti-monopoles. Inspired by \cite{Bazeia-2018-Small}, we consider a generalized magnetic permeability of the form
\begin{equation}\label{eq:mag_perm}
    M(H,r)=\frac{f(r)}{H^{\alpha}},
    \qquad f(r)>0,\quad \alpha\in\mathbb{R}.
\end{equation}
This choice introduces a nontrivial coupling between the scalar field profile and the effective medium.

A particularly important case arises for $\alpha=1$, for which the BPS equations decouple and become exactly integrable. In this case, the system (\ref{eq:1st-order}) reduces to
\begin{subequations}
    \begin{align}
        H' &= (1-K^2)\frac{H}{r^2 f(r)},
        \\
        K' &= -f(r)\,K.
    \end{align}
\end{subequations}
The second equation can be integrated directly, resulting in
\begin{align}\label{eq:generalsol1}
    K(r)=\exp\!\big[-I_1(r)\big],
    \qquad
    I_1(r)\equiv \int_0^r f(t)\,\dd t.    
\end{align}
Substituting this result into the first equation and integrating, one obtains
\begin{equation}\label{eq:generalsol2}
    H(r)=\exp\!\big[I_2(r)\big],
    \qquad
    I_2(r)\equiv \int_{\infty}^r\left(1-\mathrm{e}^{-2I_1(t)}\right)\frac{\dd t}{t^2 f(t)}.
\end{equation}

Therefore, along the line $\alpha=1$, the monopole profiles are completely determined once the function $f(r)$ is specified. This provides a nice example in which the BPS equations admit closed-form solutions in a nontrivial inhomogeneous background, allowing us to construct analytical monopole configurations without resorting to numerical methods.

\section{Power-law inhomogeneous coupling}\label{sec:pfunc}

Let us consider a power-law profile for the inhomogeneous coupling,
\begin{equation}\label{eq:power_law}
    f(r)=r^{\beta},
    \qquad
    \beta\in\mathbb{R}.
\end{equation}
This choice provides a simple one-parameter family of radially inhomogeneous media and is sufficiently flexible to produce a wide range of monopole profiles. In particular, the parameter $\beta$ controls how strongly the effective magnetic permeability varies with the distance from the monopole center. Positive values of $\beta$ enhance the coupling away from the origin, whereas negative values make the inhomogeneity stronger near the origin.

With this choice, the BPS equations~(\ref{eq:1st-order}) take the explicit form
\begin{subequations}\label{eq:bps}
    \begin{align}
        H'&=
        (1-K^2)\frac{H^{\alpha}}{r^{\beta+2}},
        \label{eq:bpsH}
        \\
        K'&=
        -K\,\frac{r^{\beta}}{H^{\alpha-1}},
        \label{eq:bpsK}
    \end{align}
\end{subequations}
and the corresponding energy density, from Eq.~(\ref{eq:eps2}), becomes
\begin{equation}\label{eq:energydensity}
    \mathcal{E}=
    (1-K^2)^2\frac{H^{\alpha}}{r^{\beta+4}}
    +2K^2\frac{r^{\beta-2}}{H^{\alpha-2}}.
\end{equation}

\begin{figure}[H]
    \centering
    \includegraphics[width=0.5\textwidth]{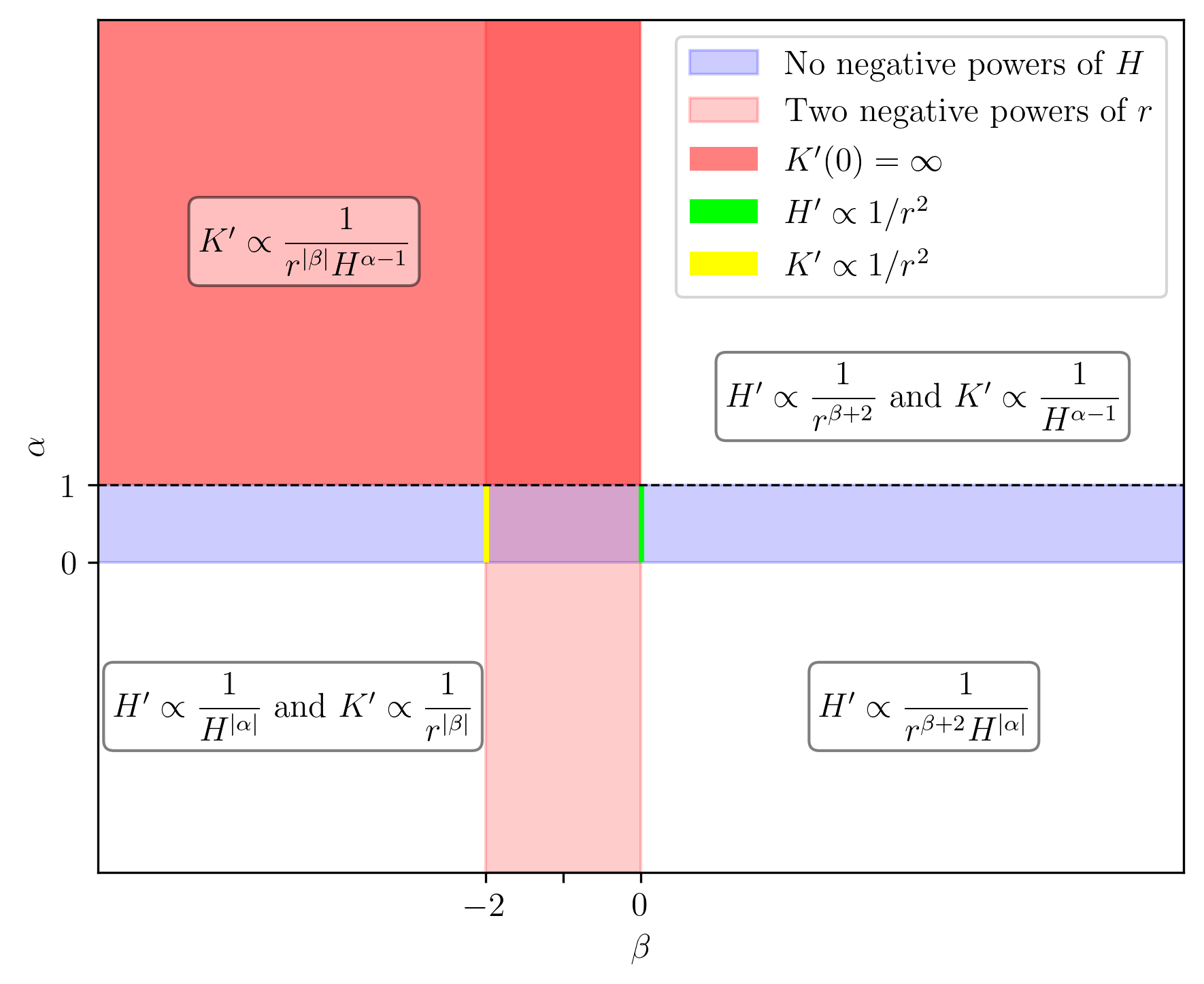}
        \caption{Regularity map of the BPS equations~(\ref{eq:bps}) in the $(\alpha,\beta)$ plane, indicating the singular behavior of the system at the origin across different regions of the plane.}
    \label{fig:regularity_map}
\end{figure}

The system~(\ref{eq:bps}) is generically singular at the origin $r=0$. This singular behavior has two distinct sources. The first is the explicit appearance of negative powers of $r$, which may generate pole-type divergences in both equations. The second, present only in part of the parameter space, is the appearance of negative powers of $H$. This is especially relevant because regular monopole solutions must satisfy
\[
H(0)=0,
\qquad
K(0)=1,
\]
so that inverse powers of $H$ may become singular as $r\to0$.
It is useful to analyze the two equations separately. In Eq.~(\ref{eq:bpsH}), the potentially singular factor is $r^{-(\beta+2)}$. However, since $K(0)=1$, the factor $1-K^2$ vanishes at the origin. Therefore, the singularity in $H'$ may be softened or even removed if $1-K^2$ approaches zero sufficiently rapidly as $r\to0$. In this sense, the regularity of the first equation depends not only on the explicit power of $r$, but also on the detailed asymptotic behavior of the gauge profile $K(r)$ near the origin.
The situation is more restrictive in Eq.~(\ref{eq:bpsK}). When $\alpha>1$, the denominator contains the factor $H^{\alpha-1}$, which introduces a negative power of $H$. Since $H(0)=0$, this term generically causes a divergence in $K'$ at the origin. Such a divergence can only be avoided if the vanishing of $H^{\alpha-1}$ is sufficiently mild to be compensated by the factor $r^{\beta}$. In particular, this requires $\beta>0$ together with a sufficiently slow vanishing of $H^{\alpha-1}$ as $r\to0$. For $\beta\leq0$, the singular behavior of Eq.~(\ref{eq:bpsK}) at the origin is unavoidable whenever $\alpha>1$.
The different types of singular behavior exhibited by the system~(\ref{eq:bps}) are summarized in Fig.~\ref{fig:regularity_map}.

In the interval $0\leq\alpha\leq1$, the system contains no negative powers of $H$, and the singularity structure is controlled entirely by the powers of $r$. In this region, the singularities are therefore of pole type, and their order is determined by the value of $\beta$. In particular, for $-2<\beta<0$, negative powers of $r$ appear in both equations, so both $H'$ and $K'$ may become singular at the origin. By contrast, in the region $\alpha>1$ and $\beta\leq0$, the divergence of $K'$ is unavoidable because it is driven simultaneously by a nonpositive power of $r$ and a negative power of $H$.

Within the admissible blue band shown in Fig.~\ref{fig:regularity_map}, the values $\beta=-2$ and $\beta=0$ are especially favorable. In these cases, the system exhibits at most a second-order pole in only one of the equations, which makes the construction of regular solutions considerably more controlled. 

\begin{figure}[H]
    \centering
    \includegraphics[width=0.45\textwidth]{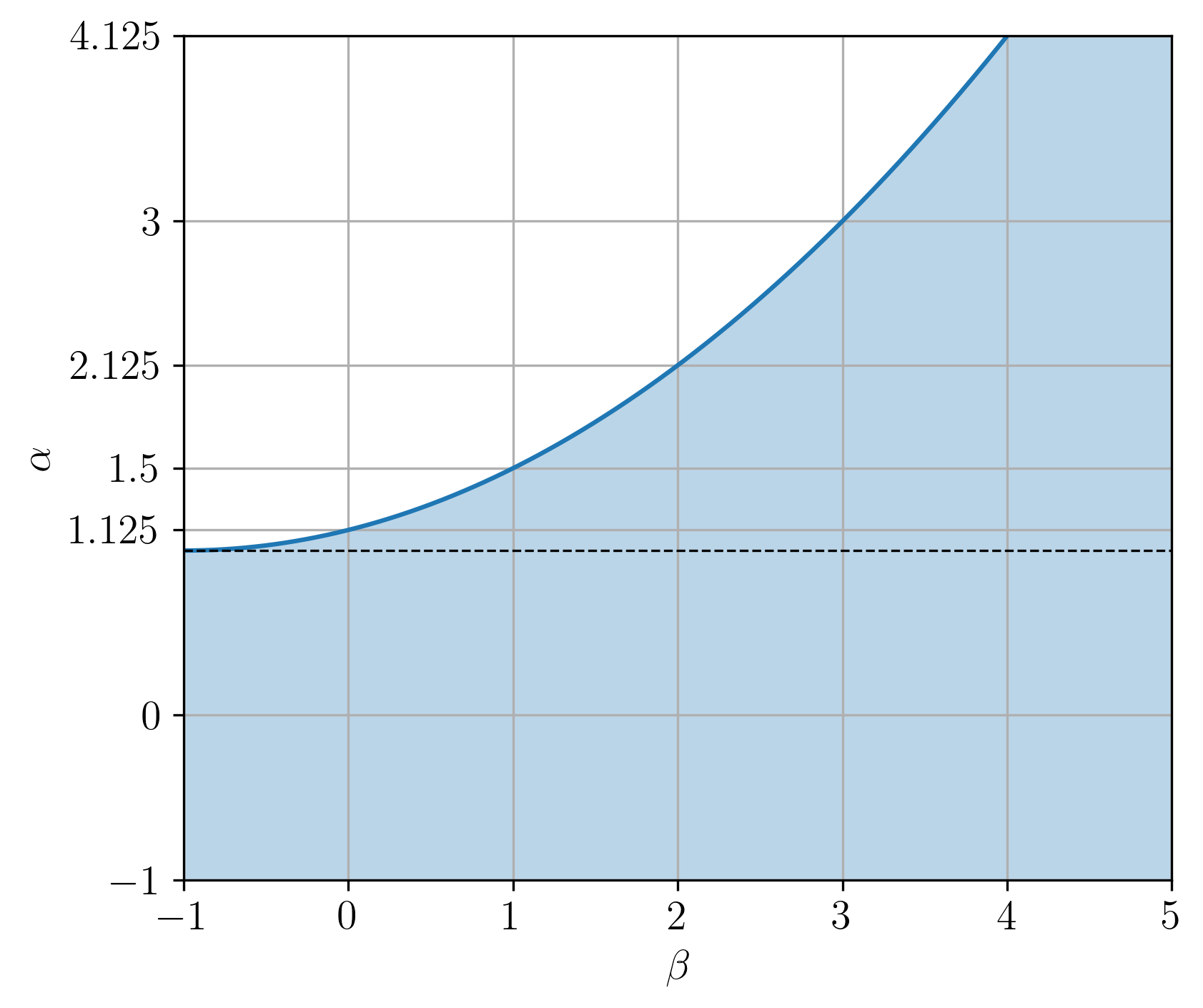}
    \caption{Admissible region in the $(\alpha,\beta)$ plane. The blue curve corresponds to the parabola $\alpha(\beta)=1+(\beta+1)^2/8$ and the dashed line to $\alpha=1$.}
    \label{fig:domain}
\end{figure}

A more precise characterization of the allowed parameter region can be obtained from the asymptotic analysis presented in Appendix~\ref{apx:asymp}. There we show that regular BPS solutions can only exist in the domain of the $(\alpha,\beta)$ plane defined by two conditions. The first is
\[
\alpha(\beta)\leq 1+\frac{(\beta+1)^2}{8},
\]
that is, the allowed region lies below the parabola
\[
\alpha(\beta)=1+\frac{(\beta+1)^2}{8}.
\]
The second condition is $\beta>-1$.
Together, these constraints define the admissible domain displayed in Fig.~\ref{fig:domain}.

More explicitly, the domain is left-bounded and its structure changes across the line $\alpha=1$. For $\alpha>1$, the admissible region is closed and satisfies
\[
\beta+1\geq\sqrt{8(\alpha-1)}.
\]
For $\alpha\leq1$, on the other hand, the allowed region is open and satisfies
\[
\beta+1>0.
\]
In this way, the line $\beta=-1$ acts as a limiting boundary for all values of $\alpha$, while the parabola determines the upper boundary of the parameter space compatible with regular solutions.

\subsection{The line \texorpdfstring{$\alpha=1$}{alpha=1}}

For $\alpha=1$, the general solution derived in Sec.~\ref{sec:model} can be evaluated analytically. Using Eqs.~(\ref{eq:generalsol1}) and (\ref{eq:generalsol2}), the integrals reduce to
\begin{equation}
    I_1(r)=\int_0^r t^{\beta}\,\dd t = \frac{r^{\beta+1}}{\beta+1},
\end{equation}
which is finite provided $\beta+1>0$, precisely the condition already identified in the asymptotic analysis as necessary for the existence of regular solutions along the line $\alpha=1$, and
\begin{align}
    I_2(r)&= \int\limits_{\infty}^r\left(1 - \mathrm{e}^{-2t^{\beta+1}/(\beta+1)}\right)\frac{\dd t}{t^{\beta+2}}
   \nonumber \\
    &= C\int\limits_{x(\infty)}^{x(r)}\frac{(1-\mathrm{e}^{-x})}{x^2}\,\dd x,
\end{align}
where we have introduced the constant $C\equiv 2/(\beta+1)^2$ and the change of variables $x(t)\equiv 2t^{\beta+1}/(\beta+1)>0$.
Evaluating the integral yields
\begin{equation}
    I_2(r) = C\left.\left(\Ei(-x) + \frac{\e^{-x}-1}{x}\right)\right|_{x(\infty)}^{x(r)}.
\end{equation}
Since $\beta>-1$, one has $x(\infty)=\infty$, and therefore
\[
\Ei(-x)\to 0,
\qquad
\frac{\e^{-x}-1}{x}\to 0,
\qquad
\text{as }x\to\infty.
\]
Thus the contribution from the lower limit vanishes, and the monopole profiles take the closed form
\begin{subequations}\label{eq:sol1}
    \begin{align}
        H(r)
        &=
        \exp\left[
        C\left(
        \Ei(-x(r))
        +\frac{\e^{-x(r)}-1}{x(r)}
        \right)
        \right],
        \\
        K(r)
        &=
        \e^{-x(r)/2}.
    \end{align}
\end{subequations}
Therefore, for every $\beta>-1$, the monopole profiles along the line $\alpha=1$ are known analytically.

\medskip

We now examine the properties of these analytical BPS monopole solutions for representative values of $\beta$ along the line $\alpha=1$. Figure~\ref{fig:(1,beta)} displays the scalar profile $H(r)$ and the gauge profile $K(r)$ for several values of $\beta$, together with the corresponding energy density $\mathcal{E}(r)$ obtained from Eq.~(\ref{eq:energydensity}). The left panels show that $H(r)$ increases monotonically from its vanishing value at the origin toward its asymptotic vacuum value, while $K(r)$ decreases monotonically from $K(0)=1$ to zero. The right panels show the corresponding radial energy density, together with a cross section of its three-dimensional spherical distribution, which helps visualize how the energy is distributed in space.

\begin{figure}[H]
    \centering
    \renewcommand{\arraystretch}{1.8}
    \begin{tabular}{@{} c @{\hspace{0.01\textwidth}} c @{\hspace{0.04\textwidth}} c @{\hspace{0.01\textwidth}} c @{}}
        \includegraphics[width=0.235\textwidth]{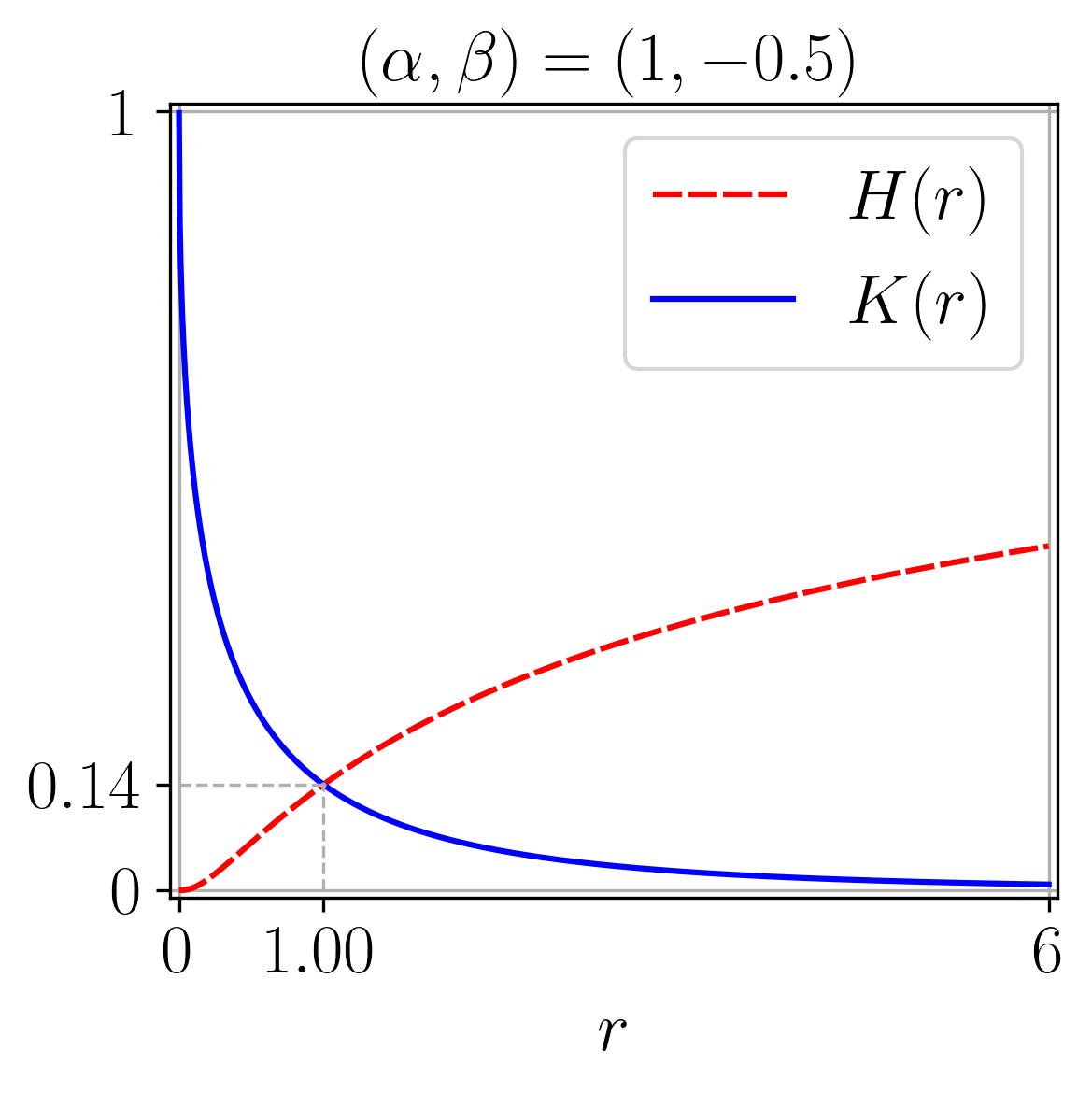} &
        \includegraphics[width=0.235\textwidth]{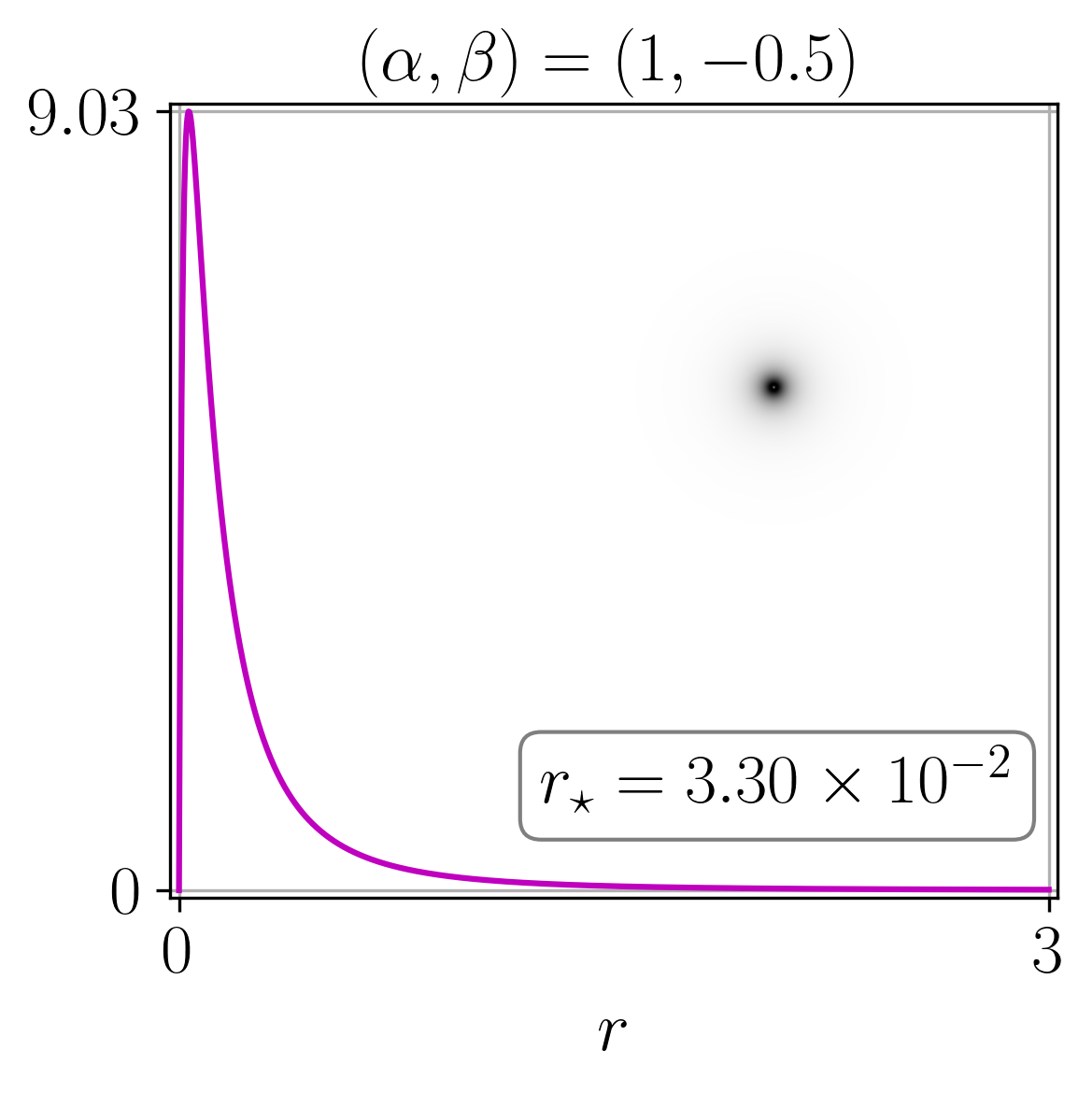} &
        \includegraphics[width=0.235\textwidth]{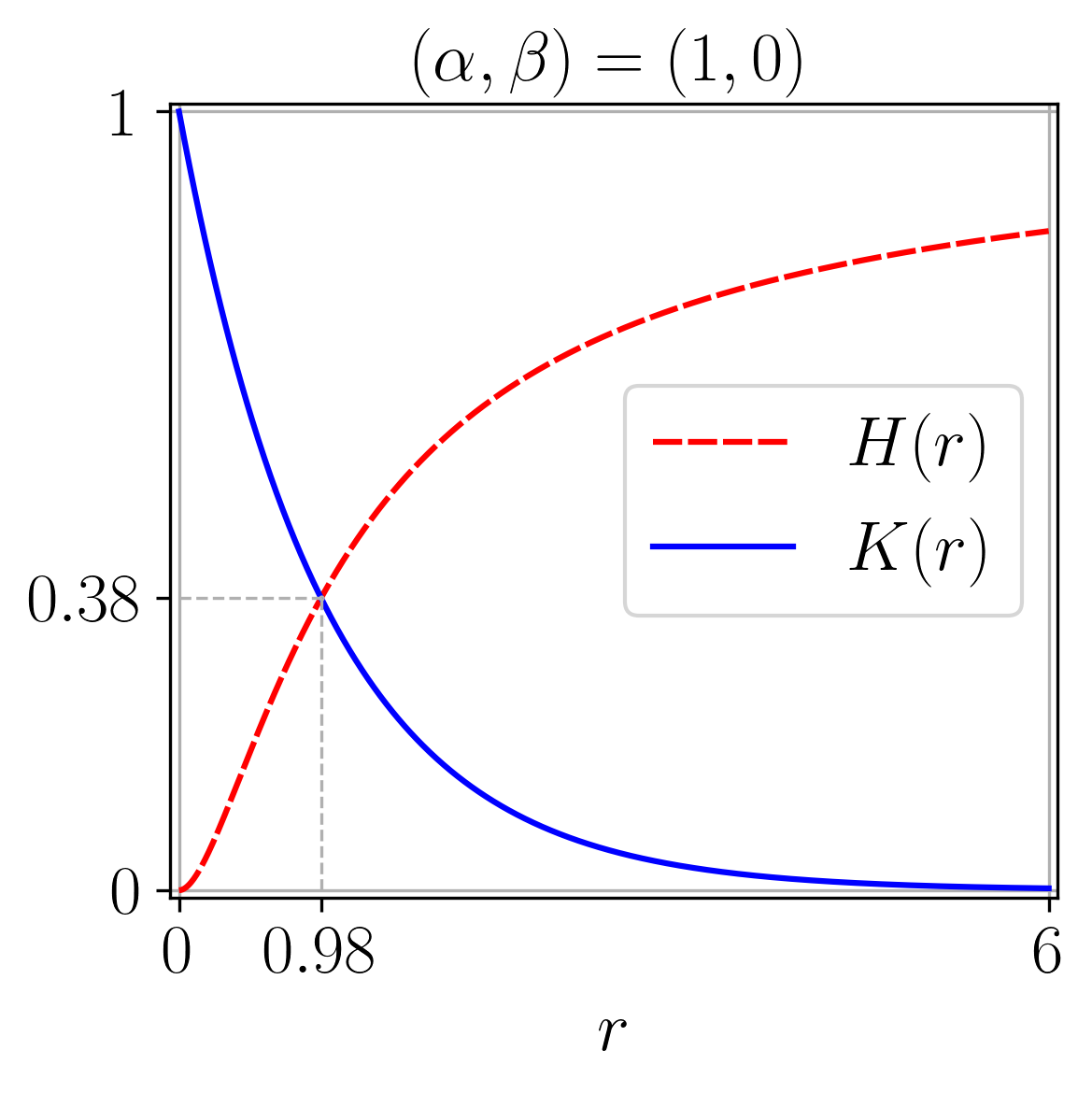} &
        \includegraphics[width=0.235\textwidth]{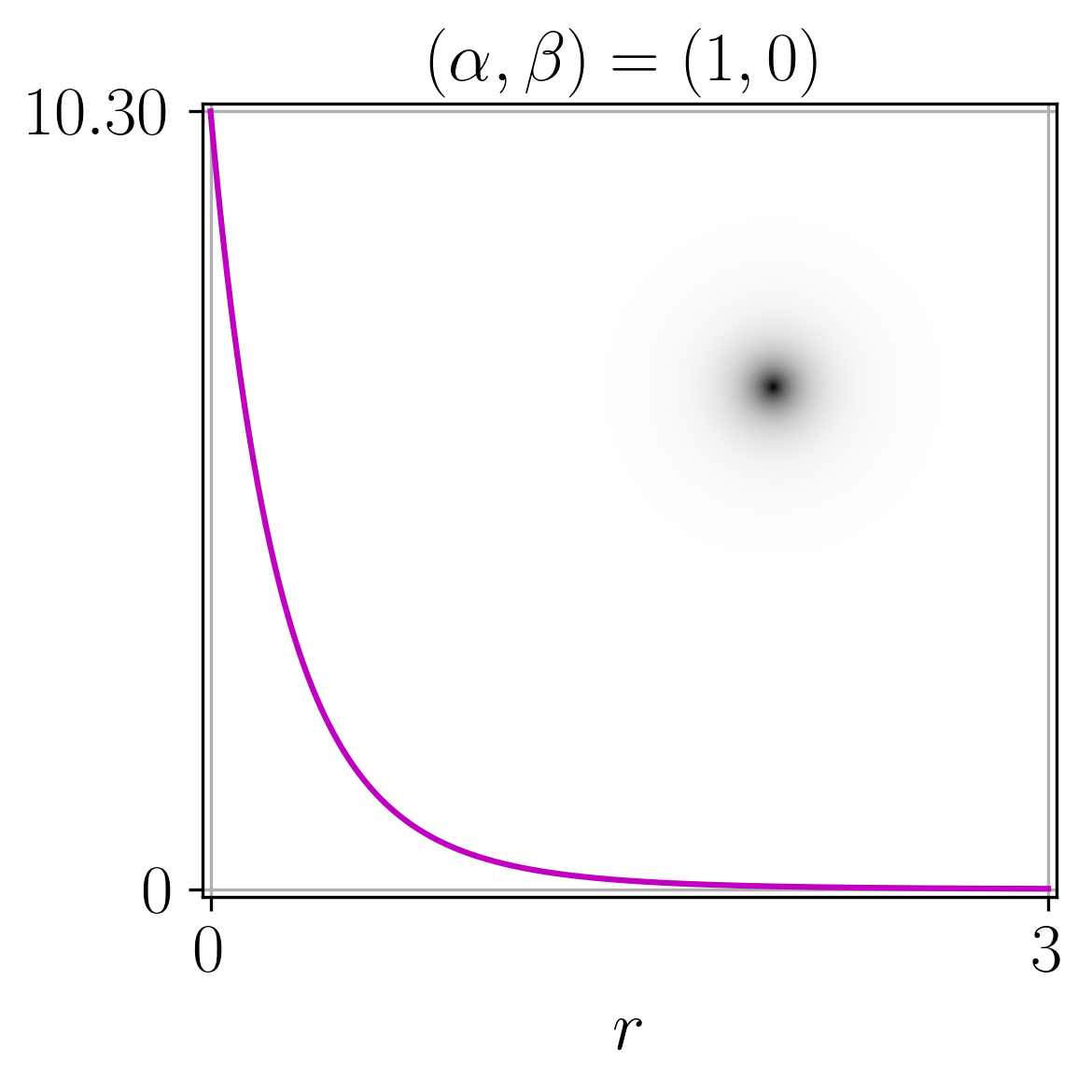} \\
        
        \includegraphics[width=0.235\textwidth]{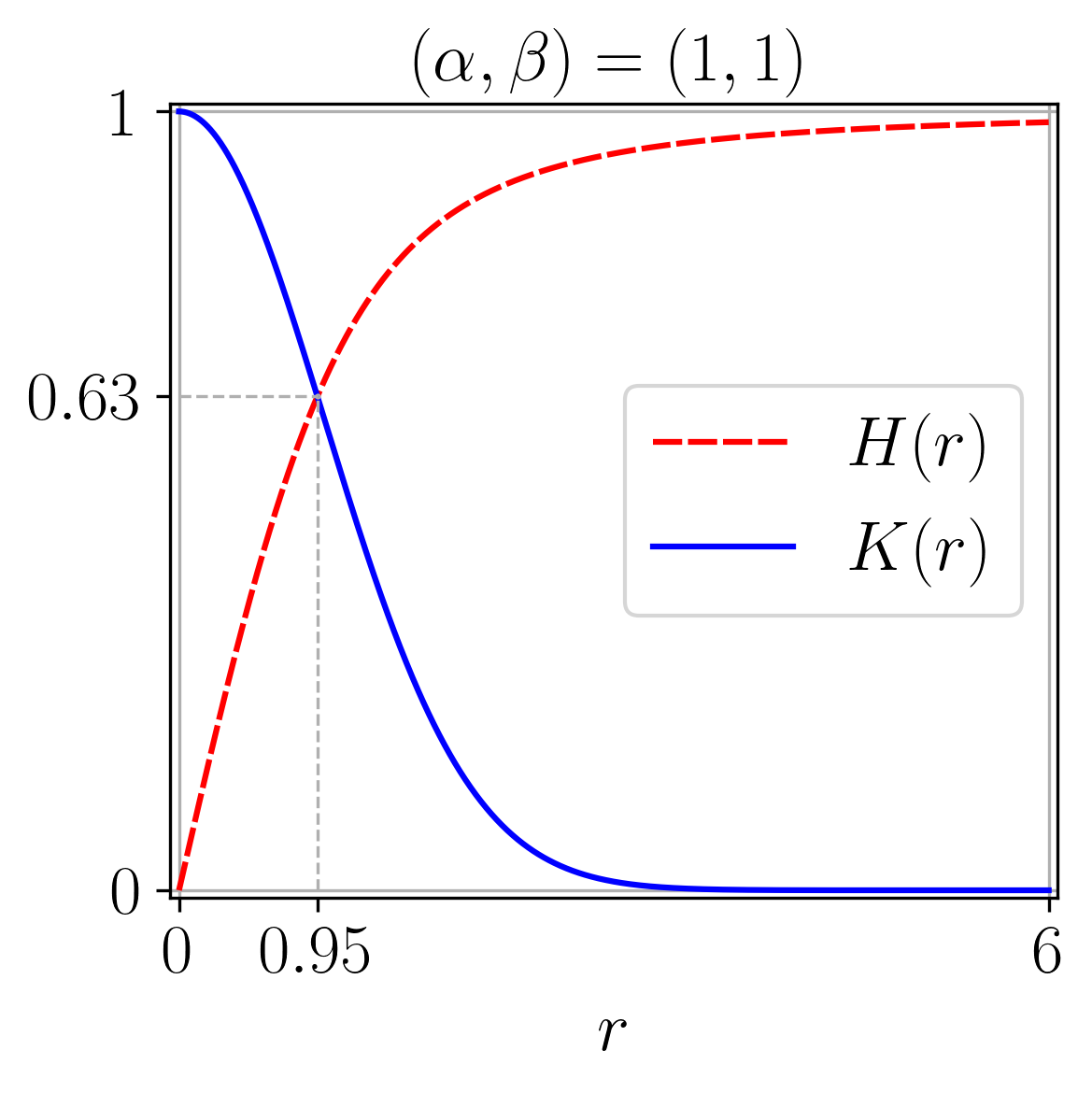} &
        \includegraphics[width=0.235\textwidth]{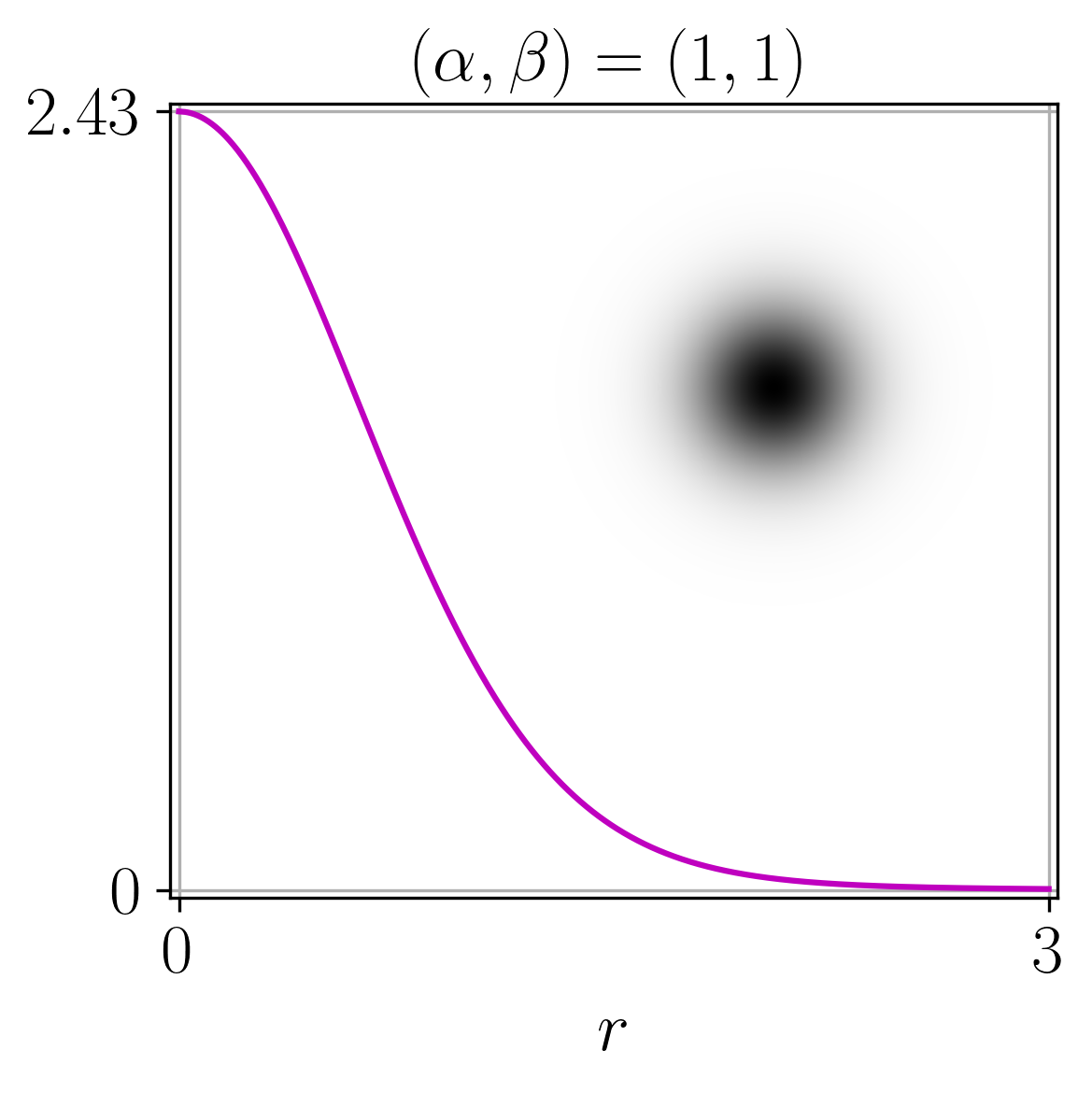} &
        \includegraphics[width=0.235\textwidth]{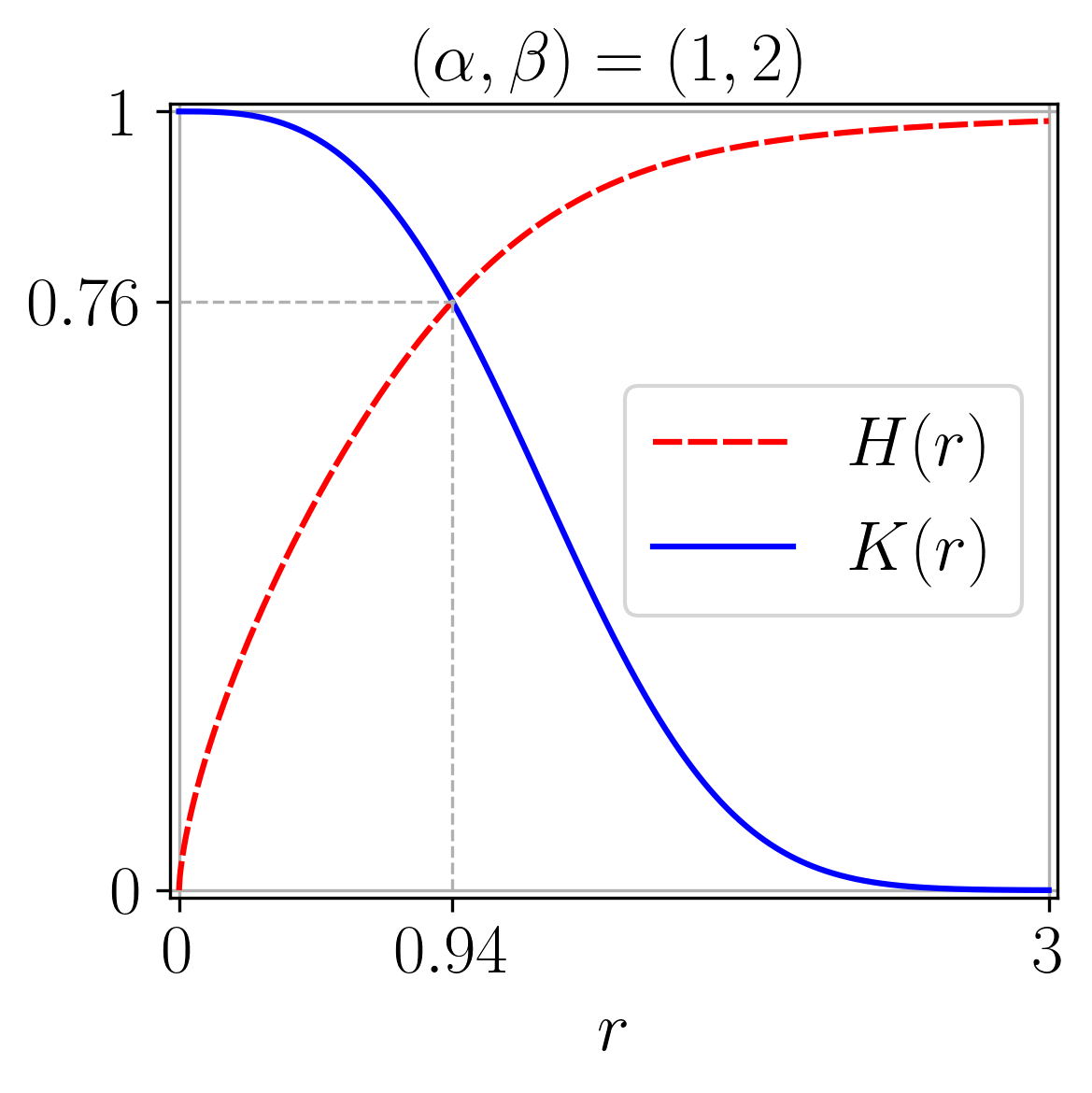} &
        \includegraphics[width=0.235\textwidth]{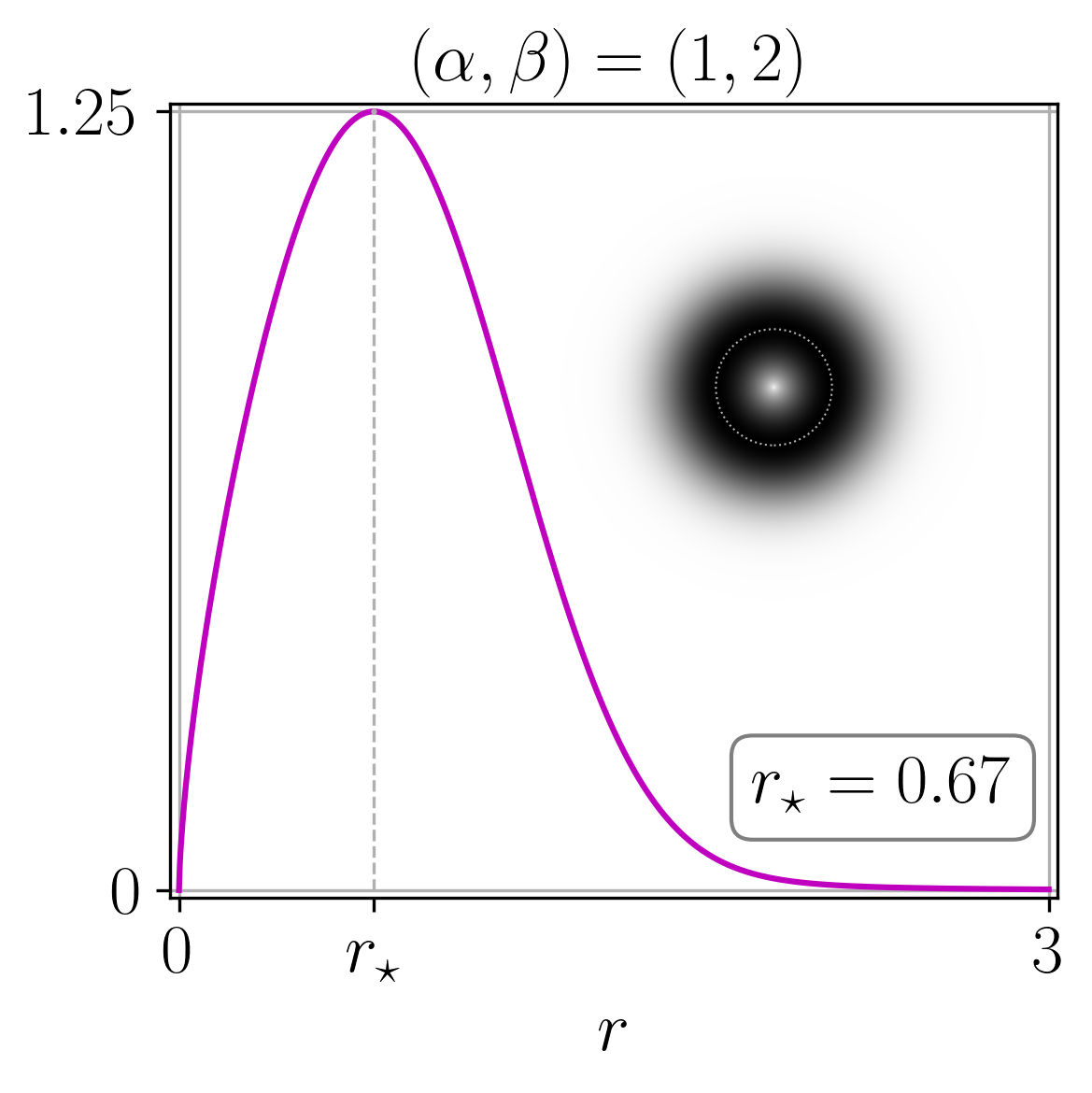}
    \end{tabular}

    \caption{BPS monopole solutions along the analytical line $\alpha=1$ for representative values of $\beta$, with the profile functions $H(r)$ and $K(r)$ on the left and the corresponding energy density $\mathcal{E}(r)$ on the right for each $\beta$. The cases $\beta=-0.5$, $0$, $1$, and $2$ illustrate different possible monopole configurations. The configuration with $\beta=0$ was called ``small monopole" in Ref.~\cite{Bazeia-2018-Small}.}
    \label{fig:(1,beta)}
\end{figure}

The behavior of the solutions changes significantly as $\beta$ varies.
For $\beta<0$, and especially as $\beta$ approaches $-1$, both $H(r)$ and $K(r)$ become increasingly suppressed. Indeed, in the limit $\beta\to-1^+$ one has
\[
x(r)=\frac{2r^{\beta+1}}{\beta+1}\to\infty
\qquad
\text{for every fixed } r>0,
\]
so that the explicit solution~(\ref{eq:sol1}) approaches the trivial configuration
\[
H(r)\to0,
\qquad
K(r)\to0.
\]
At the same time, the energy density undergoes a qualitative redistribution. Its maximum, initially located at the origin, shifts to a finite small radius $r_\star>0$ as $\mathcal{E}(0)$ decreases to zero. This indicates the formation of a small central cavity, with the energy concentrated in a narrow shell around $r_\star$. As $\beta\to-1$, the cavity radius shrinks, and the energy becomes increasingly localized, resulting in an effectively point-like monopole.

For nonnegative values of $\beta$, a different pattern emerges. As $\beta$ increases from zero, the intersection point of the two profile functions $H(r)$ and $K(r)$ moves slowly inward and its value increases. In the interval $\beta\in[0,1]$ no cavity is present: the energy density is maximal at the origin and decreases monotonically to zero, although its concavity changes gradually as $\beta$ varies. In this range, the monopole core becomes more spatially extended, so that the configuration may be interpreted as a larger monopole.

As we continue increasing $\beta$, the central cavity reappears and grows as the energy accumulates around $r_{\star}$, ultimately forming a highly localized, shell-like monopole. We can see in Fig.~\ref{fig:(1,beta_large)} that for sufficiently large $\beta$, the energy density develops a pronounced maximum at a finite radius. The monopole then acquires a shell-like structure, with the energy concentrated around $r_\star$, as illustrated in  Fig.~\ref{fig:(1,10)}.

\begin{figure}[H]
    \centering
    \begin{subfigure}[c]{0.315\textwidth}
        \centering
        \includegraphics[width=\textwidth]{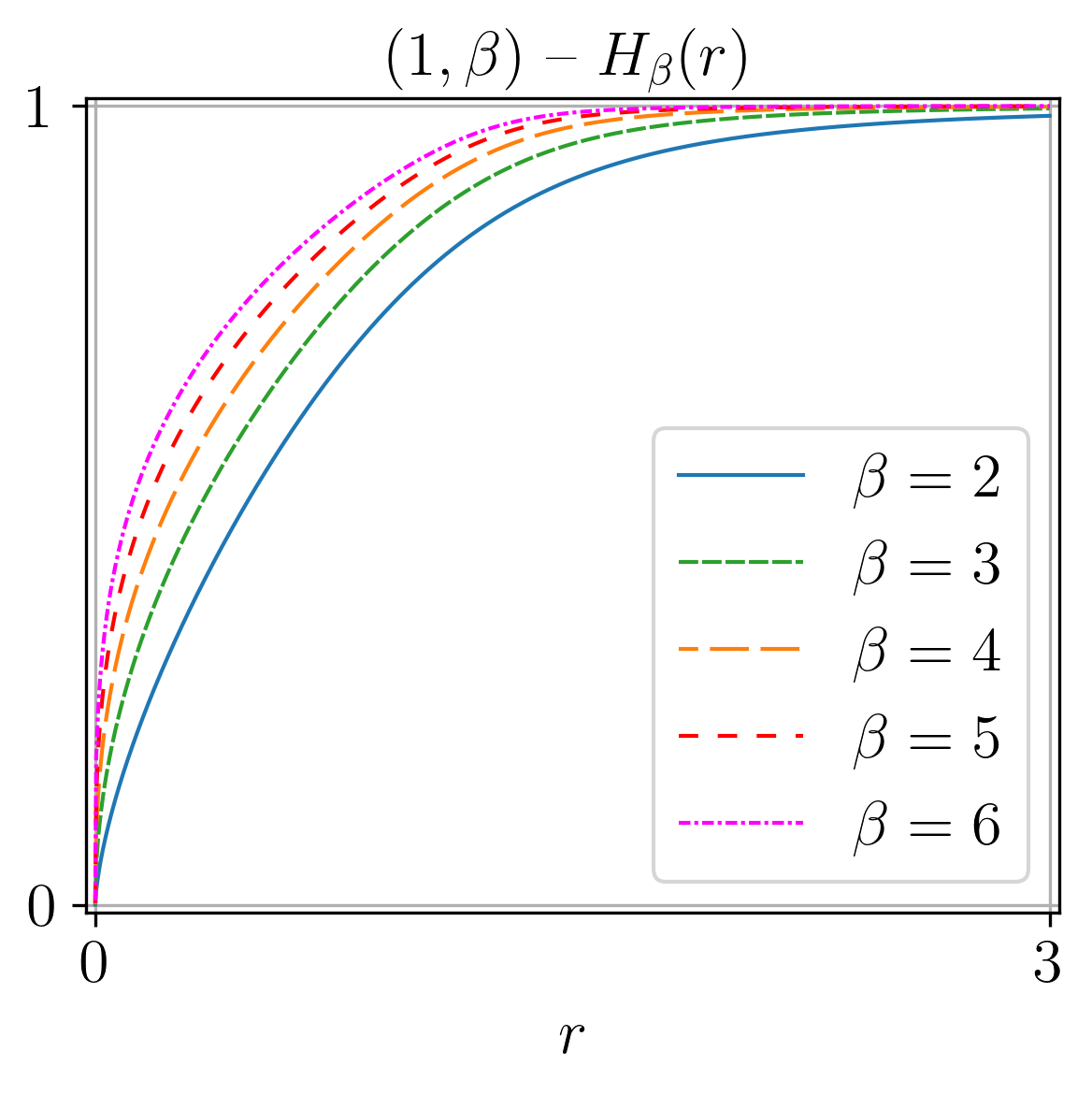}
    \end{subfigure}
    \hspace{6pt}
    \begin{subfigure}[c]{0.315\textwidth}
        \centering
        \includegraphics[width=\textwidth]{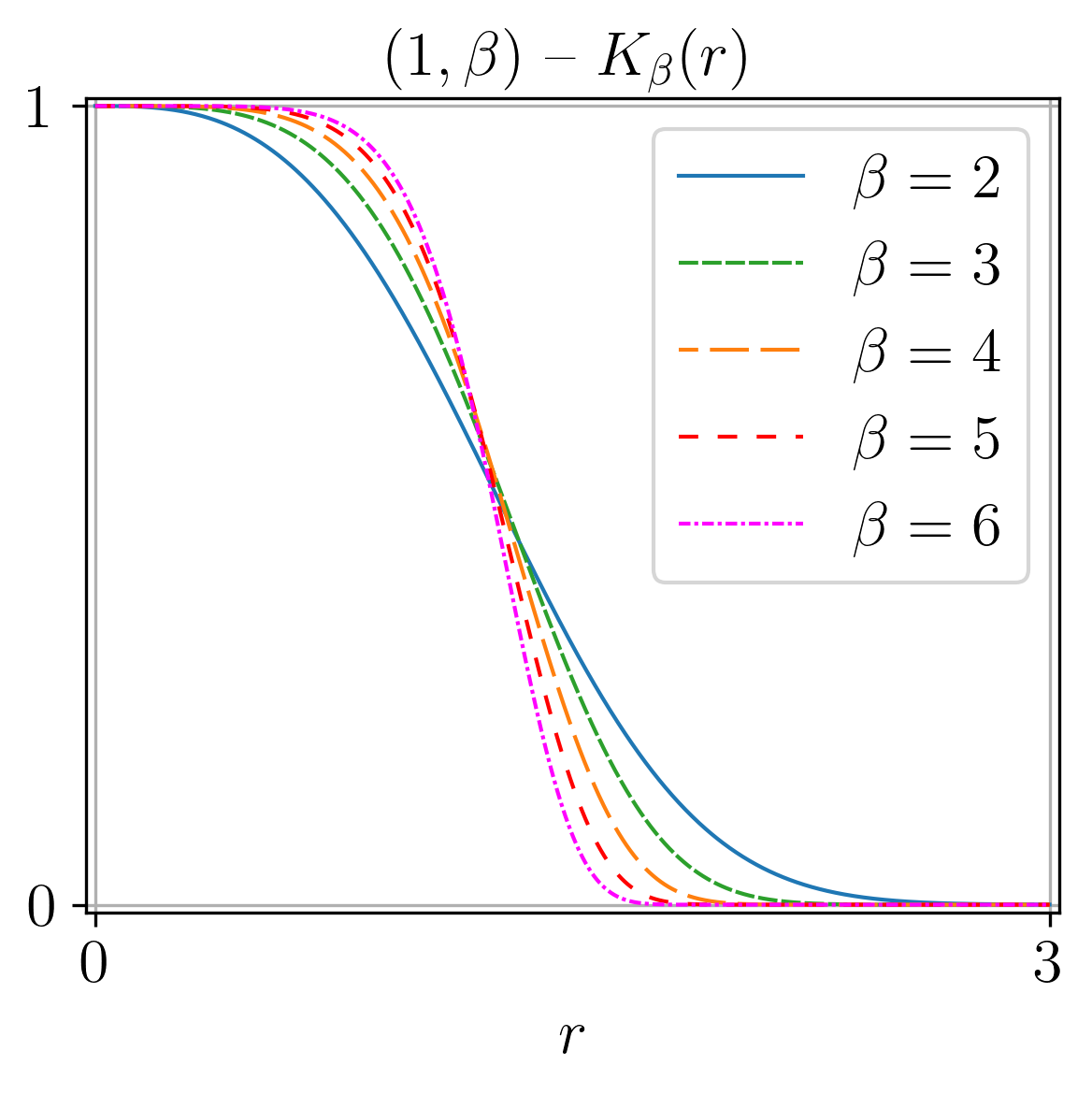}
    \end{subfigure}
    \hspace{6pt}
    \begin{subfigure}[c]{0.315\textwidth}
        \centering
        \includegraphics[width=\textwidth]{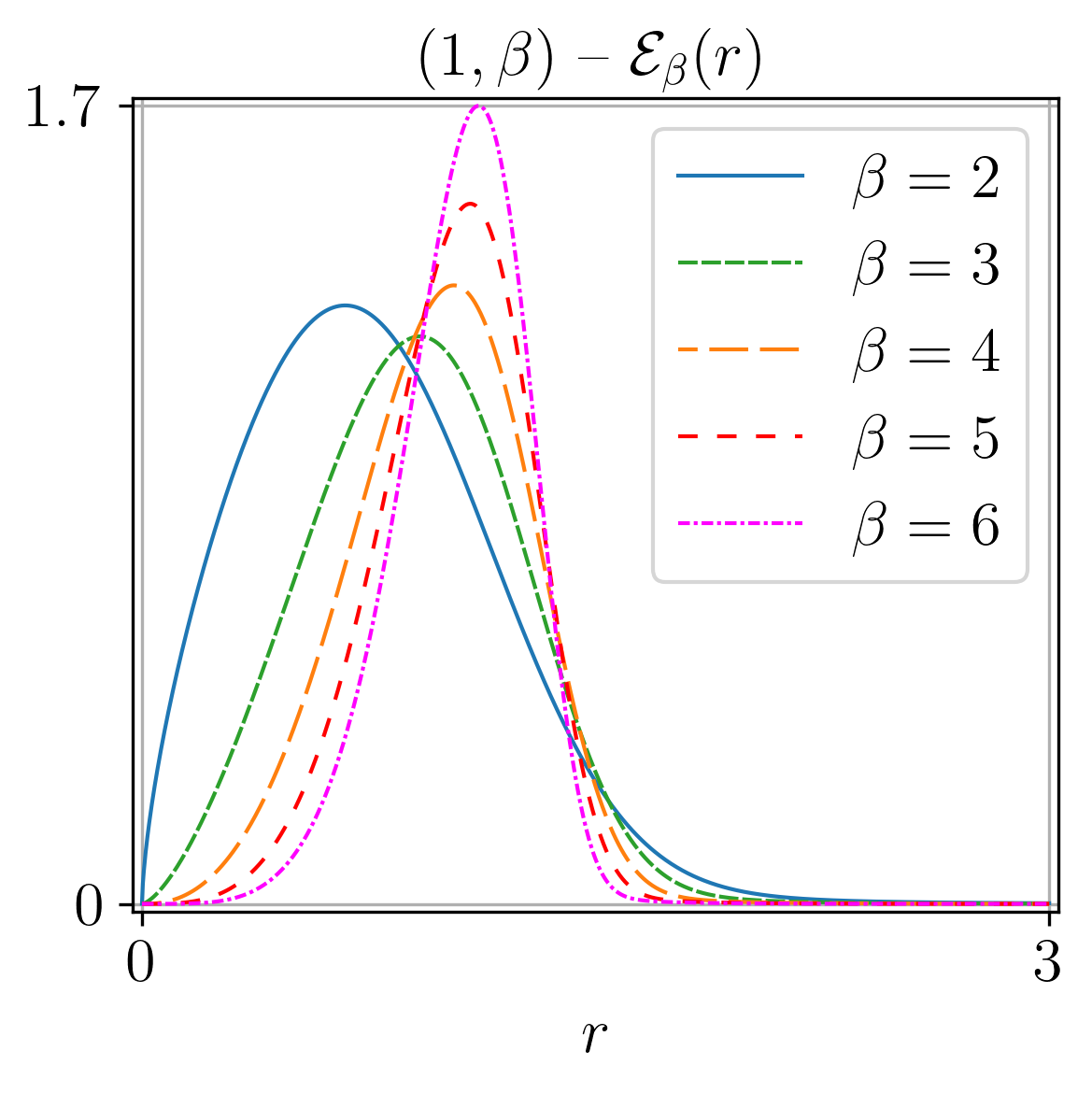}
    \end{subfigure}

       \caption{Field profiles and energy density for $\alpha=1$ and $\beta\in[2,6]$. From left to right: scalar profile $H(r)$, gauge profile $K(r)$, and energy density $\mathcal{E}(r)$.}
           \label{fig:(1,beta_large)}
\end{figure}

\begin{figure}[H]
    \centering
    \begin{subfigure}[c]{0.39\textwidth}
        \centering
        \includegraphics[width=0.9\textwidth]{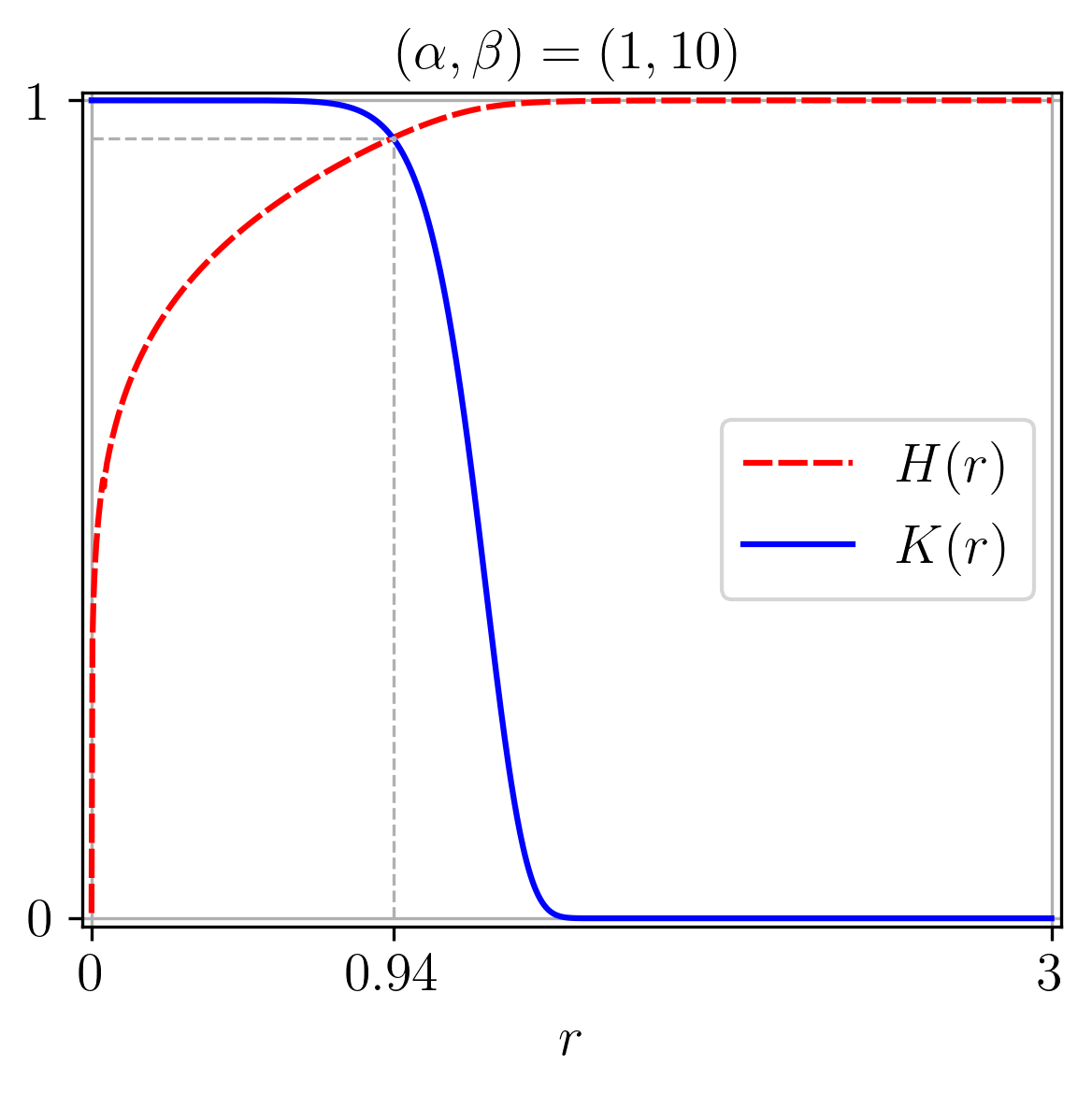}
    \end{subfigure}
    \hspace{0.5cm}
    \begin{subfigure}[c]{0.39\textwidth}
        \centering
        \includegraphics[width=0.9\textwidth]{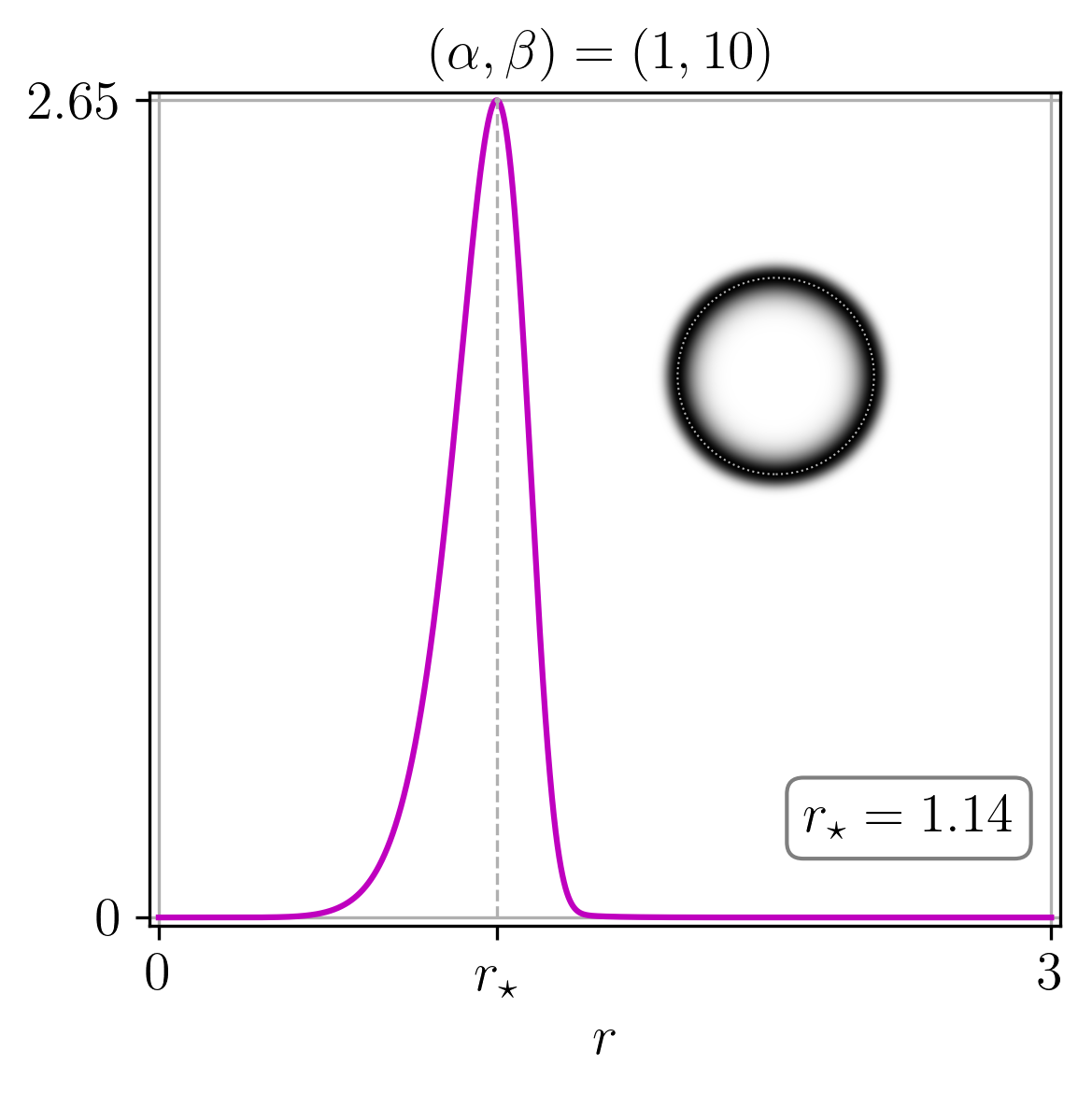}
    \end{subfigure}
    \caption{A shell-like monopole obtained along the analytical line $\alpha=1$ for a large value of $\beta$.}
    \label{fig:(1,10)}
\end{figure}

\begin{figure}[H]
    \centering
    \includegraphics[width=0.48\textwidth]{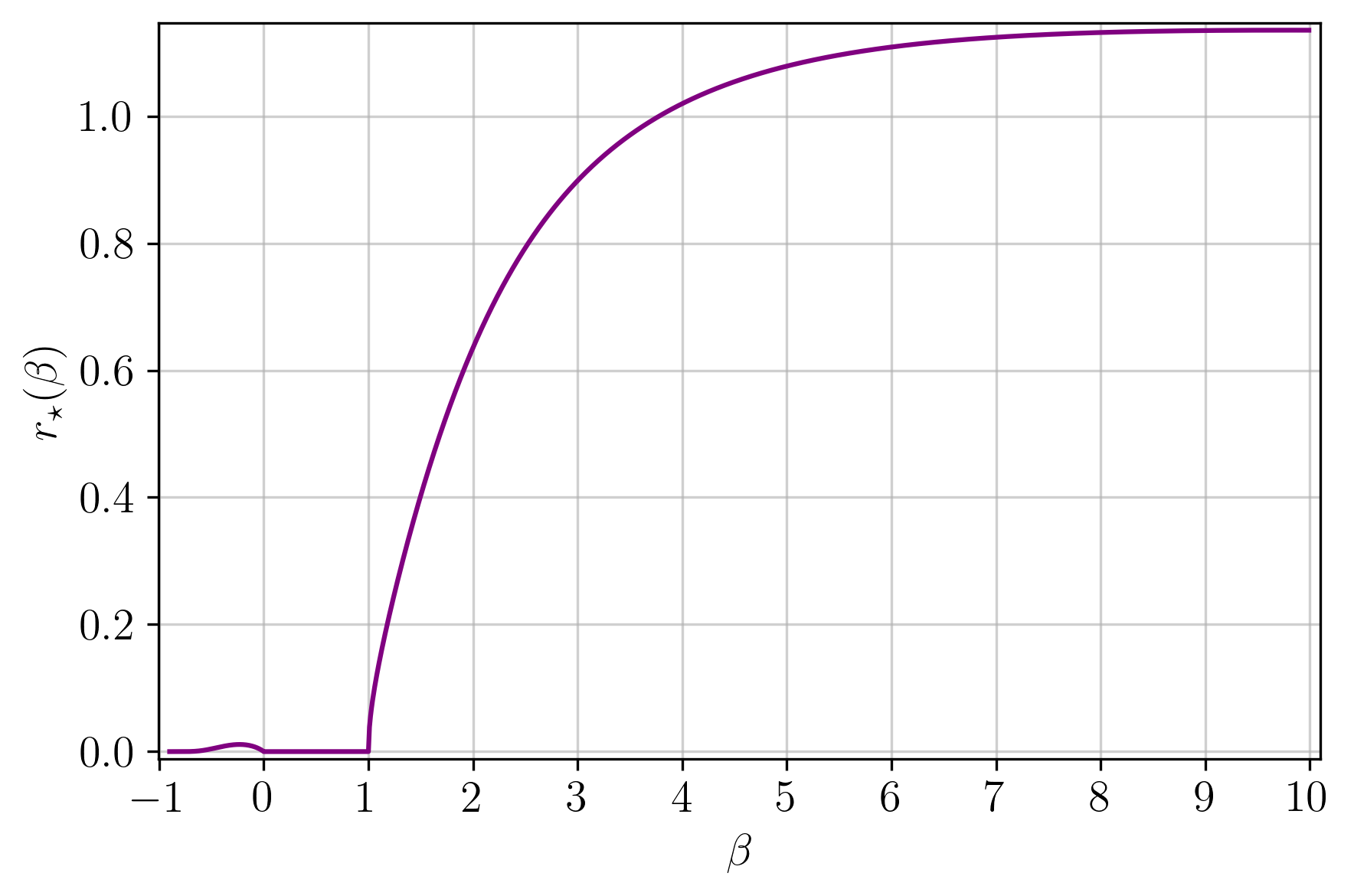}
    \caption{ The cavity radius $r_\star$ as a function of $\beta$.}
    \label{fig:radius}
\end{figure}

The overall behavior of the cavity size is summarized in Fig.~\ref{fig:radius}, which shows the radius $r_{\star}$ of the energy-density maximum as a function of $\beta$. One observes three regimes. First, for negative values of $\beta$, there is a small local maximum of $r_{\star}$. Second, in the interval $\beta\in[0,1]$, the graph becomes flat, reflecting the absence of a cavity and the persistence of core-like monopoles. Third, for sufficiently large positive $\beta$, the radius $r_{\star}$ grows asymptotically, forming increasingly extended hollow shells.

Thus, even within the analytically solvable line $\alpha=1$, the model displays a wide variety of monopole configurations. Depending on the value of $\beta$, one finds compact core-like monopoles, broadened monopoles, and hollow shell-like monopoles. The parameter $\beta$ therefore controls how the energy is redistributed in space, interpolating between configurations concentrated near the origin and configurations whose energy is localized on a spherical shell.
 
To explore the remaining constant-$\alpha$ lines, however, one must solve the BPS equations numerically. This requires solving the coupled non-autonomous first-order system~(\ref{eq:bps}), subject to the boundary conditions~(\ref{eq:bc}) at the origin and at infinity. Since these conditions are inherited from the underlying second-order Euler-Lagrange equations, the resulting first-order problem is effectively overdetermined. Moreover, the singular behavior of the equations at $r=0$ obstructs a naive numerical implementation. In Appendix~\ref{apx:asymp}, we explain how this difficulty can be overcome by replacing the original boundary conditions with effective asymptotic data imposed near the endpoints, while in Appendix~\ref{apx:axl_func}, we introduce auxiliary functions that are regular at the origin, allowing the BPS system to be reformulated in a way that its singular behavior reduces to a simple pole, thus making it suitable for numerical integration.

\subsection{Lines \texorpdfstring{$\alpha>1$}{alpha>1}}

\begin{figure}[H]
    \centering
    \renewcommand{\arraystretch}{1.8}
    \begin{tabular}{@{} c @{\hspace{0.01\textwidth}} c @{\hspace{0.04\textwidth}} c @{\hspace{0.01\textwidth}} c @{}}
        \includegraphics[width=0.235\textwidth]{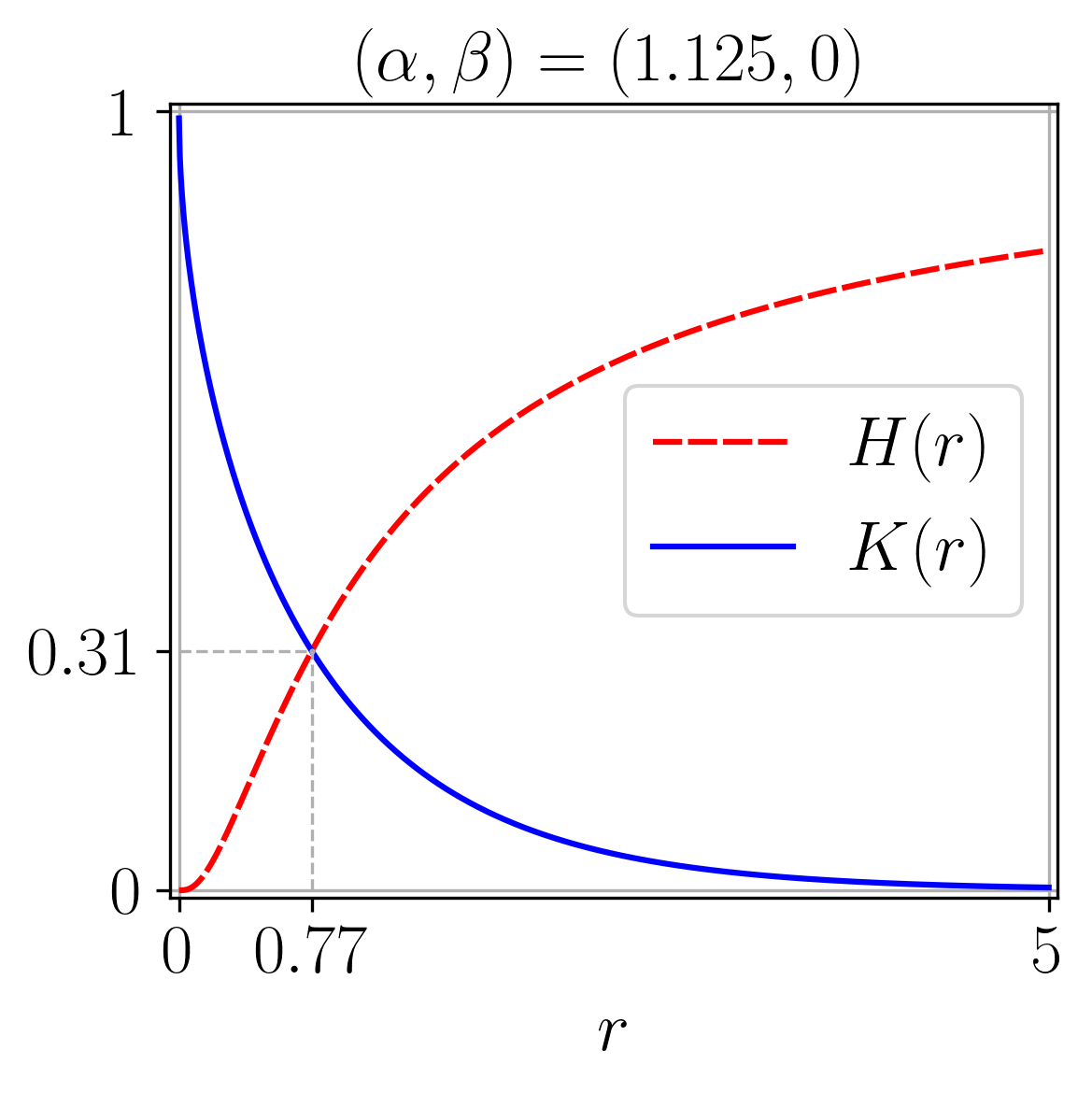} &
        \includegraphics[width=0.235\textwidth]{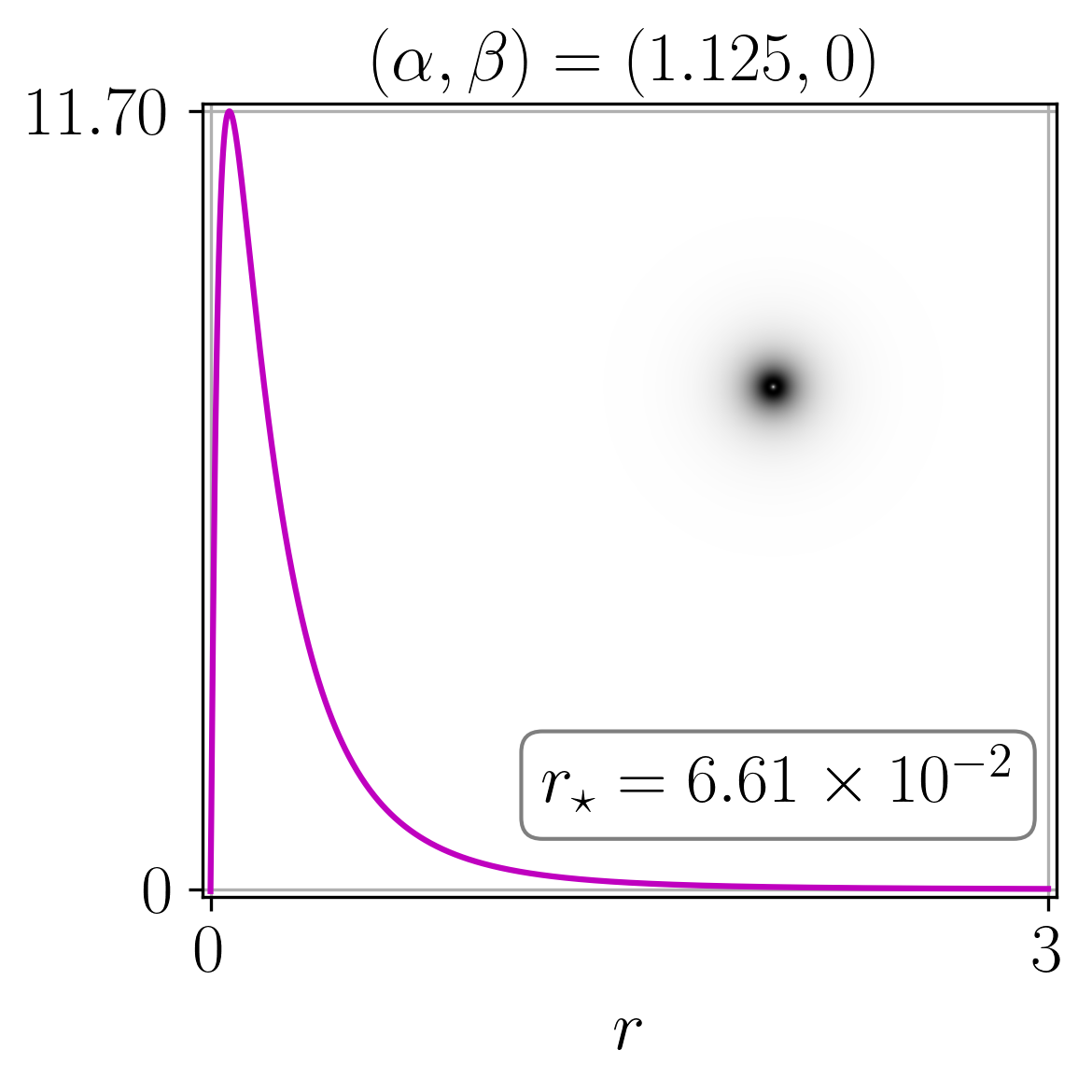} &
        \includegraphics[width=0.235\textwidth]{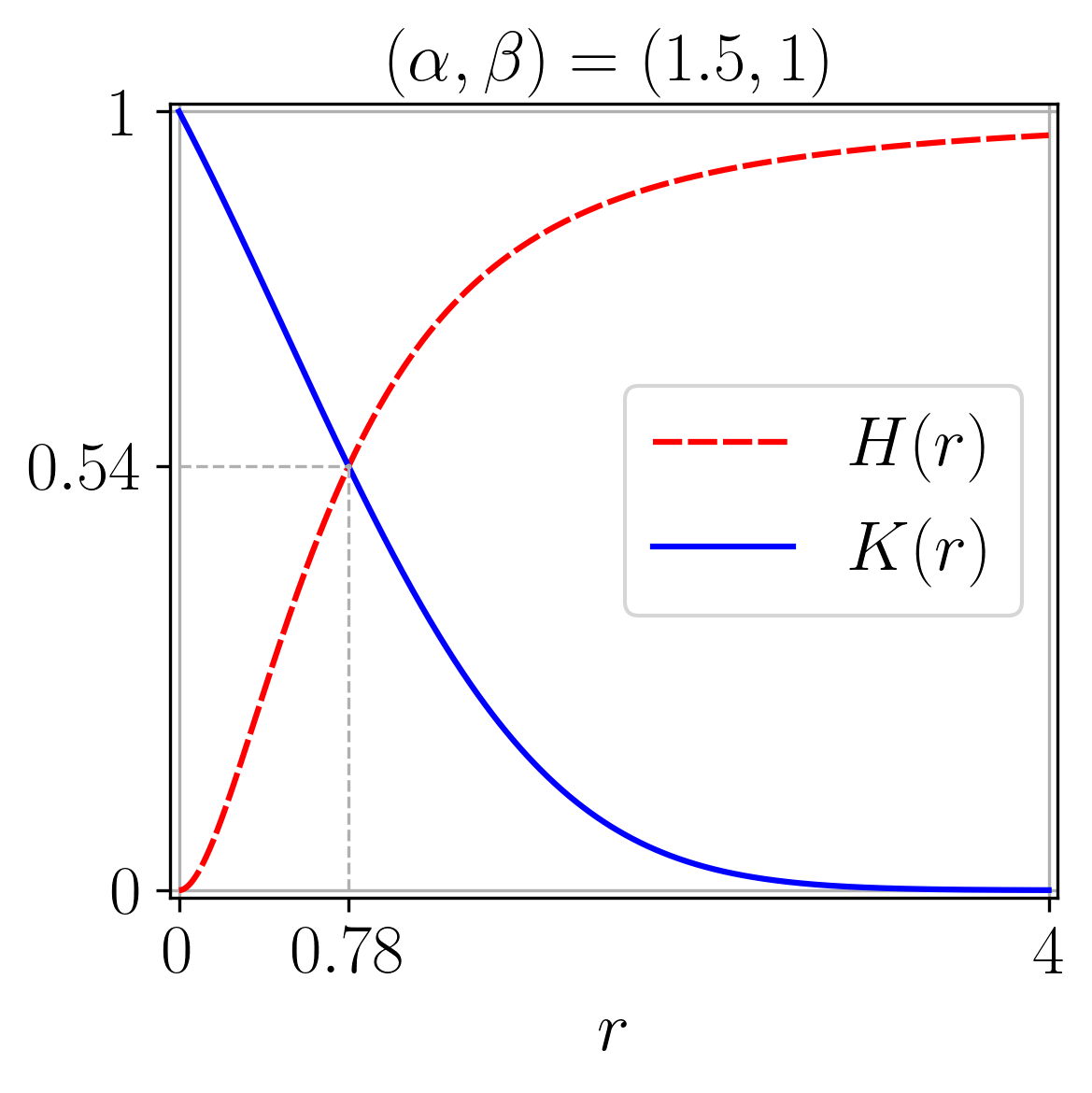} &
        \includegraphics[width=0.235\textwidth]{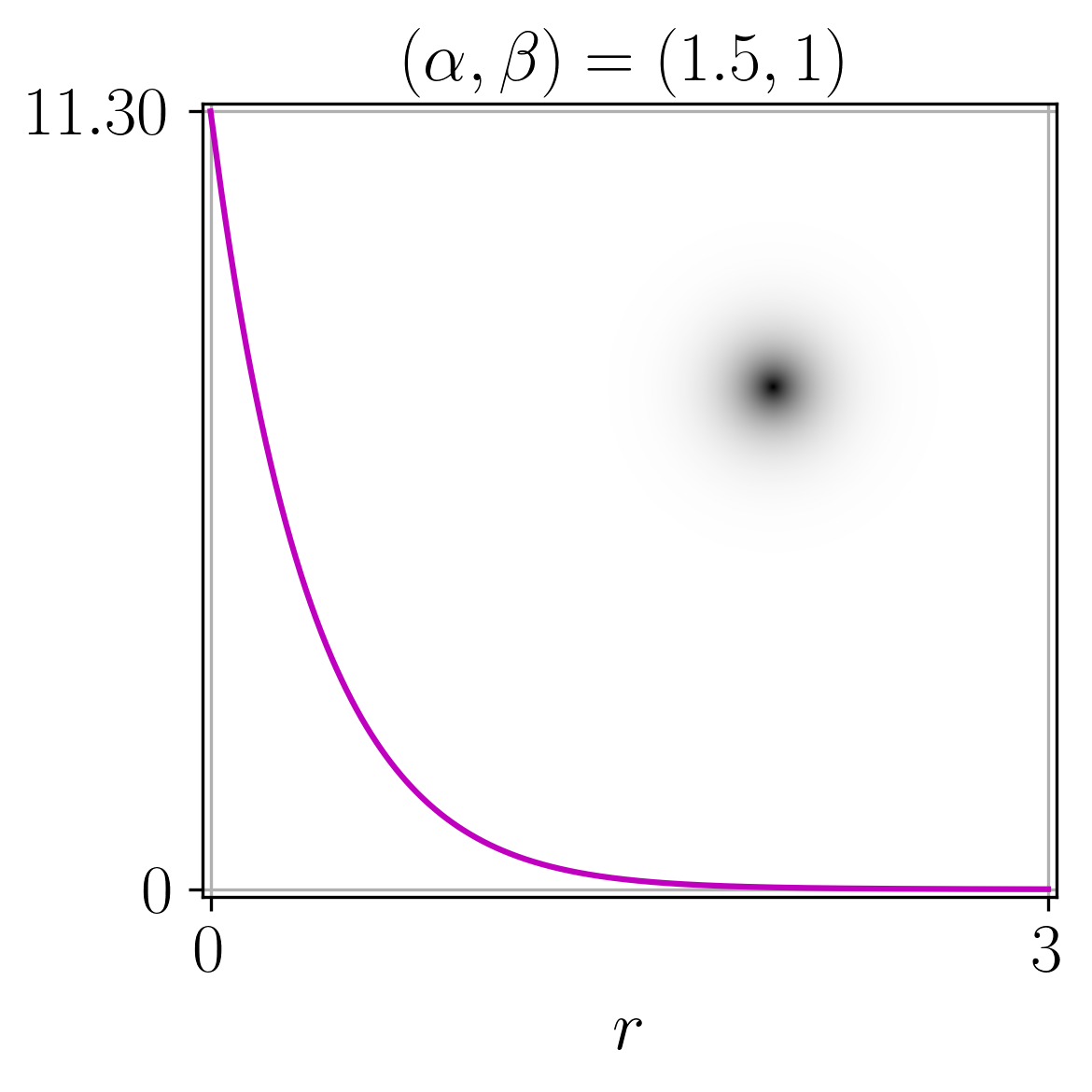} \\
        
        \includegraphics[width=0.235\textwidth]{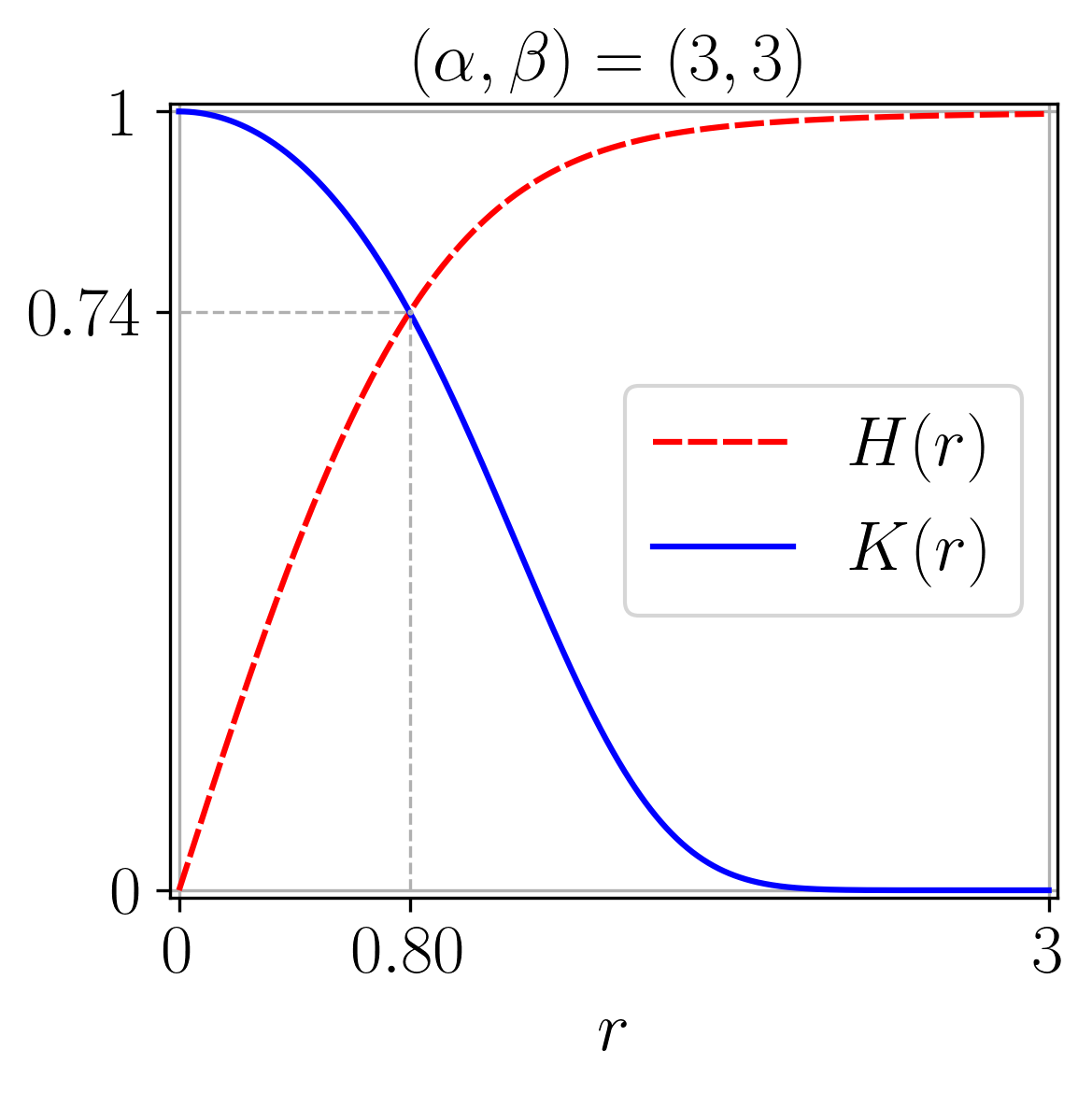} &
        \includegraphics[width=0.235\textwidth]{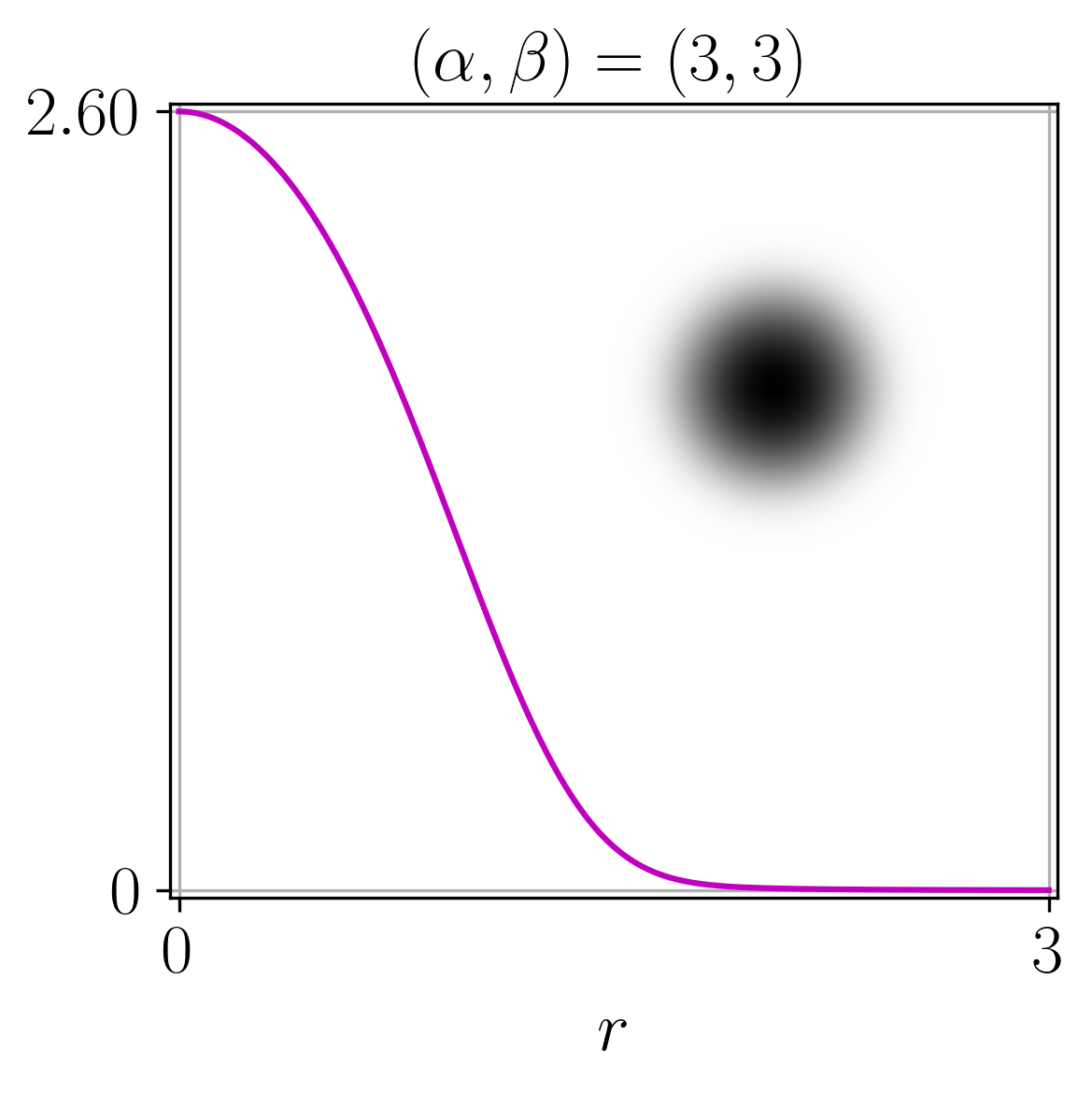} &
        \includegraphics[width=0.235\textwidth]{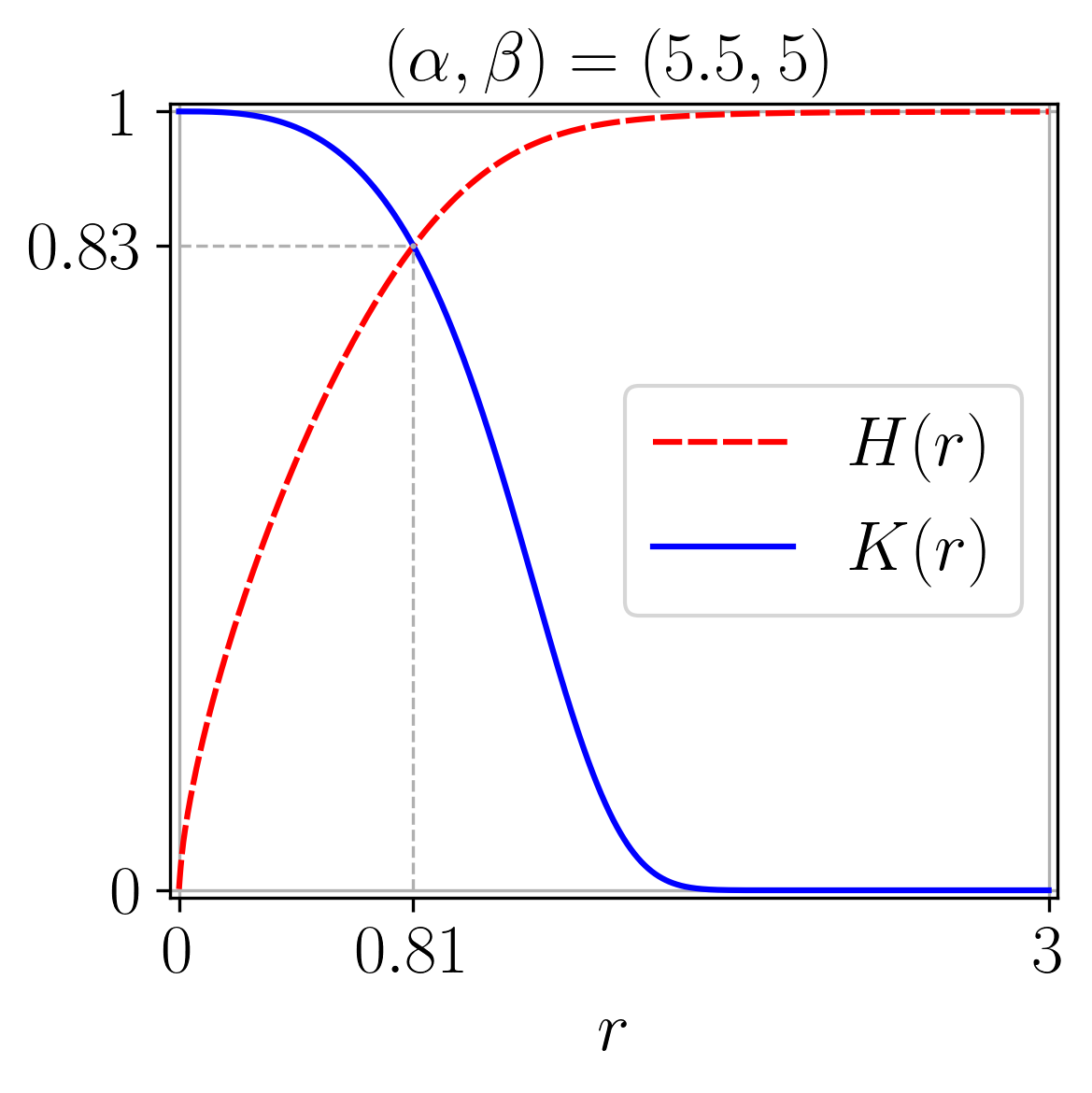} &
        \includegraphics[width=0.235\textwidth]{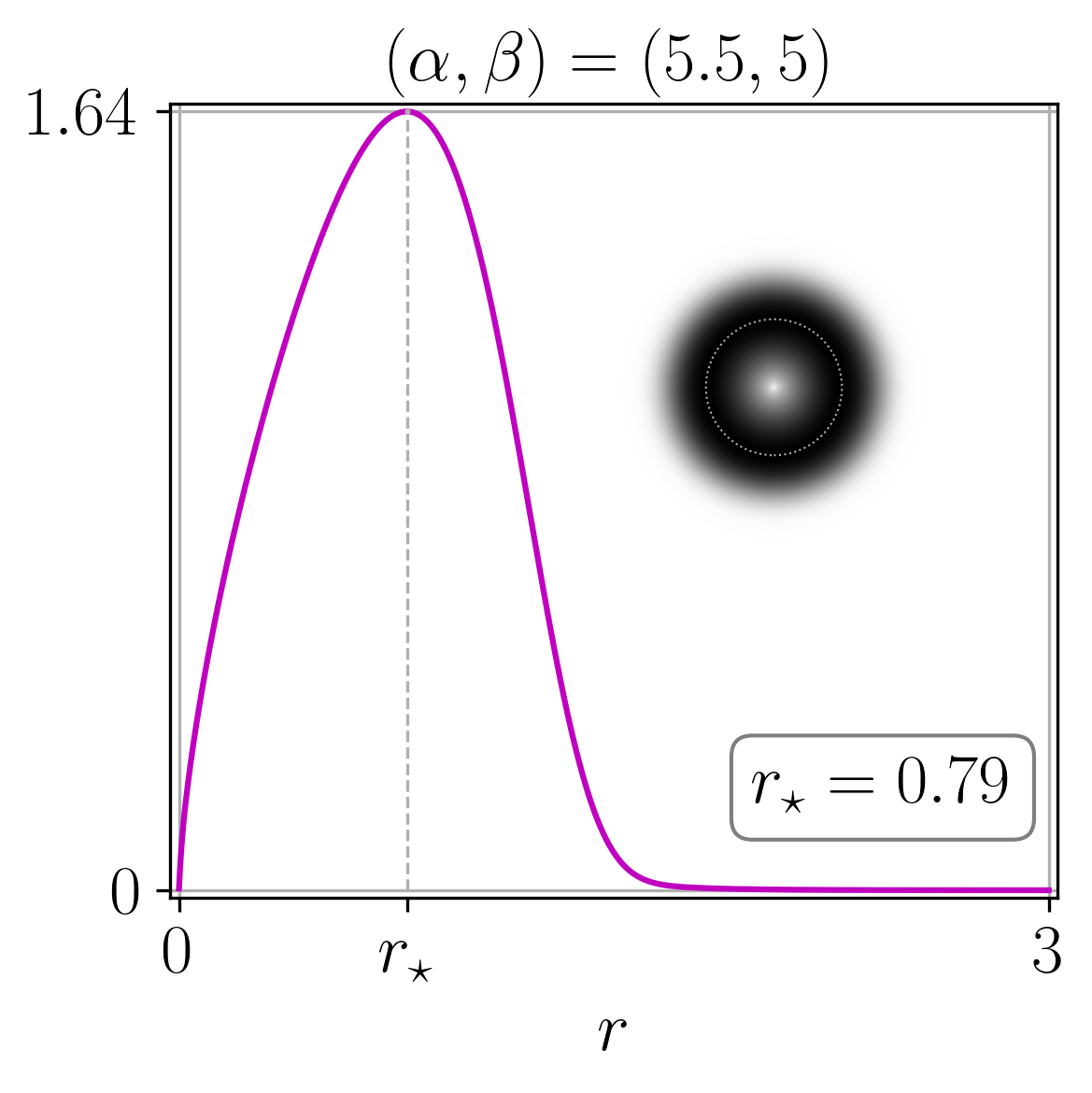}
    \end{tabular}

    \caption{BPS monopole solutions along the parabola $\alpha(\beta)=1+(\beta+1)^2/8$ for $\beta=0,1,3,$ and $5$.}
    \label{fig:along_parabola}
\end{figure}

Above the analytically solvable line $\alpha=1$, the BPS equations no longer decouple and must be solved numerically. Nevertheless, valuable insight into the structure of the solutions can be gained by exploring the admissible region of the $(\alpha,\beta)$ plane identified in Sec.~\ref{sec:pfunc}. In particular, it is natural to begin by following the boundary parabola
\[
\alpha(\beta)=1+\frac{(\beta+1)^2}{8},
\]
which forms the upper edge of the domain shown in Fig.~\ref{fig:domain}. Figure~\ref{fig:along_parabola} displays representative monopole configurations at selected points along this parabola, illustrating how the structure of the solutions evolves as one moves away from the line $\alpha=1$ into the region $\alpha>1$.

The qualitative similarity with the behavior observed along the line $\alpha=1$ (see Fig.~\ref{fig:(1,beta)}) is remarkable. For instance, starting from the point $(\alpha,\beta)=(1.125,0)$, which lies slightly above $\alpha=1$, one observes the presence of a small central cavity. As $\beta$ increases, this cavity is gradually filled, giving rise to compact, core-like monopole configurations.  Upon further increasing $\beta$, the monopoles become more spatially extended,  until a central cavity reappears, giving rise to hollow configurations.

To explore the behavior away from the boundary parabola, let us fix a point $(\alpha(\beta_0),\beta_0)$ on the parabola and consider the horizontal line
\[
\alpha(\beta)=\alpha(\beta_0),
\]
parametrized by $\beta$. In other words, we move along a constant-$\alpha$ line extending into the interior of the admissible region.
For all values of $\alpha(\beta_0)$ that we have examined, the BPS solutions along such constant-$\alpha$ lines exhibit the same qualitative pattern described above. Namely, as $\beta$ increases, the system evolves from configurations with a small or absent cavity to increasingly hollow configurations, eventually approaching shell-like monopoles. This behavior is illustrated in Fig.~\ref{fig:(1.5,beta)}, which shows the evolution of the profiles for $\alpha=1.5$.

\begin{figure}[H]
    \centering
    \begin{subfigure}[c]{0.315\textwidth}
        \centering
        \includegraphics[width=\textwidth]{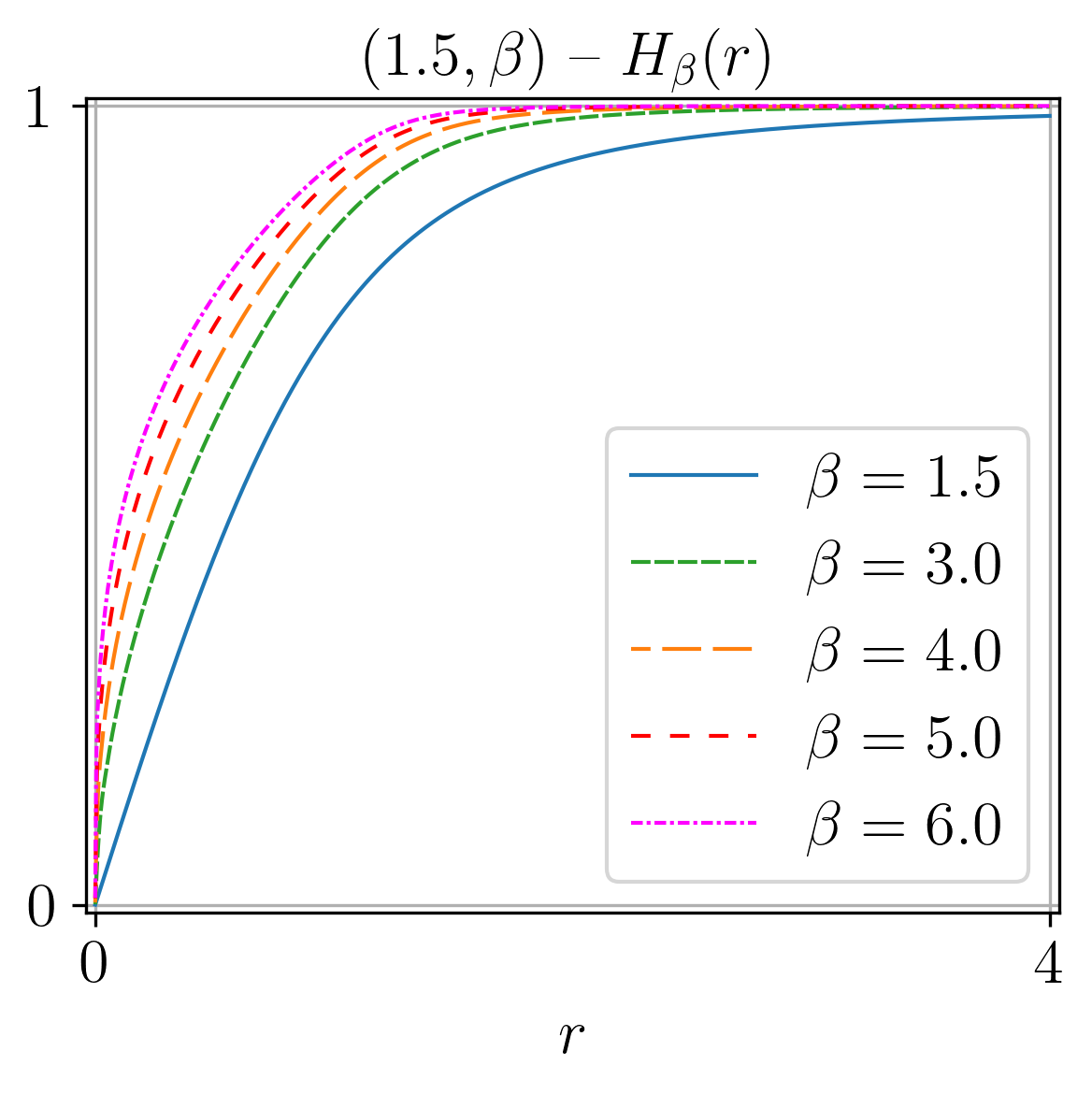}
    \end{subfigure}
    \hspace{6pt}
    \begin{subfigure}[c]{0.315\textwidth}
        \centering
        \includegraphics[width=\textwidth]{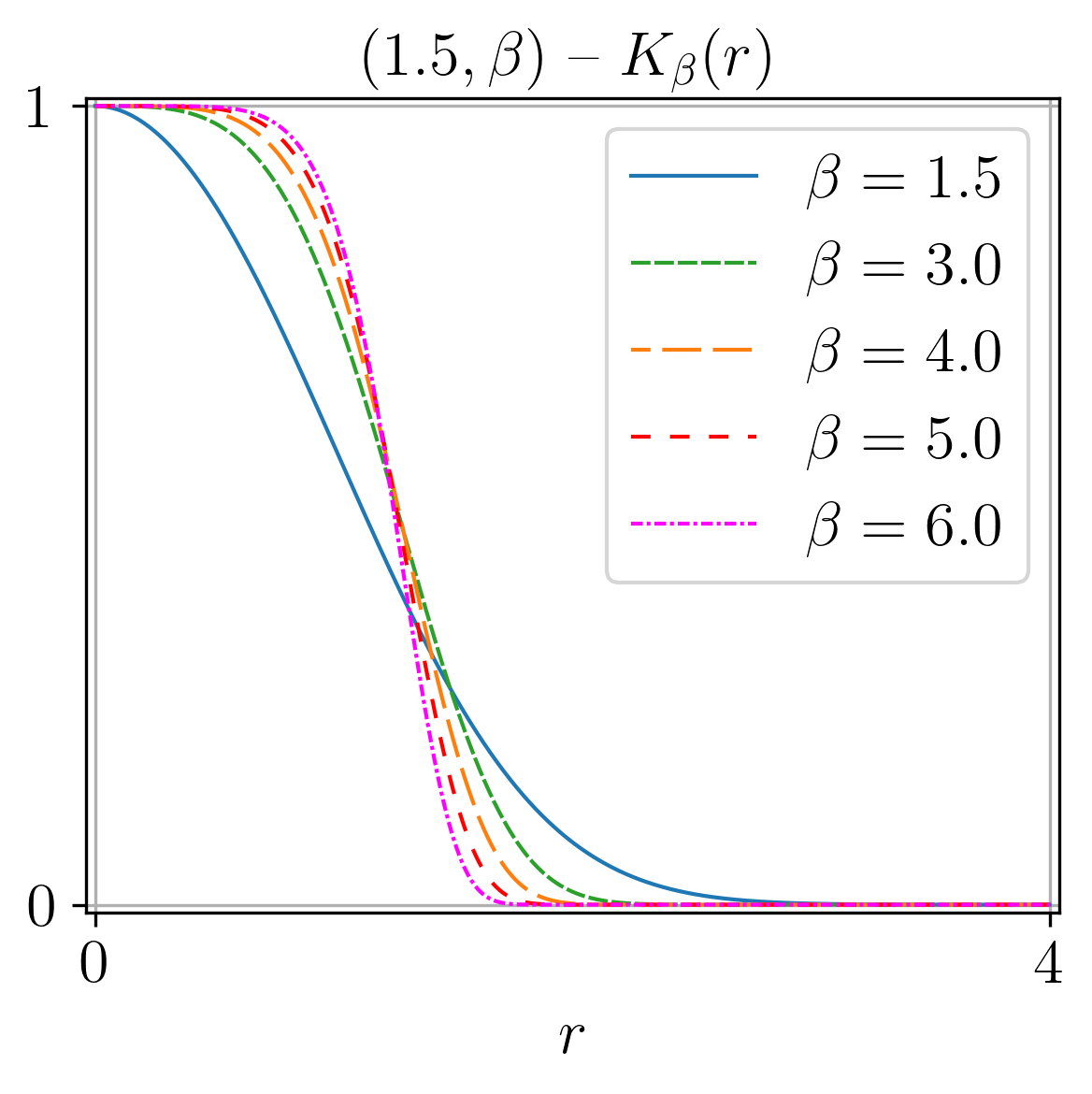}
    \end{subfigure}
    \hspace{6pt}
    \begin{subfigure}[c]{0.315\textwidth}
        \centering
        \includegraphics[width=\textwidth]{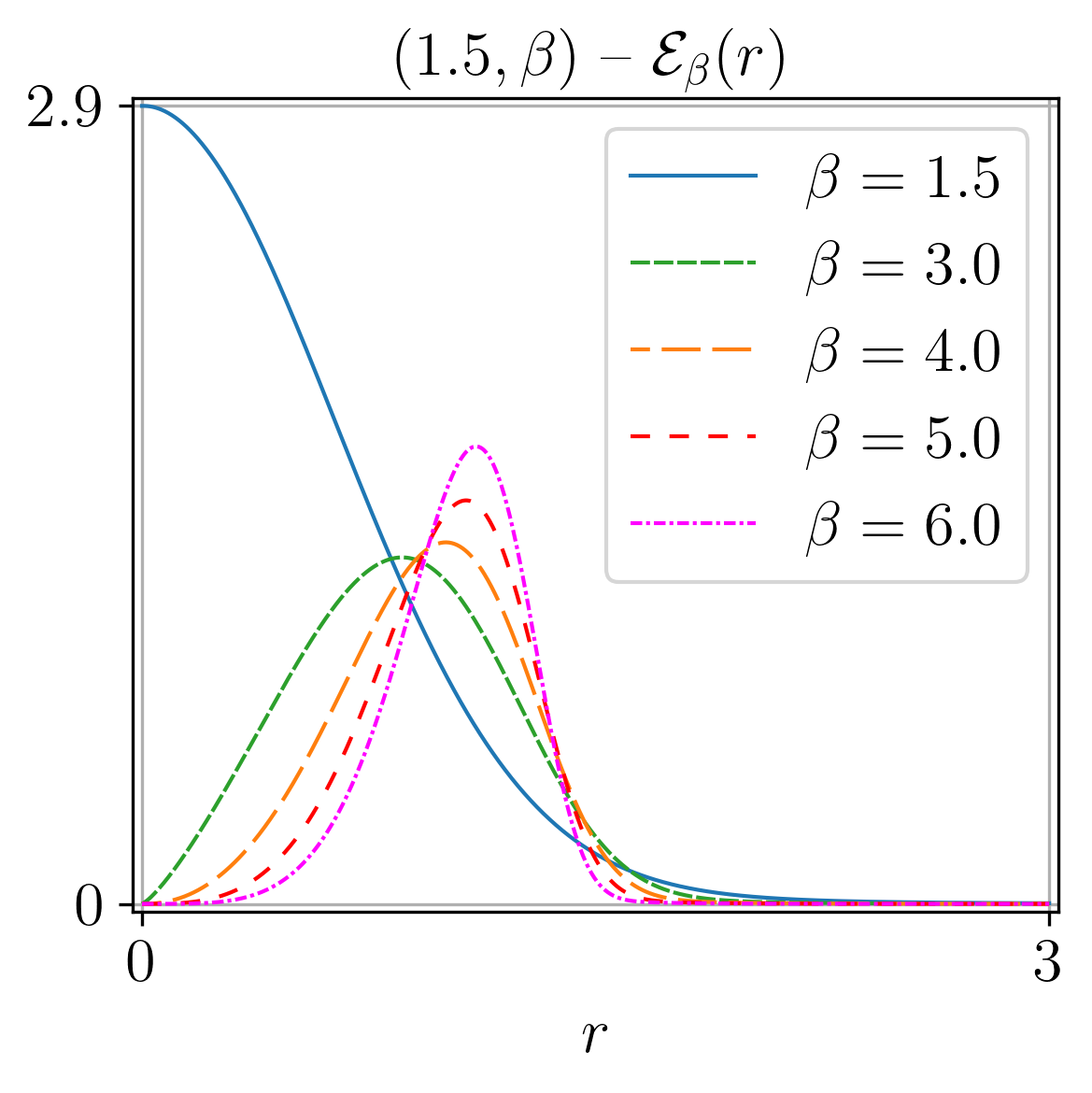}
    \end{subfigure}
        \caption{Monopole profiles and energy densities along the line $\alpha=1.5$ for several values of $\beta$.}
    \label{fig:(1.5,beta)}
\end{figure}

In Appendix~\ref{apx:asymp}, we show that, for $\alpha>1$, the BPS equations admit two distinct classes of solutions at each point of the parameter space, characterized by the exponents $m_+$ and $m_-$. Along the boundary parabola, these two branches coincide, $m_+=m_-$, and the distinction between them disappears. Away from the parabola, however, the two branches lead to solutions with quantitatively different behaviors. In particular, they differ in how the energy density redistributes and concentrates around the characteristic radius $r_{\star}$ as $\beta$ increases.

\begin{figure}[H]
    \centering
    \begin{subfigure}[c]{0.315\textwidth}
        \centering
        \includegraphics[width=\textwidth]{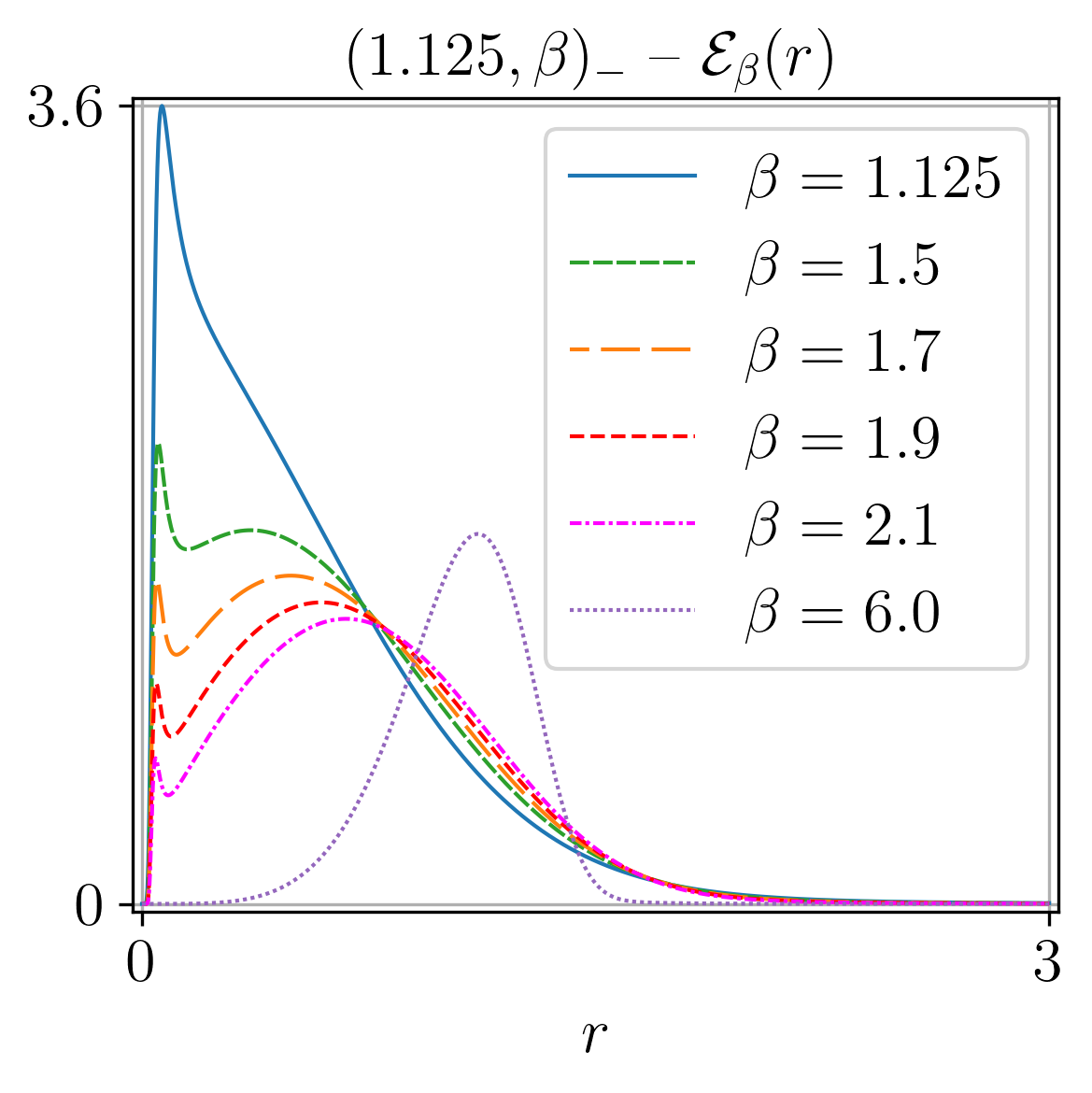}
    \end{subfigure}
    \hspace{6pt}
    \begin{subfigure}[c]{0.315\textwidth}
        \centering
        \includegraphics[width=\textwidth]{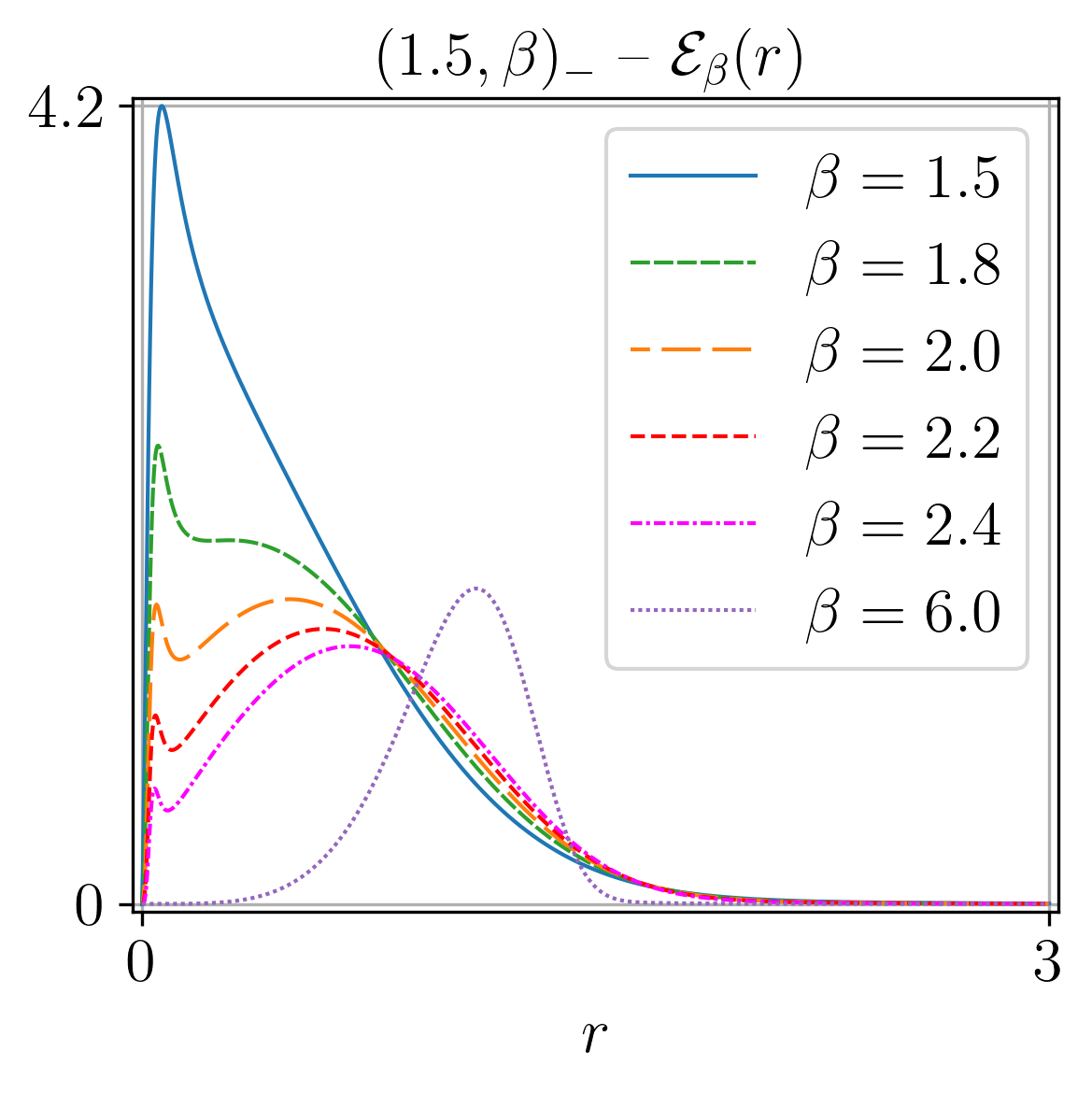}
    \end{subfigure}

    \caption{Comparison of energy densities along constant-$\alpha$ lines for $\alpha=1.125$ and $\alpha=1.5$.}
    \label{fig:alpha_minus_1.125_1.5}
\end{figure}

The solutions obtained from $m_+$ behave as described above. By contrast, the $m_-$ branch leads to a different pattern at smaller values of $\beta$, as illustrated in Fig.~\ref{fig:alpha_minus_1.125_1.5}, which shows sequences of energy density profiles along the lines $\alpha=1.125$ and $\alpha=1.5$.

\begin{figure}[H]
    \centering
    \begin{subfigure}[c]{0.31\textwidth}
        \centering
        \includegraphics[width=\textwidth]{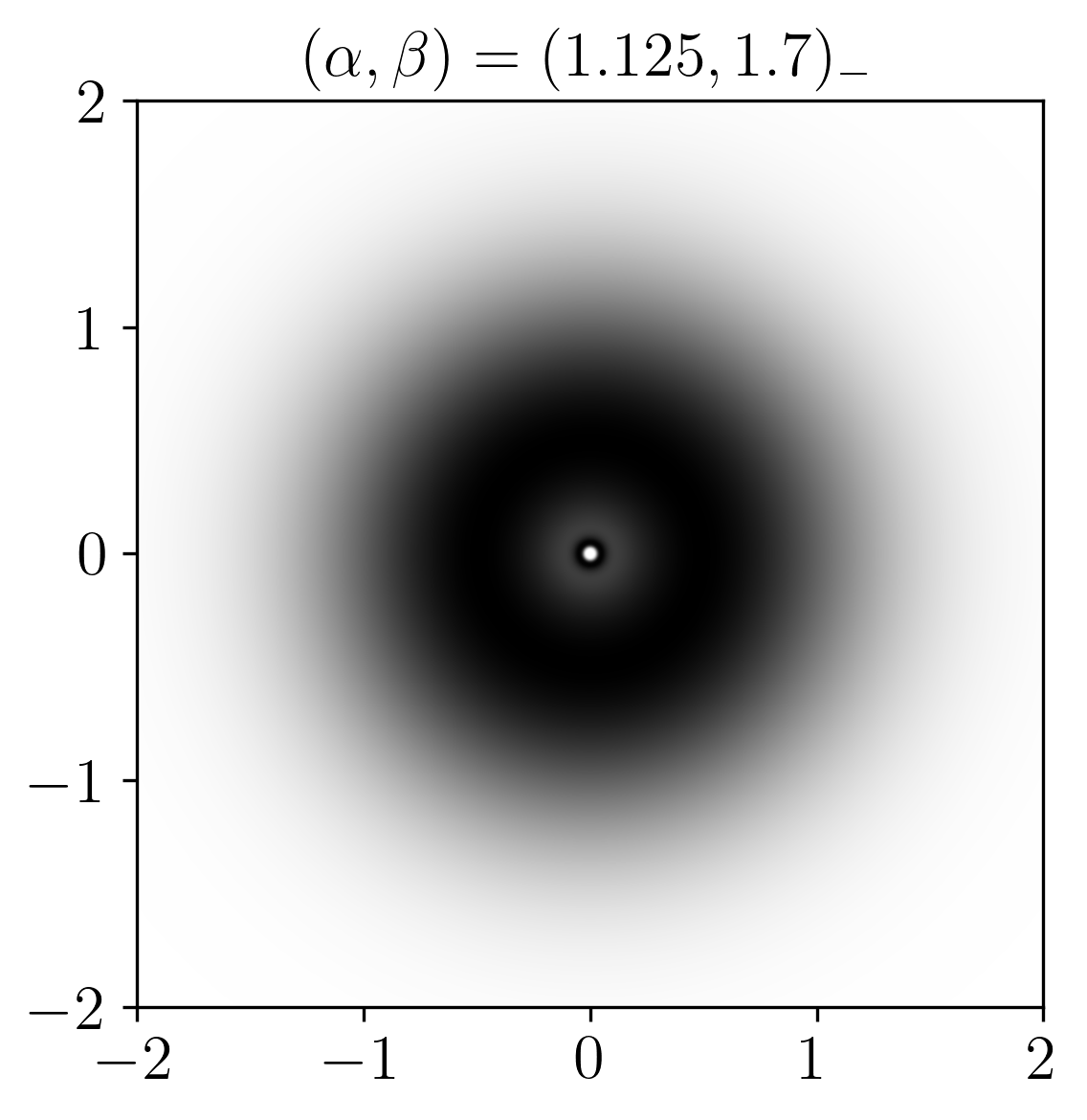}
    \end{subfigure}
    \hspace{6pt}
    \begin{subfigure}[c]{0.31\textwidth}
        \centering
        \includegraphics[width=\textwidth]{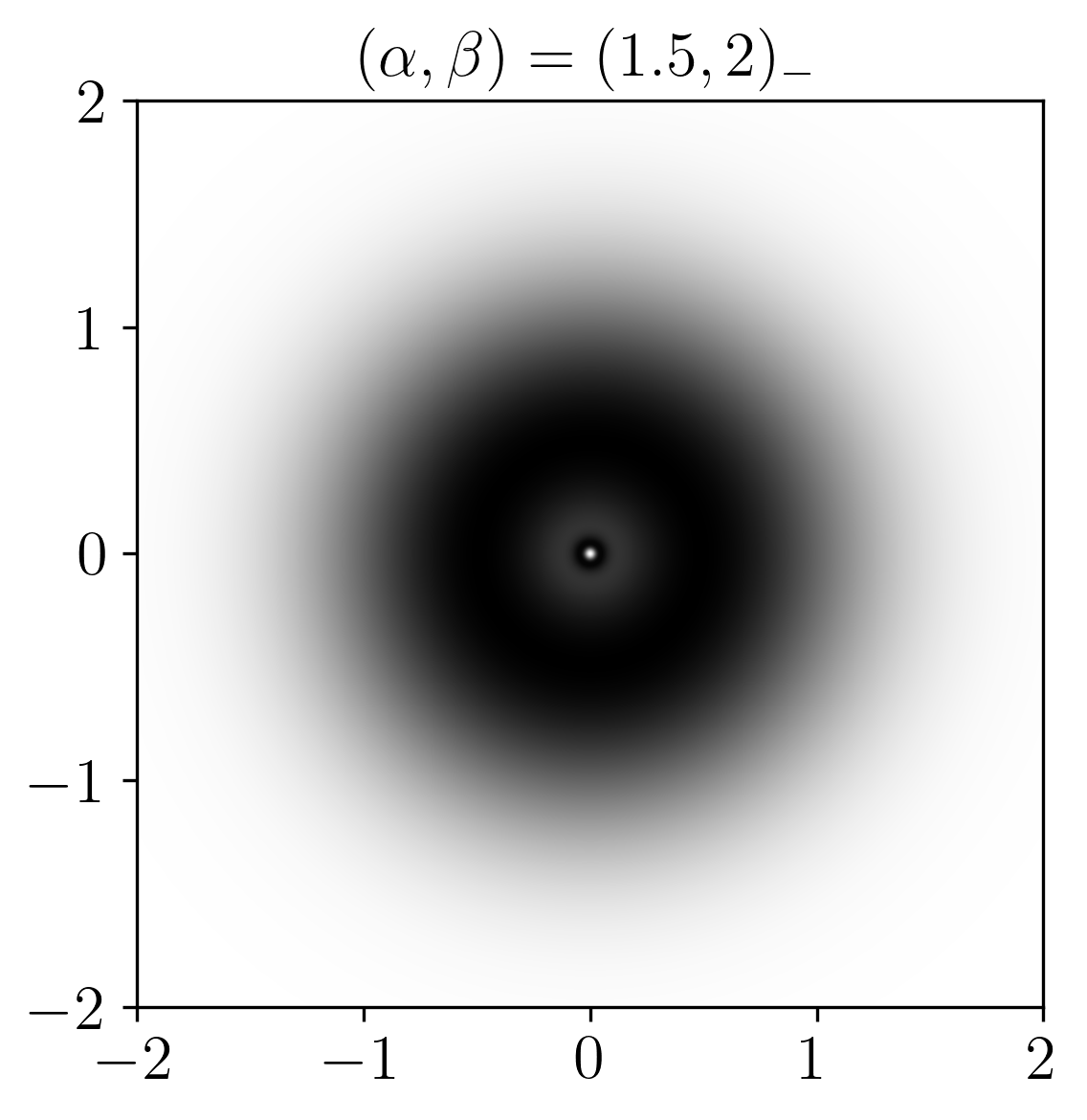}
    \end{subfigure}

    \caption{Energy densities for two examples of multi-shell monopoles, one with $\alpha=1.125$ and $\beta=1.7$, and the other with $\alpha=1.5$ and $\beta=2$.}
    \label{fig:energies_2D}
\end{figure}

In this case, one observes the emergence of configurations in which a secondary local maximum develops in the tail of the energy density, at $r_{\star}+\Delta r$. As $\beta$ increases, this local maximum eventually becomes the global one. Consequently, the $m_-$ solutions enrich the space of configurations by introducing monopoles with a more intricate internal structure, characterized by the presence of multiple peaks in the energy density, as illustrated in Fig.~\ref{fig:energies_2D}.

At larger values of $\beta$, however, the distinction between the $m_+$ and $m_-$ branches becomes progressively less pronounced. In this case the results point to a robust and universal pattern for both classes of solutions: any trajectory in the admissible region that moves away from the vertex $(1,-1)$ of the parabola leads, for sufficiently large $\beta$, to shell-like monopole configurations. In particular, the asymptotic regions of the parameter space, characterized by
\[
\alpha\gg1
\qquad\text{or}\qquad
\beta\gg \sqrt{8(\alpha-1)}-1,
\]
are expected to be populated exclusively by hollow monopoles whose energy density is concentrated on a spherical shell.

\subsection{Lines \texorpdfstring{$\alpha<1$}{alpha<1}}

Below the analytically tractable line $\alpha=1$, we first highlight the homogeneous-medium vertical line $\beta=0$ studied in~\cite{Bazeia-2018-Small}. It is along this line that the standard 't~Hooft--Polyakov BPS monopole is recovered, located at the point $(\alpha,\beta)=(0,0)$. At this point, the BPS equations admit the well-known analytical solution found by Prasad and Sommerfield~\cite{Prasad-1975-Exact},
 \begin{align}
        H(r) &= \frac{1}{\tanh r} - \frac{1}{r},
        \\[3pt]
        K(r) &= \frac{r}{\sinh r}.
        \label{eq:sol0}
    \end{align}
As shown in Fig.~\ref{fig:(0,0)}, this solution exhibits a compact core and a monotonically decreasing energy density.
\begin{figure}[H]
    \centering
    \begin{subfigure}[c]{0.39\textwidth}
        \centering
        \includegraphics[width=0.9\textwidth]{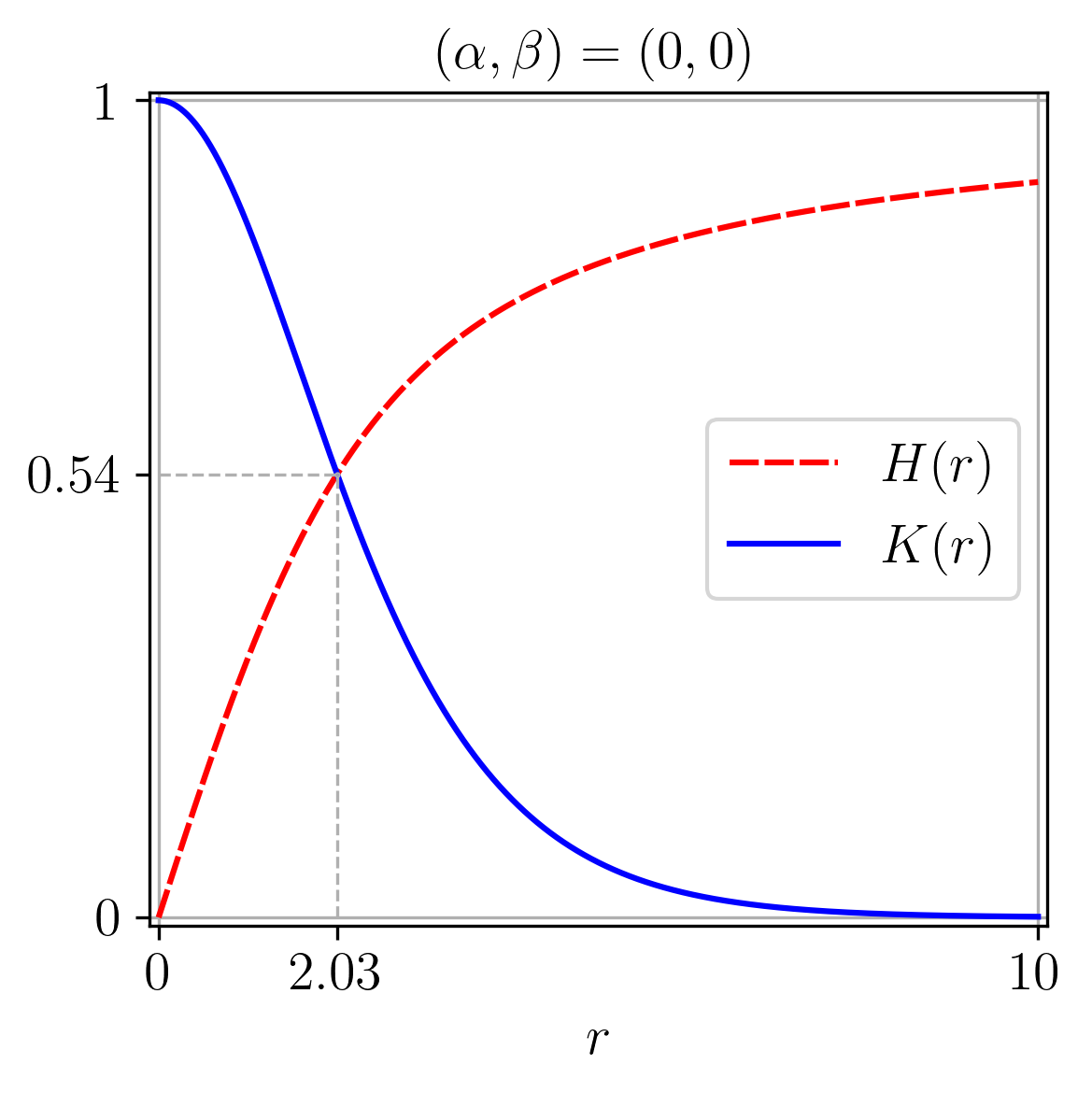}
    \end{subfigure}
    \hspace{0.5cm}
    \begin{subfigure}[c]{0.39\textwidth}
        \centering
        \includegraphics[width=0.9\textwidth]{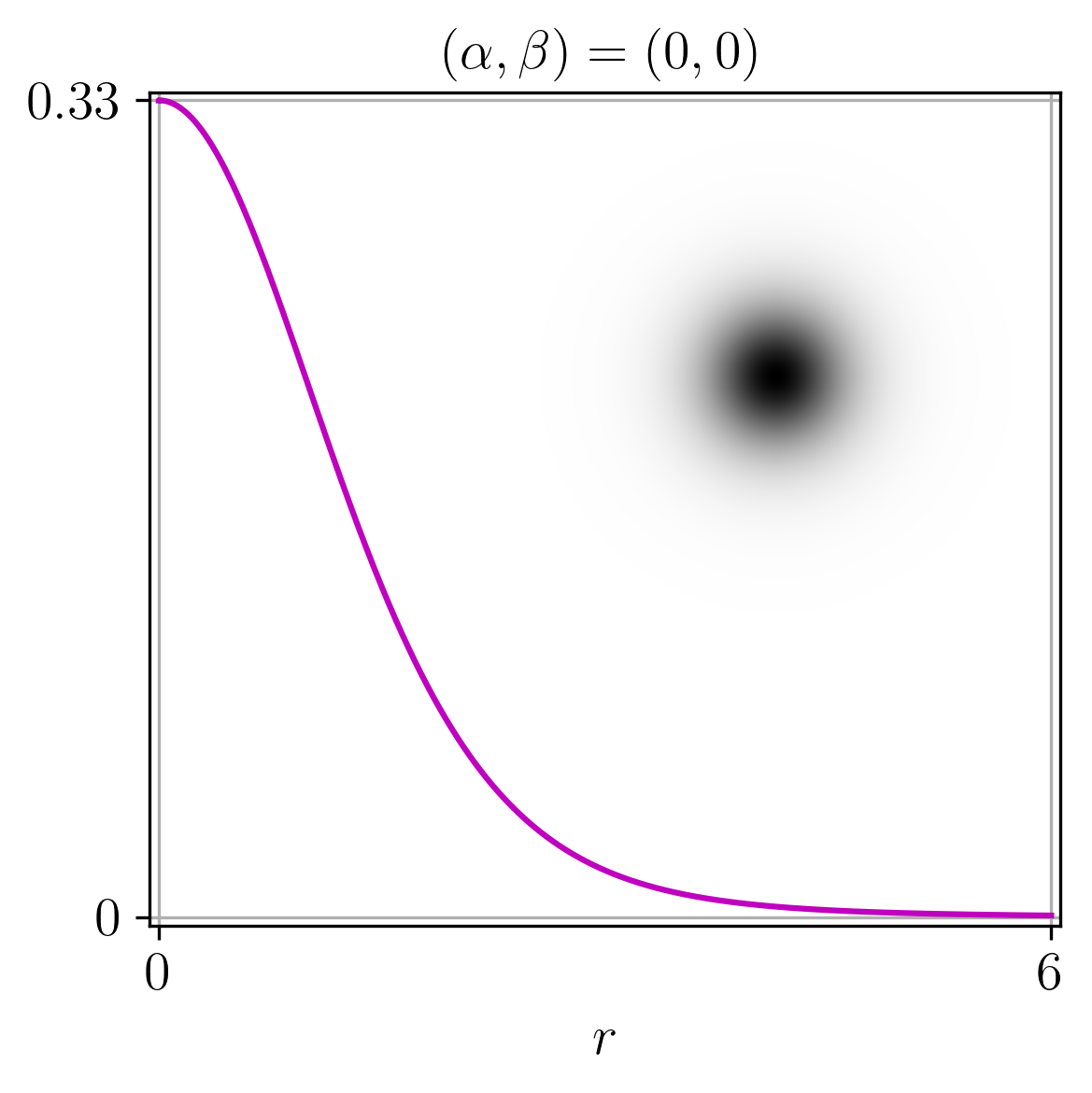}
    \end{subfigure}
    
    \caption{The standard Prasad-Sommerfield monopole at $(\alpha,\beta)=(0,0)$.}
    \label{fig:(0,0)}
\end{figure}

The $\beta=0$ line extends from the point $(1.125,0)$ on the boundary parabola, whose small-cavity monopole solution is shown in Fig.~\ref{fig:along_parabola}. For successively decreasing values of $\alpha$, the monopole configurations along this line evolve in a manner closely analogous to the $m_+$ solutions observed along constant-$\alpha$ horizontal lines in the region $\alpha>1$ as $\beta$ increases. In particular, the initial small-cavity configuration is progressively replaced by compact core-like monopoles, see $(1,0)$ in Fig.~\ref{fig:(1,beta)} and $(0,0)$ in Fig.~\ref{fig:(0,0)}, followed by the emergence of hollow configurations, which eventually dominate the line, as shown in Fig.~\ref{fig:(alpha,0)}.
There is, however, an important distinction. Although the characteristic radius $r_{\star}$ associated with hollow monopoles increases as $\alpha$ decreases, this growth is not accompanied by a corresponding concentration of energy. Instead, the energy density becomes progressively less localized around $r_{\star}$, with its maximum decreasing as it spreads over a larger spatial region.

\begin{figure}[H]
    \centering
    \begin{subfigure}[c]{0.315\textwidth}
        \centering
        \includegraphics[width=\textwidth]{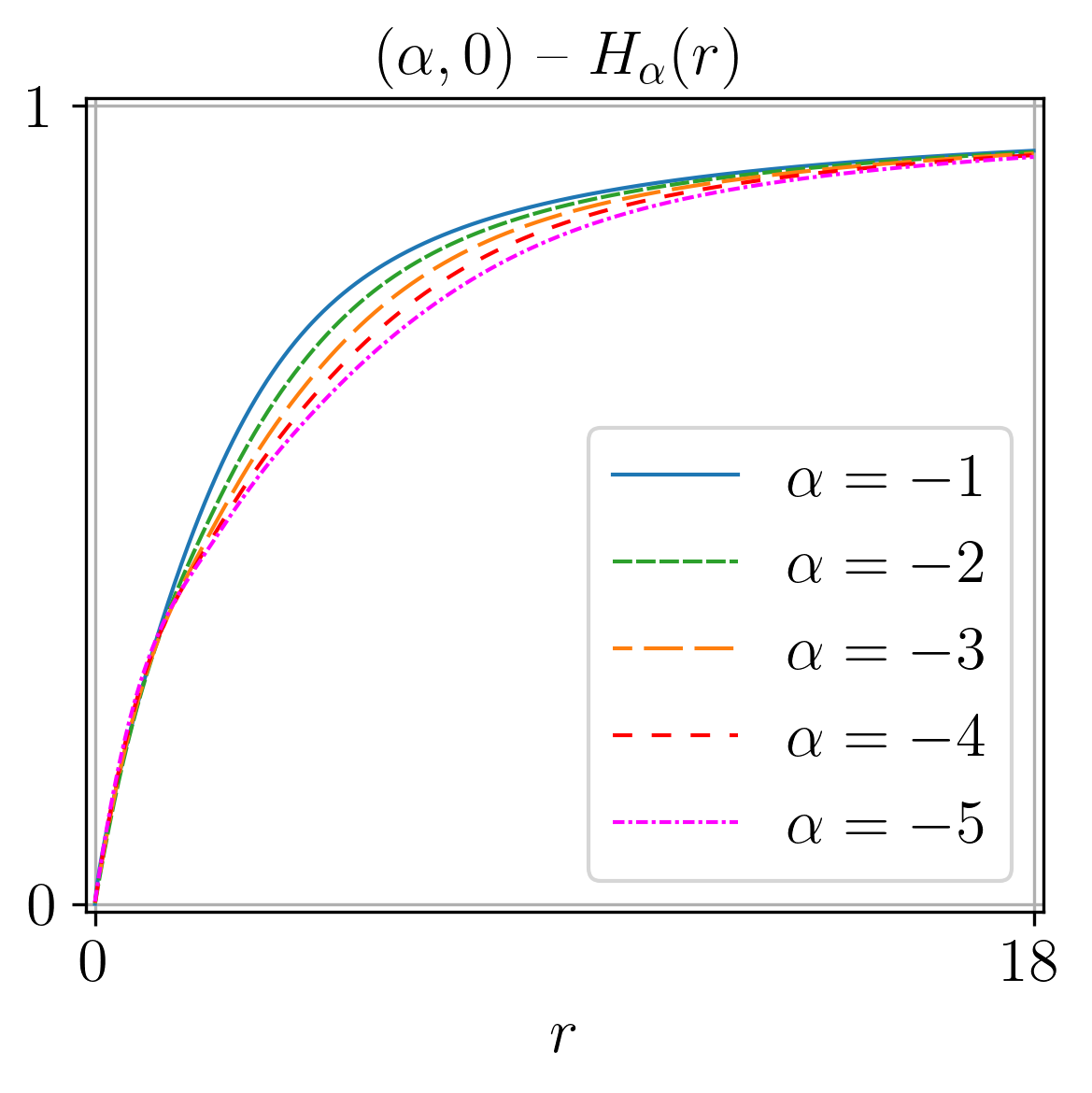}
    \end{subfigure}
    \hspace{6pt}
    \begin{subfigure}[c]{0.315\textwidth}
        \centering
        \includegraphics[width=\textwidth]{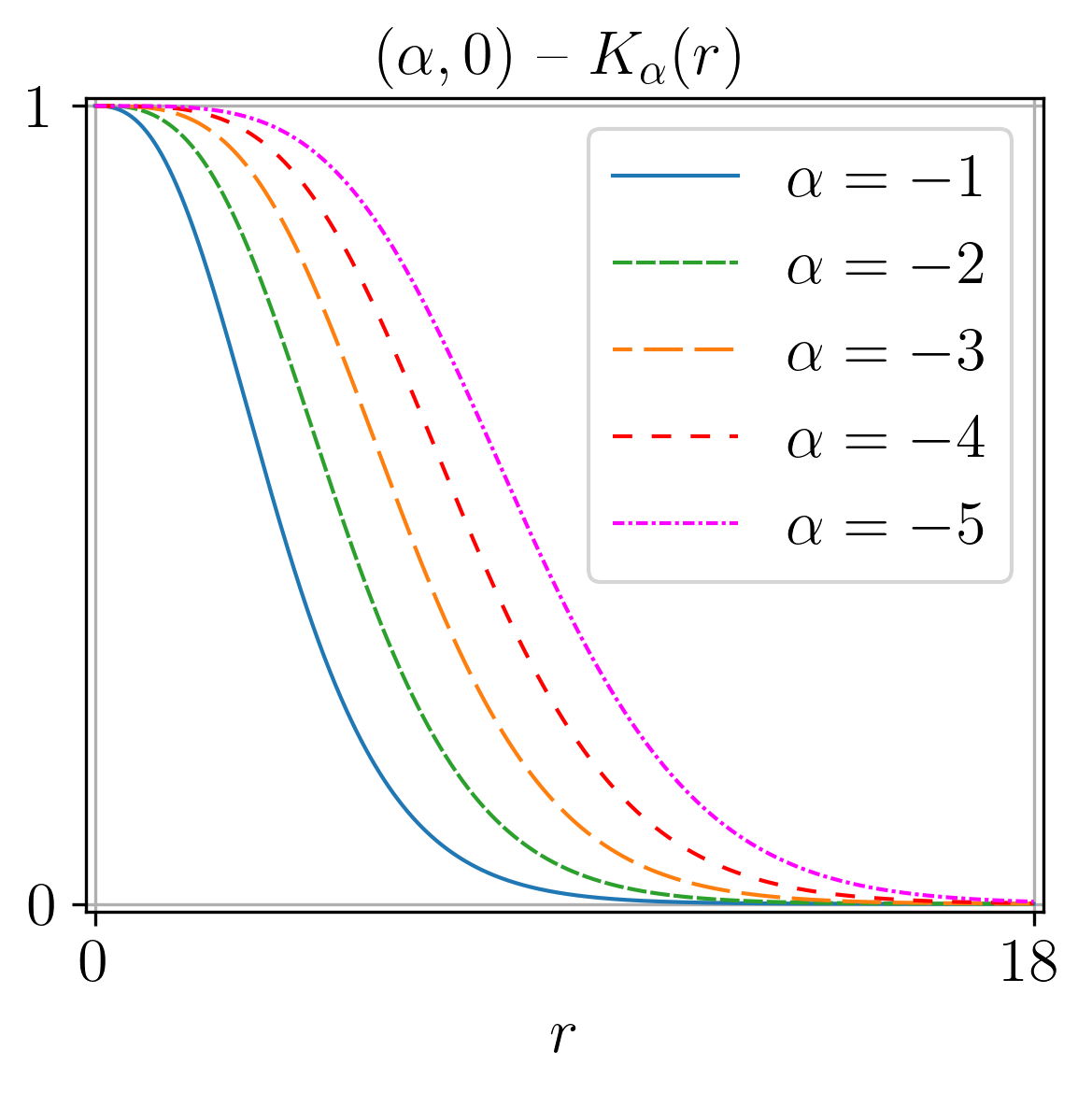}
    \end{subfigure}
    \hspace{6pt}
    \begin{subfigure}[c]{0.315\textwidth}
        \centering
        \includegraphics[width=\textwidth]{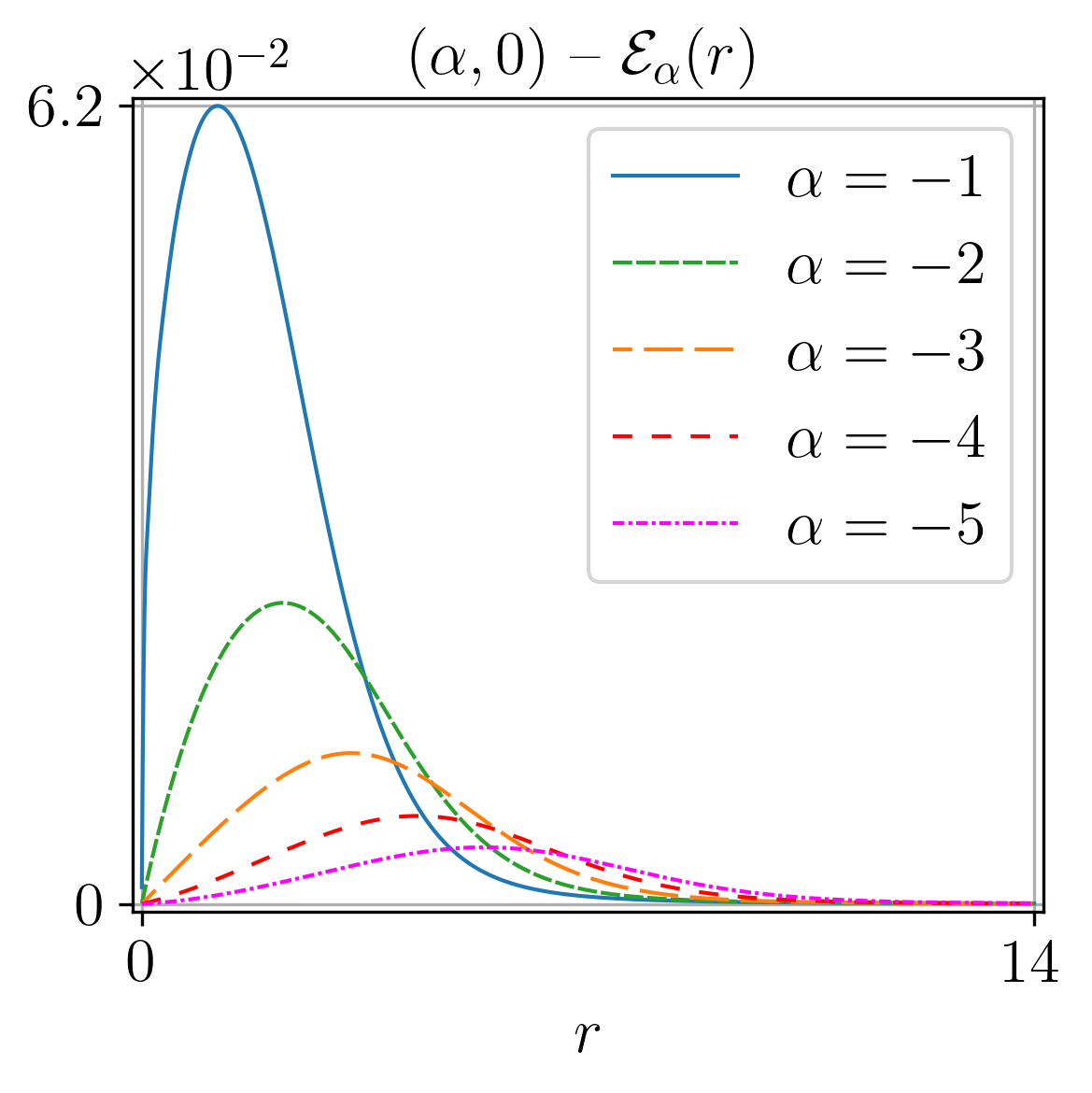}
    \end{subfigure}

    \caption{Monopole profiles and energy densities along the line $\beta=0$ for several negative values of $\alpha$.}
    \label{fig:(alpha,0)}
\end{figure}

Figure~\ref{fig:(0,beta_positive)} shows that, for positive values of $\beta$, the sequence of configurations follows the same pattern observed for the $m_+$ branch in the region $\alpha>1$: a central cavity develops and grows monotonically with $\beta$, leading to hollow, shell-like monopoles.

\begin{figure}[H]
    \centering
    \begin{subfigure}[c]{0.315\textwidth}
        \centering
        \includegraphics[width=\textwidth]{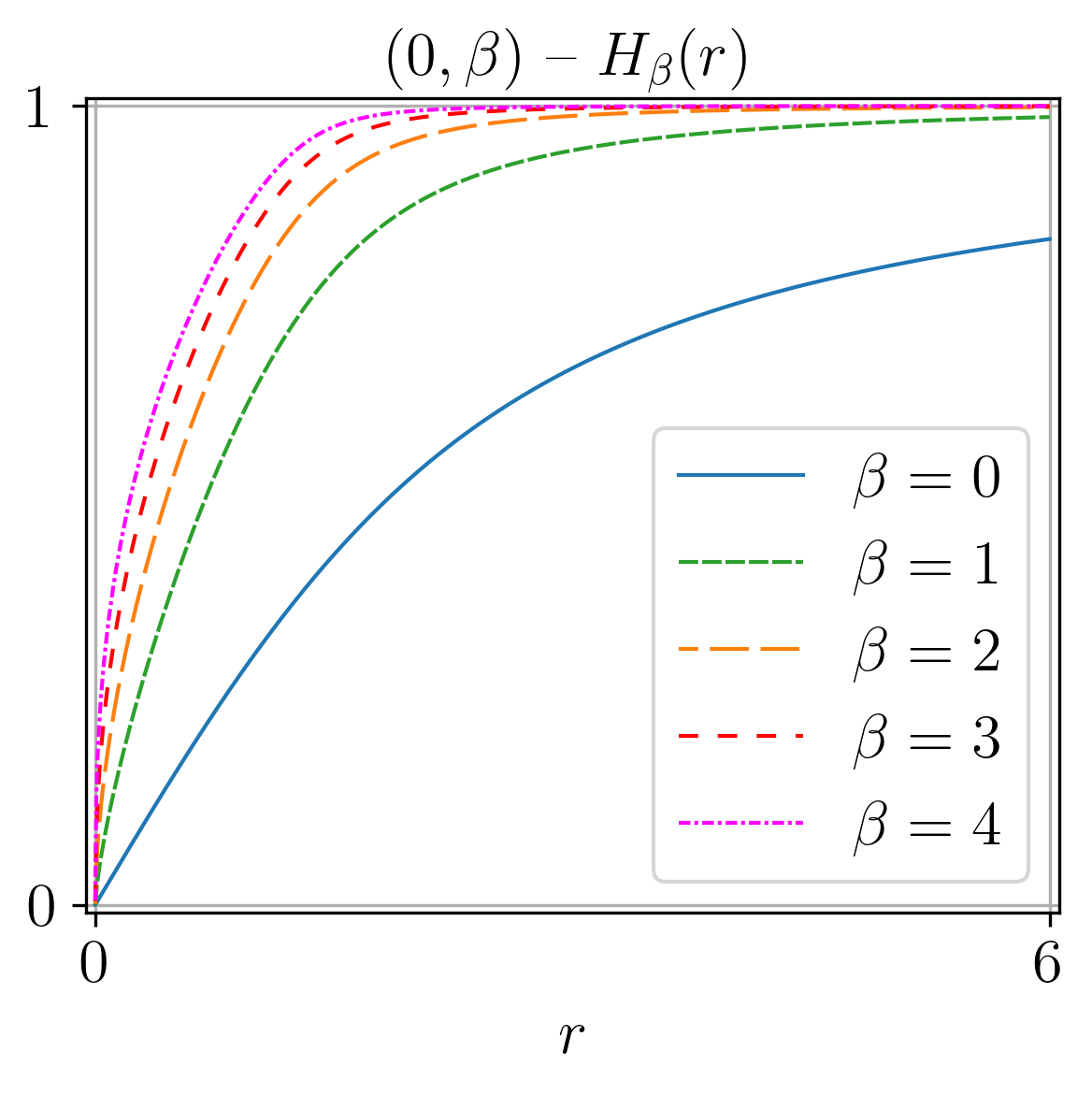}
    \end{subfigure}
    \hspace{6pt}
    \begin{subfigure}[c]{0.315\textwidth}
        \centering
        \includegraphics[width=\textwidth]{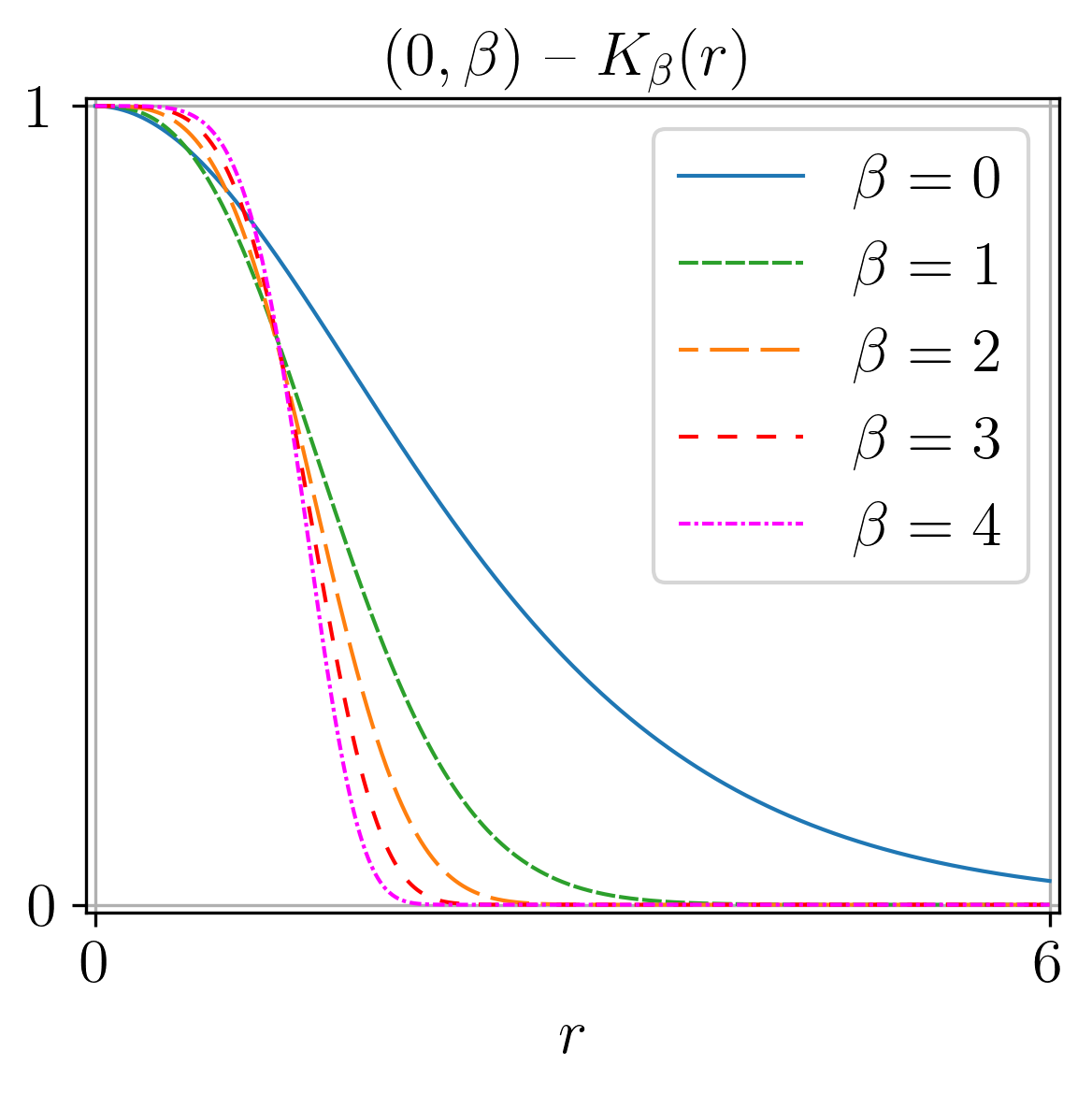}
    \end{subfigure}
    \hspace{6pt}
    \begin{subfigure}[c]{0.315\textwidth}
        \centering
        \includegraphics[width=\textwidth]{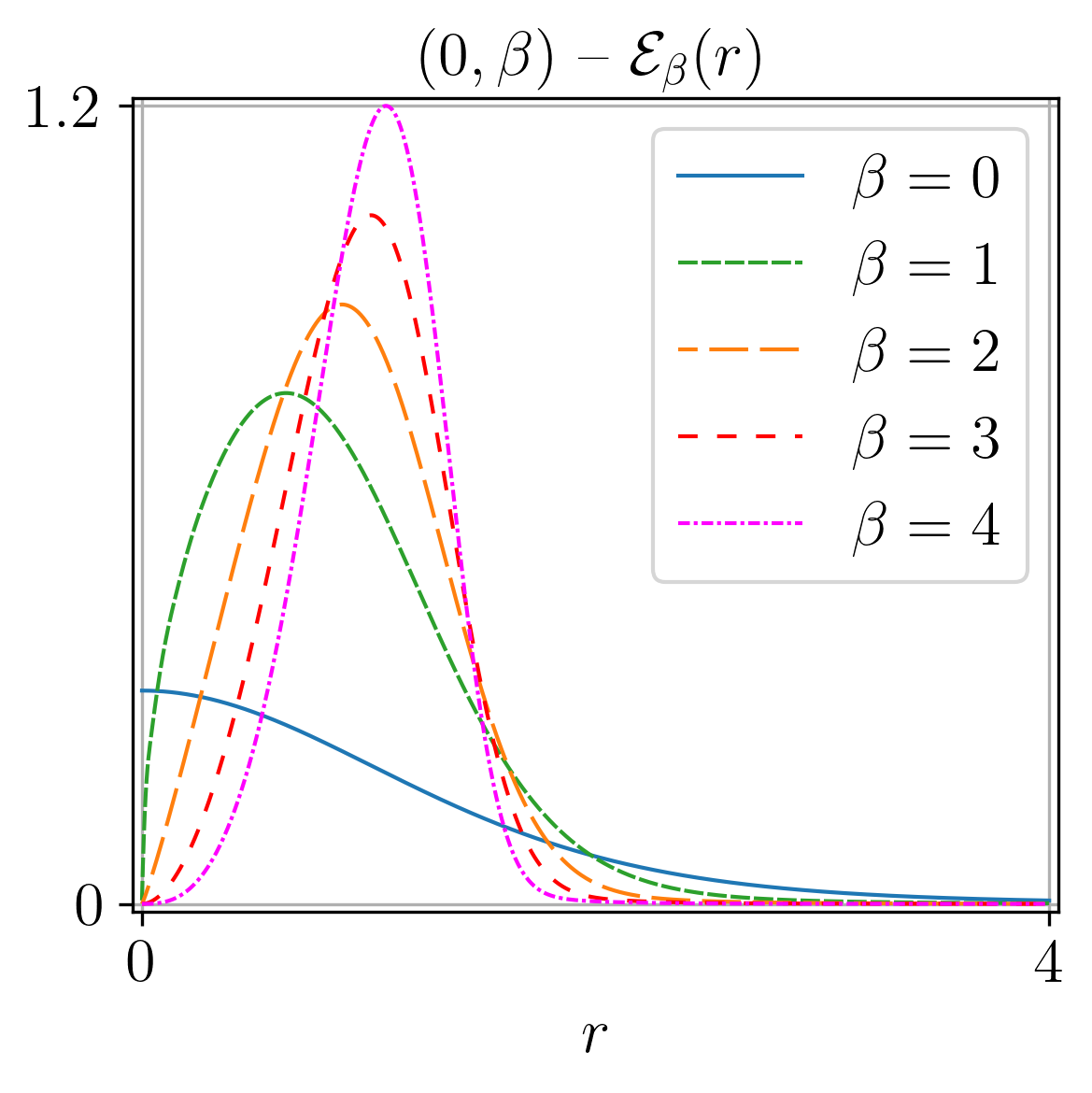}
    \end{subfigure}

    \caption{Emergence and growth of hollow-monopole configurations along the line $\alpha=0$ as $\beta$ increases.}
    \label{fig:(0,beta_positive)}
\end{figure}

For negative values of $\beta$, a qualitatively different behavior emerges. As $\beta \to -1$, the monopole profiles extend over increasingly large spatial regions, while the corresponding energy density becomes progressively more diffuse and less localized. This behavior is illustrated in Fig.~\ref{fig:(-2,beta_negative)}, which shows representative solutions for $\beta<0$.

\begin{figure}[H]
    \centering
    \begin{subfigure}[c]{0.315\textwidth}
        \centering
        \includegraphics[width=\textwidth]{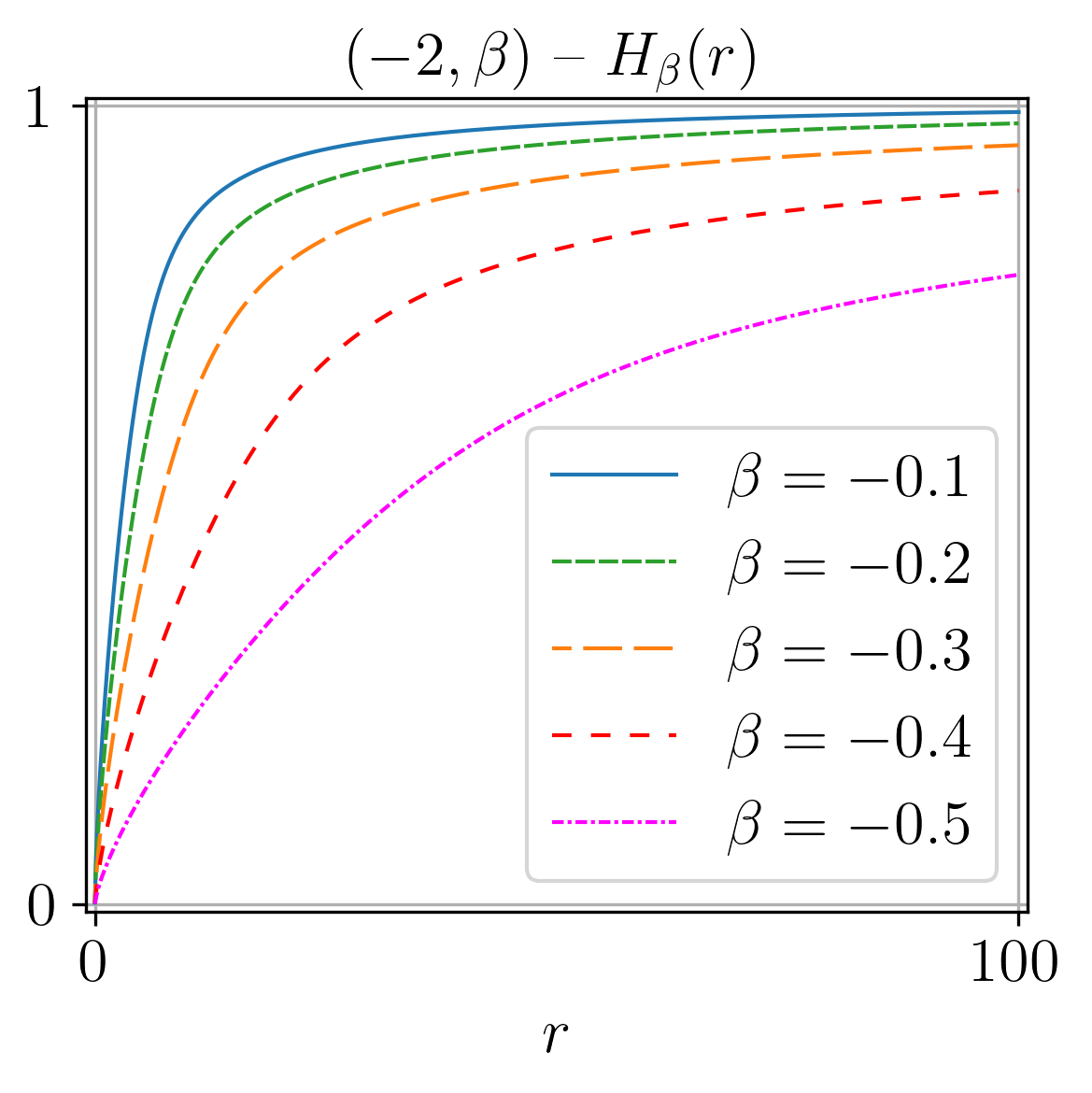}
    \end{subfigure}
    \hspace{6pt}
    \begin{subfigure}[c]{0.315\textwidth}
        \centering
        \includegraphics[width=\textwidth]{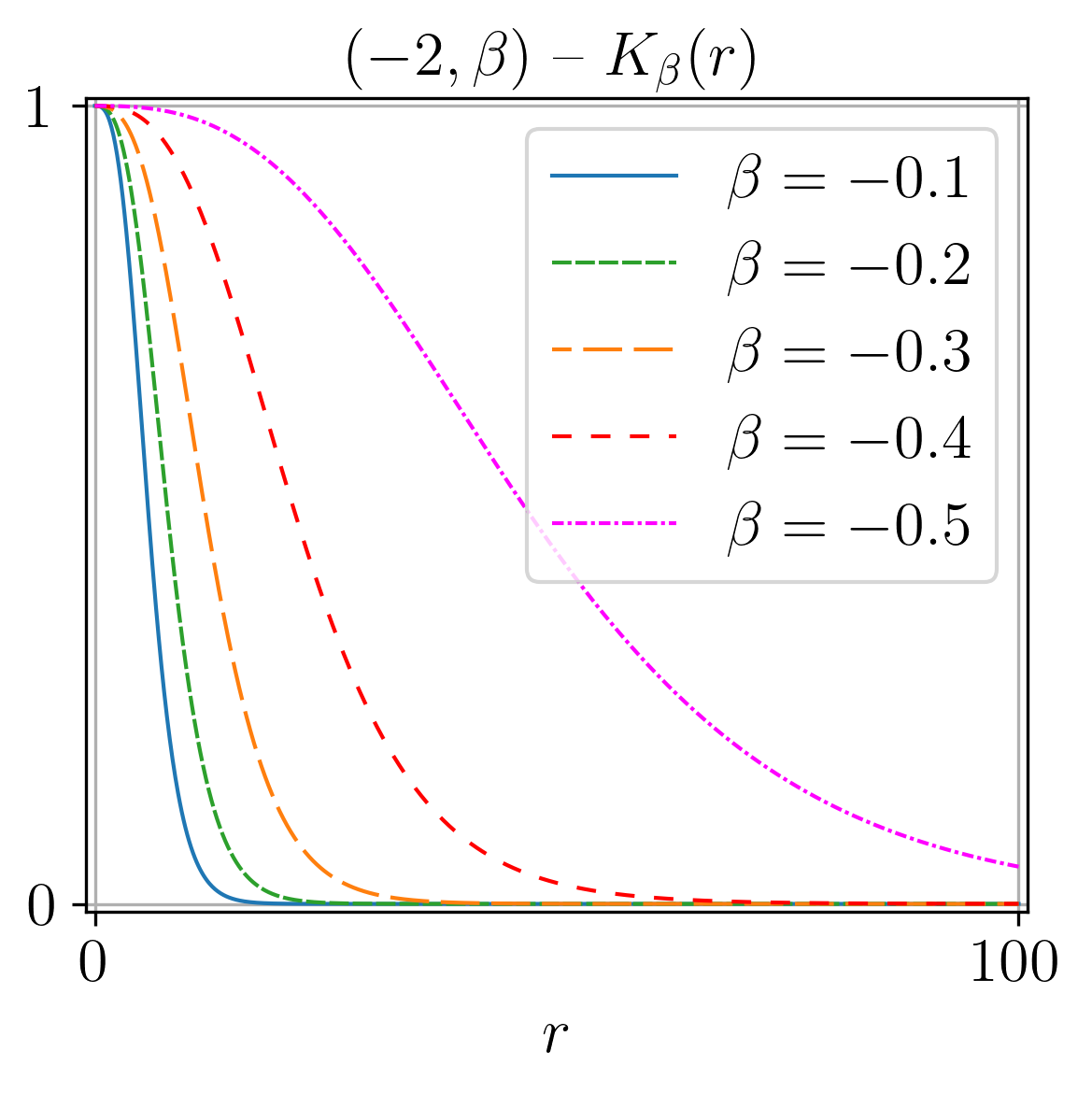}
    \end{subfigure}
    \hspace{6pt}
    \begin{subfigure}[c]{0.315\textwidth}
        \centering
        \includegraphics[width=\textwidth]{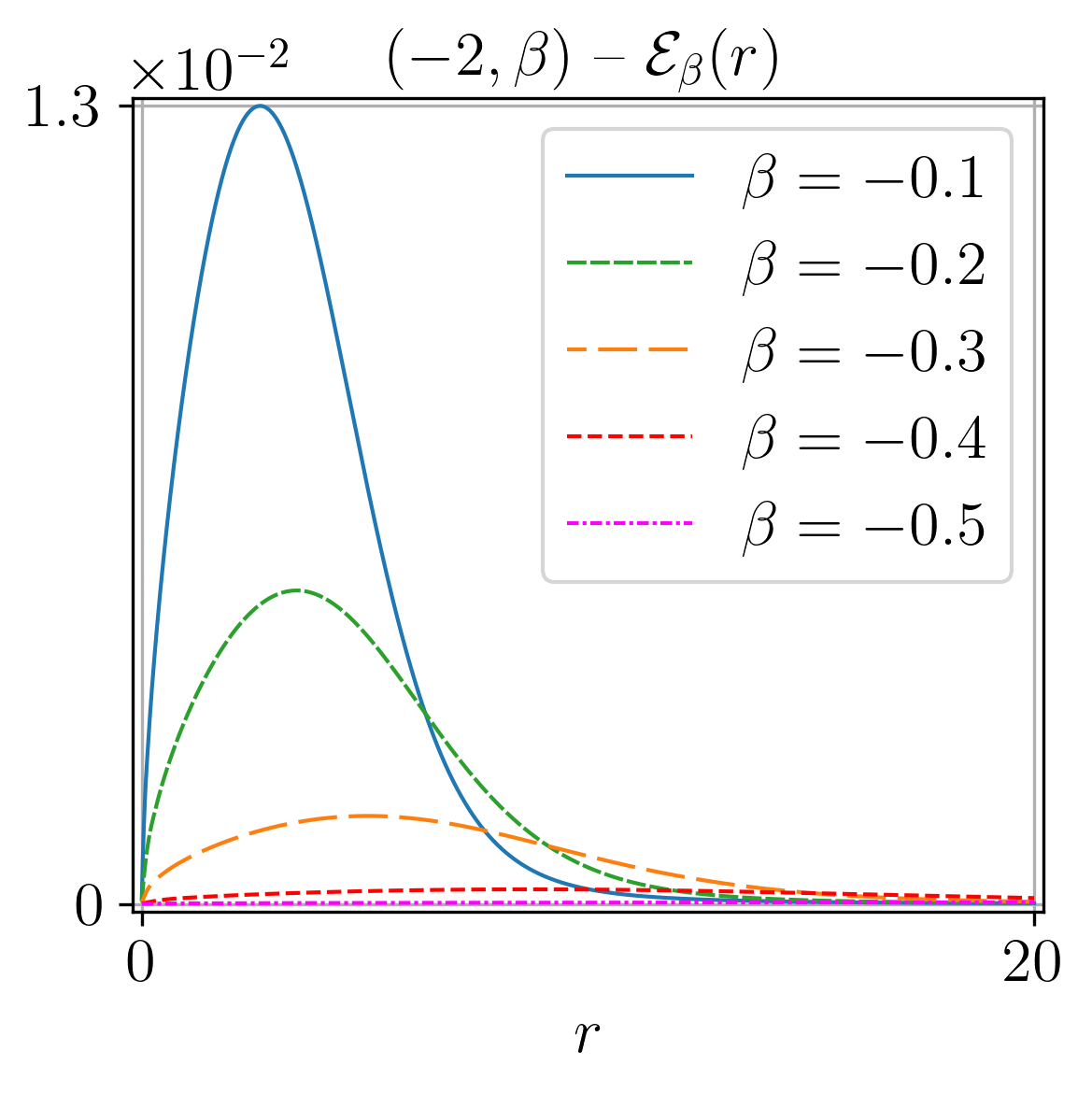}
    \end{subfigure}
    
    \caption{Monopole solutions along the line $\alpha=-2$ for negative values of $\beta$.}
    \label{fig:(-2,beta_negative)}
\end{figure}

The observed results show that as $\alpha$ decreases further below unity while $\beta$ remains above the limiting value $-1$, the monopole configurations become increasingly spatially extended and less localized. Consequently, progressively larger values of $\beta$ are required for shell-like monopoles to emerge. This behavior reinforces the global picture that hollow configurations dominate increasingly larger regions of the $(\alpha,\beta)$ parameter space as one moves away from its left boundary.

Taken together, the results support a unified picture across the admissible $(\alpha,\beta)$ parameter space. Disregarding the $m_-$ branch in the region $\alpha>1$, the monopole configurations follow a characteristic sequence as one moves away from the vertex $(1,-1)$ of the boundary parabola. This notion of ``distance" from the vertex provides a natural geometric organization of the solutions. Starting near the vertex, one typically finds monopoles with a small central cavity, followed by compact core-like configurations without a cavity. As one moves further away, the cavity reappears and grows, ultimately leading to fully developed shell-like monopoles.

\section{Conclusion}\label{sec:conclusion}

In this work, we have presented a generalized $\SU(2)$ Yang-Mills-Higgs model~(\ref{eq:lagrangian}) to describe magnetic monopoles in inhomogeneous media. The standard dynamics of the non-Abelian gauge and Higgs fields was modified through the introduction of two functions, $P(|\Phi|,r)$ and $M(|\Phi|,r)$, which depend explicitly on the radial coordinate and can be interpreted as effective electric permittivity and magnetic permeability, respectively. By imposing the constraint $P(|\Phi|,r)=M^{-1}(|\Phi|,r)$, we were able to find BPS configurations with minimum energy.

We considered static spherically symmetric solutions, with the field profile functions $H(r)$ and $K(r)$ obeying the standard monopole boundary conditions. Assuming the general form $M(H,r)=f(r)/F(H)$ for the coupling between the Higgs field and the effective medium, and adopting $F(H)=H^{\alpha}$ following Ref.~\cite{Bazeia-2018-Small}, we were able to systematically explore the properties of the minimum-energy solutions for different inhomogeneity profiles $f(r)$. The case $\alpha=1$ plays an important role, as the BPS equations decouple and admit analytical solutions for arbitrary $f(r)$.

Focusing on the power-law profile $f(r)=r^{\beta}$, we showed that the behavior of the solutions is governed by two real parameters, $\alpha$ and $\beta$, which control the field dependence and the radial inhomogeneity of the medium. Since the BPS equations exhibit singular behavior at the boundary points, we introduced a suitable regularization procedure and performed an asymptotic analysis, which allowed us to determine the admissible region of the $(\alpha,\beta)$ plane supporting regular BPS solutions (Fig.~\ref{fig:domain}).

Along the analytically solvable line $\alpha=1$, we obtained a complete characterization of the monopole configurations shown in Fig.~\ref{fig:(1,beta)}. The results showed that for $\beta\to-1^+$, the monopoles exhibit a tiny central internal cavity that becomes increasingly small, resulting in effectively point-like monopoles.  For non-negative $\beta$, the monopoles exhibit compact core-like profiles, with the energy density maximal at the origin and decreasing monotonically outward. As $\beta$ increases further, a central cavity reappears and grows, leading to configurations in which the energy density localizes around a finite radius (Fig.~\ref{fig:(1,beta_large)}), eventually resulting in shell-like monopoles (Fig.~\ref{fig:(1,10)}).

For $\alpha>1$, where analytical solutions are no longer available, we explored the structure of the solutions numerically within the admissible region. Along the boundary parabola $\alpha(\beta)=1+(\beta+1)^2/8$, the monopole configurations follow a sequence closely analogous to that observed for $\alpha=1$, interpolating between small-cavity, core-like, and hollow configurations, and converging to shell-like structures as $\beta$ increases. Extending the analysis to constant-$\alpha$ lines inside the admissible region, we identified two distinct branches ($m_+$ and $m_-$). The first reproduces the same qualitative sequence observed along the parabola, while the second reveals a richer structure, including configurations with multiple shells (Figs.~\ref{fig:alpha_minus_1.125_1.5} and~\ref{fig:energies_2D}).

To complement this analysis, we investigated the region $\alpha<1$. Along the homogeneous-medium line $\beta=0$, which includes the standard BPS solution at $(\alpha,\beta)=(0,0)$, the solutions evolve from small-cavity configurations near the boundary parabola to compact core-like monopoles, and subsequently to hollow configurations as $\alpha$ decreases. In contrast to the $\alpha>1$ region, however, the growth of the characteristic radius of hollow monopoles is accompanied by a progressive delocalization of the energy density, whose maximum decreases as it spreads over a larger spatial region (Fig.~\ref{fig:(alpha,0)}).

Overall, except for $m_-$ branch for $\alpha>1$, our results reveal a unified picture of monopole configurations in inhomogeneous media. The interplay between the parameters $\alpha$ and $\beta$ governs the radial redistribution of energy, interpolating between localized core-like monopoles and extended shell-like structures. In particular, shell-like configurations dominate the asymptotic regions of the parameter space, emerging as the natural endpoint of the solution space as one moves away from the vertex $(\alpha,\beta)=(1,-1)$ of the boundary parabola.

 The study developed here opens several directions for future work. It would be interesting to investigate the dynamical properties and stability of these generalized monopoles, as well as their interactions. Moreover, extensions to other forms of inhomogeneity, higher-rank gauge groups, or the inclusion of additional fields may provide further insight into the role of spatially varying media in non-Abelian soliton physics.

\section*{Acknowledgements}

AM acknowledges financial support from CNPq (Conselho Nacional de Desenvolvimento Científico e Tecnológico), Grant No. 306295/2023-7, and CAPES (Coordenação de Aperfeiçoamento de Pessoal de Nível Superior). FRS acknowledges financial support from Fundação de Amparo à Ciência e Tecnologia do Estado de Pernambuco (FACEPE). This work used AI-based tools for language editing. All content was carefully reviewed and verified by the authors.


\appendix

\section{Asymptotics}\label{apx:asymp}

To study the asymptotic behavior of the BPS solutions, it is convenient to map the semi-infinite radial domain $r\in[0,\infty)$ onto a finite interval. We therefore introduce the compactified radial coordinate
\begin{equation}
    r \equiv \frac{\xi}{1-\xi},
    \qquad
    \xi\in[0,1),
\end{equation}
so that the origin $r=0$ is mapped to $\xi=0$, while spatial infinity $r\to\infty$ corresponds to $\xi\to1$. In terms of this new variable, we define the rescaled profile functions
\begin{equation}
    \widetilde{H}(\xi)\equiv H\bigl(r(\xi)\bigr),
    \qquad
    \widetilde{K}(\xi)\equiv K\bigl(r(\xi)\bigr).
\end{equation}
In these variables, the BPS equations~(\ref{eq:bps}) take the form
\begin{subequations}\label{eq:bps_xi}
    \begin{align}
        \widetilde{H}' &=
        (1-\widetilde{K}^2)\widetilde{H}^{\alpha}
        \frac{(1-\xi)^{\beta}}{\xi^{\beta+2}},
        \\[4pt]
        \widetilde{K}' &=
        -\widetilde{K}\widetilde{H}^{1-\alpha}
        \frac{\xi^{\beta}}{(1-\xi)^{\beta+2}},
    \end{align}
\end{subequations}
where the prime now denotes differentiation with respect to $\xi$. The boundary conditions~(\ref{eq:bc}) translate into
\begin{subequations}\label{eq:bc_xi}
    \begin{align}
        \widetilde{H}(0)=0
        &\quad\mathrm{and}\quad
        \widetilde{K}(0)=1,
        \\
        \lim_{\xi\to1}\widetilde{H}(\xi)=1
        &\quad\mathrm{and}\quad
        \lim_{\xi\to1}\widetilde{K}(\xi)=0.
    \end{align}
\end{subequations}

Although the compactification brings the problem to a finite interval, the system~(\ref{eq:bps_xi}) remains singular at both endpoints, $\xi=0$ and $\xi=1$, and therefore cannot be imposed there directly in a numerical implementation. To handle this difficulty, we analyze the local behavior of the solutions near each endpoint and replace the exact boundary conditions~(\ref{eq:bc_xi}) by effective asymptotic boundary conditions imposed at nearby points $\xi_a>0$ and $\xi_b<1$, respectively.

\subsection{Near \texorpdfstring{$\xi=0$}{xi=0}}

Near the origin, it is convenient to introduce the deviations
\begin{equation}\label{eq:deviations}
    \delta_{\widetilde{H}}(\xi)\equiv \big|\widetilde{H}(0)-\widetilde{H}(\xi)\big|,
    \qquad
    \delta_{\widetilde{K}}(\xi)\equiv \big|\widetilde{K}(0)-\widetilde{K}(\xi)\big|,
\end{equation}
which vanish at $\xi=0$ and grow toward unity as $\xi\to1$, consistent with~(\ref{eq:bc_xi}). Since $\widetilde{H}(0)=0$ and $\widetilde{K}(0)=1$, we may write
\[
\widetilde{H}(\xi)=\delta_{\widetilde{H}}(\xi),
\qquad
\widetilde{K}(\xi)=1-\delta_{\widetilde{K}}(\xi).
\]
For regular monopole solutions, both $\widetilde{H}(\xi)$ and $\widetilde{K}(\xi)$ must approach their boundary values smoothly, without divergences or oscillatory behavior. This motivates a Frobenius analysis near $\xi=0$, for which we assume the expansions
\begin{equation}
    \delta_{\widetilde{H}}(\xi)=\xi^{m}\sum_{s=0}^{\infty}a_s\xi^s,
    \qquad
    \delta_{\widetilde{K}}(\xi)=\xi^{n}\sum_{s=0}^{\infty}b_s\xi^s,
\end{equation}
with $m,n>0$. For numerical purposes, it is sufficient to retain only the leading-order behavior,
\begin{equation}
    \widetilde{H}(\xi)\approx a_0\xi^{m},
    \qquad
    \widetilde{K}(\xi)\approx 1-b_0\xi^{n},
\end{equation}
where $a_0$ and $b_0$ are positive constants. Substituting these leading-order expressions into the BPS system~(\ref{eq:bps_xi}) and matching the dominant powers of $\xi$, we obtain the algebraic relations
    \begin{align}
        &(1-\alpha)m^2 + (\beta+1)m - 2 = 0, \nonumber
        \\
        &n = \frac{2}{m}, \nonumber
        \\ 
        &b_0 = \frac{m a_0^{1-\alpha}}{2}.
    \end{align}
The first equation determines the exponent $m$, while the remaining two relations fix $n$ and $b_0$ once $m$ and $a_0$ are known.

Solving the quadratic equation for $m$ gives
\begin{equation}\label{eq:m}
    m=
    \begin{cases}
        \dfrac{2}{\beta+1}, &\alpha=1,
        \\[8pt]
        \dfrac{-(\beta+1)\pm\sqrt{\Delta}}{2(1-\alpha)}, &\alpha\neq1,
    \end{cases}
\end{equation}
where $\Delta=(\beta+1)^2+8(1-\alpha)$.
A necessary condition for $m$ to be real is $\Delta\geq0$, which excludes the region above the parabola
$\alpha(\beta)=1+(\beta+1)^2/8$ with upwards concavity and vertex $(1, -1)$ on the $(\alpha,\beta)$ plane. Reality of $m$, however, is not sufficient. Regularity at the origin also requires $m>0$. The allowed possibilities are therefore as follows. For $\alpha>1$, both roots
\[
m_{\pm}=\frac{-(\beta+1)\pm\sqrt{\Delta}}{2(1-\alpha)}
\]
are positive whenever $\beta+1\geq\sqrt{8(\alpha-1)}$,
so both branches are admissible in this region, with $m_-\geq m_+$.  For $\alpha=1$ we have $\beta+1>0$ so the vertex $(1,-1)$ is not included in the admissible region. For $\alpha<1$, only the positive branch $m_+$ satisfies $m>0$. We may therefore summarize the near-origin result as
\begin{equation}
    m=
    \begin{cases}
        m_{\pm}, & \alpha>1 \ \text{and}\ \beta+1\geq\sqrt{8(\alpha-1)},
        \\[4pt]
        \dfrac{2}{\beta+1}, & \alpha=1 \ \text{and}\ \beta+1>0,
        \\[4pt]
        m_+, & \alpha<1.
    \end{cases}
\end{equation}
It is important to note that, for $\alpha<1$, the near-origin analysis alone imposes no additional restriction on $\beta$ beyond the requirement that $m_+$ be positive. The stronger condition $\beta+1>0$
will emerge from the asymptotic analysis near $\xi=1$, and therefore applies uniformly throughout the full parameter space. Combining both endpoint analyses then leads to the admissible domain shown in Fig.~\ref{fig:domain}.

\begin{figure}[H]
    \centering
    \begin{subfigure}[c]{0.49\textwidth}
        \centering
        \includegraphics[width=\textwidth]{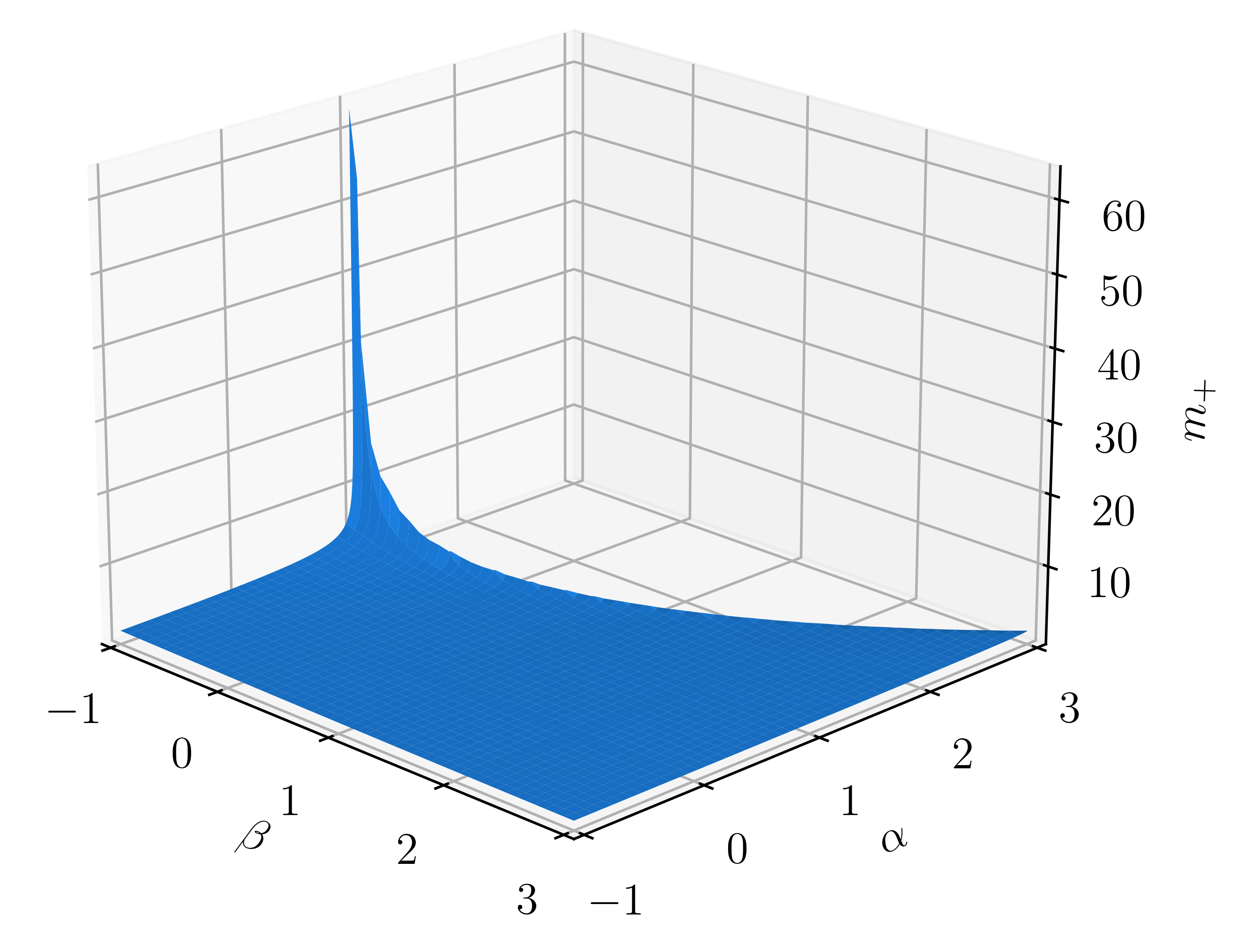}
    \end{subfigure}
    \hspace{2pt}
    \begin{subfigure}[c]{0.49\textwidth}
        \centering
        \includegraphics[width=\textwidth]{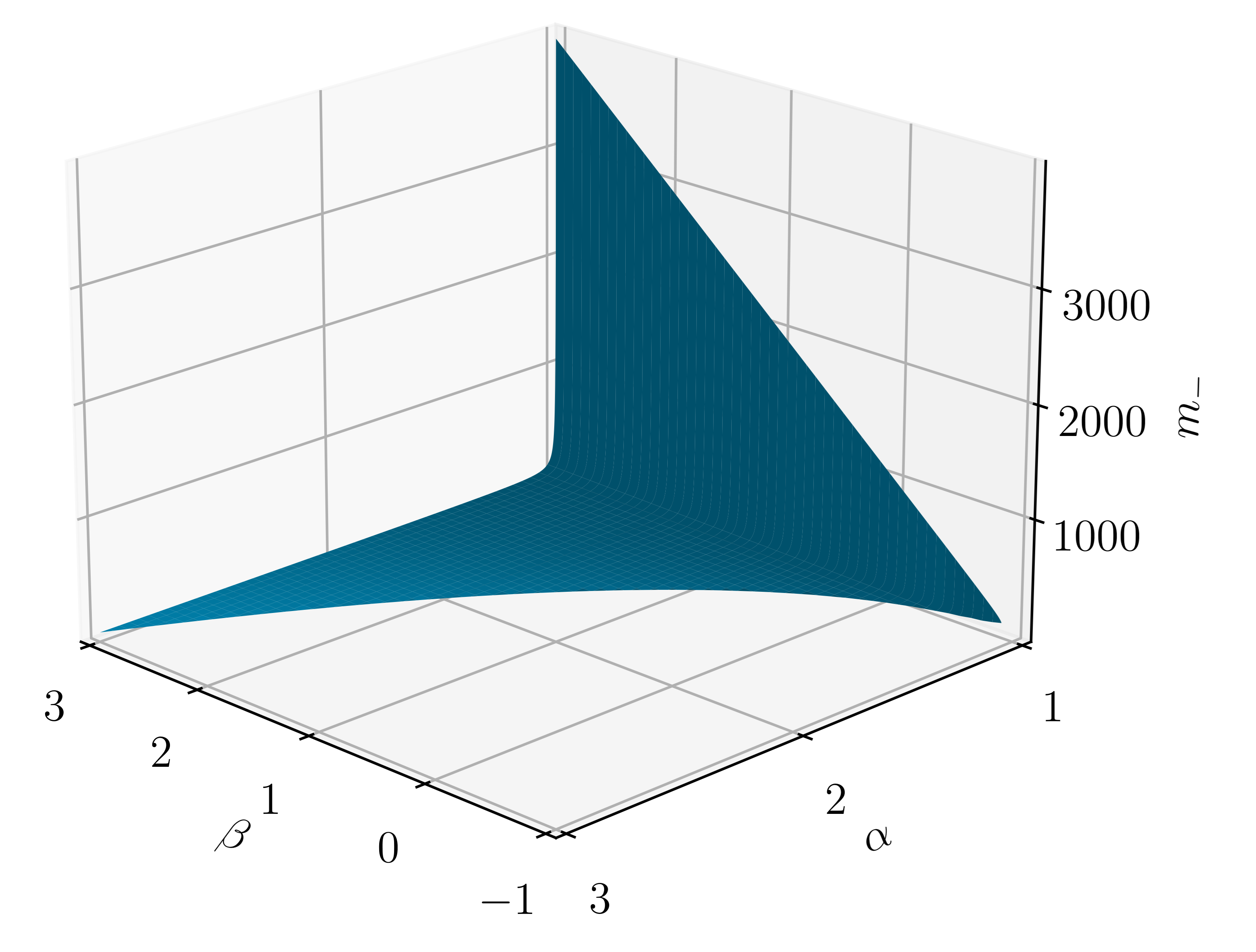}
    \end{subfigure}
    \caption{Left: surface $m_+(\alpha,\beta)$, which grows towards $\alpha=1$ and negative values of $\beta$, but due to the boundary $\beta=-1$ acquires the form of a cusp accumulating near the vertex $(1,-1)$. Right: surface $m_-(\alpha,\beta)$, which grows towards $\alpha=1$ and positive values of $\beta$.}
    \label{fig:exp_m}
\end{figure}

For the numerical implementation, we regularize the problem near $\xi=0$ by introducing a small cutoff $\xi_a\ll1$ and replacing the exact boundary conditions at the origin by their leading asymptotic form,
    \begin{align}
        \widetilde{H}(\xi_a) = a_0 \xi_a^{m},
        \qquad
        \widetilde{K}(\xi_a) = 1-\frac{m a_0^{1-\alpha}}{2} \, \xi_a^{2/m},
        \label{eq:init_cond}
    \end{align}
as the first effective boundary conditions. The remaining free coefficient $a_0$ can then be determined in two equivalent ways. One may solve the problem as a boundary value problem (BVP), treating $a_0$ as an additional unknown parameter, or alternatively as an initial value problem (IVP) using a shooting method, in which $a_0$ serves as the shooting parameter. In either approach, the solution obtained from the near-origin data must be matched to the appropriate asymptotic behavior as $\xi\to1$.

\subsection{Near \texorpdfstring{$\xi=1$}{xi=1}}

To analyze the behavior of the solutions near spatial infinity, we introduce the variable
\begin{equation}
    \tau \equiv 1-\xi,
\end{equation}
so that $\tau\to0$ corresponds to $r\to\infty$. We further define
\begin{equation}
    \bar{H}(\tau)\equiv \widetilde{H}(\xi(\tau)),
    \qquad
    \bar{K}(\tau)\equiv \widetilde{K}(\xi(\tau)).
\end{equation}

\noindent In terms of these variables, the BPS equations can be written as
\begin{subequations}
    \begin{align}
        \bar{K}(\tau)
        &=
        \exp\!\left[
        \int_{1}^{\tau}t^{-(\beta+2)}(1-t)^{\beta}\,\bar{H}^{\,1-\alpha}(t)\,\dd t
        \right],
        \label{eq:bpstau1}
        \\[4pt]
        \bar{K}^2
        &=
        1+\tau^{-\beta}(1-\tau)^{\beta+2}\,\bar{H}^{-\alpha}\,\bar{H}',
        \label{eq:bpstau2}
    \end{align}
\end{subequations}
where the prime denotes differentiation with respect to $\tau$. The boundary conditions become
    \begin{align}
        \bar{H}(1)=0
        &\quad\mathrm{and}\quad
        \bar{K}(1)=1, \nonumber
        \\
        \lim_{\tau\to0}\bar{H}(\tau)=1
        &\quad\mathrm{and}\quad
        \lim_{\tau\to0}\bar{K}(\tau)=0.
    \end{align}
We parametrize deviations from the asymptotic values by defining $\delta_{\bar{H}}(\tau)\equiv\abs{\bar{H}(0)-\bar{H}(\tau)}$ respecting the boundary conditions $\delta_{\bar{H}}(0)=0$ and $\delta_{\bar{H}}(1)=1$. This way we have 
$\bar{H}(\tau)=1-\delta_{\bar{H}}(\tau)$.
Substituting this form into Eq.~(\ref{eq:bpstau1}) and expanding the integrand around $\tau=0$, we get
\begin{align}
    \bar{K}(\tau)
    &= \exp\left\{\int\limits_1^{\tau} t^{-(\beta+2)}\left[\sum_{k=0}^{\infty}\binom{\beta}{k}(-t)^k\right]\left(\sum_{n=0}^{\infty}\frac{(\bar{H}^{1-\alpha})^{(n)}(0)}{n!}t^n\right)\dd t\right\} \nonumber
    \\[4pt]
    &= \prod_{k,n=0}^{\infty}\exp\left\{\binom{\beta}{k}(-1)^k\frac{(\bar{H}^{1-\alpha})^{(n)}(0)}{n!}\left[\frac{\tau^{n+k-(\beta+1)}-1}{n+k-(\beta+1)}\right]\right\},
\end{align}
where $n+k\neq\beta+1$, and $\binom{\beta}{k}$ is the binomial coefficient and $(\bar{H}^{1-\alpha})^{(n)}(0)=\left.\dv*[n]{(1-\delta_{\bar{H}}(\tau))^{1-\alpha}}{\tau}\right|_{\tau=0}$. Near $\tau=0$, the dominant contribution to $\bar{K}(\tau)$ arises from the leading-order term of the integrand with $n=k=0$. Provided $\beta\neq-1$, this yields the asymptotic behavior
\begin{equation}
    \bar{K}(\tau)
    \sim
    \exp\!\left[-\frac{\tau^{-(\beta+1)}}{\beta+1}\right],
\end{equation}
which shows that $\bar{K}(\tau)$ decays exponentially fast as $\tau\to0$. We also assume
\begin{equation}
    \bar{H}(\tau)\approx 1-c_0\,\tau^{p},
    \qquad
    c_0>0,\; p>0,
\end{equation}
and determine the parameters $p$ and $c_0$ from the equation (\ref{eq:bpstau2}). Retaining the leading-order contribution and noting that $\bar{K}^2$ is exponentially suppressed near $\tau=0$, Eq.~(\ref{eq:bpstau2}) reduces to
\begin{equation}
    0 = 1 - p\,c_0\,\tau^{p-(\beta+1)},
\end{equation}
which fixes
\begin{equation}
    p=\beta+1,
    \qquad
    c_0=\frac{1}{\beta+1}.
\end{equation}

\noindent We thus obtain the asymptotic forms
\begin{equation}\label{eq:asymp_tails}
    \bar{H}(\tau)=
    1-\frac{\tau^{\beta+1}}{\beta+1},
    \qquad
    \bar{K}(\tau)=
    \exp\!\left[-\frac{\tau^{-(\beta+1)}}{\beta+1}\right],
\end{equation}
which are valid for any $\alpha$ and $\beta>-1$.

The cutoff $\tau_b$ is determined by specifying a value $error_b$ and choosing
\begin{equation}
    \tau_b=
    \left(\frac{error_b}{c_0}\right)^{1/p},
\end{equation}
so that $\delta_{\bar{H}}(\tau_b)=error_b$. Then, we use Eq.~(\ref{eq:target}) to define our second effective boundary conditions:
\begin{equation}\label{eq:target}
    \bar{H}(\tau_b)=
    1-\frac{\tau_b^{\beta+1}}{\beta+1},
    \qquad
    \bar{K}(\tau_b)=
    \exp\!\left[-\frac{\tau_b^{-(\beta+1)}}{\beta+1}\right].
\end{equation}

Finally, with these results we can perform the numerical implementation to properly solve the BPS equations~(\ref{eq:bps}). When solving it as a BVP, the presence of the unknown parameter $a_0$ requires the use of three boundary conditions instead of two, such that one can tie the solution at $r=r_a$ using~(\ref{eq:init_cond}), and pick one of the tails~(\ref{eq:asymp_tails}) for the remaining boundary residual at $r=r_b$.

On the other hand, when solving it as an IVP via shooting method, once the target values $(H_{\mathrm{tar}},K_{\mathrm{tar}})$ given by Eq.~(\ref{eq:target}) are fixed, the numerical problem reduces to determining the optimal shooting parameter $a_0^{\star}$. One way to proceed is defining this parameter as the root of a function $\mathrm{shooting}(a_0)$, constructed as follows: for a given trial value of $a_0$, the BPS system is integrated from the initial conditions $(H_a,K_a)$ at $r=r_a$, specified by Eq.~(\ref{eq:init_cond}), up to $r=r_b$. The function $\mathrm{shooting}(a_0)$ then returns the mismatch
\begin{equation}
    \mathrm{shooting}(a_0) = H_{\mathrm{tar}} - H_b,
\end{equation}
where $H_b$ denotes the numerical value of $H$ at $r=r_b$.

By employing a root-finding method for $\mathrm{shooting}(a_0)$ within a sign-changing interval $[a_{\mathrm{lower}},a_{\mathrm{upper}}]$, one obtain the optimal coefficient $a_0^{\star}$ and can compute the total shooting error,
\begin{equation}
    \delta =
    \sqrt{(H_{\mathrm{tar}}-H_b^{\star})^2+(K_{\mathrm{tar}}-K_b^{\star})^2}.
\end{equation}

\section{Auxiliary regular functions}\label{apx:axl_func}

The deviations introduced in Eq.~(\ref{eq:deviations}) capture the behavior of the fields near the origin, but the same idea can be extended to the whole radial domain. Motivated by the leading asymptotic behavior obtained in Appendix~\ref{apx:asymp}, we introduce two auxiliary functions $h(r)$ and $k(r)$ through
\[
H(r)=r^{m}h(r),
\qquad
K(r)=1-r^{n}k(r),
\]
where $m$ and $n$ are the exponents determined in the near-origin analysis. By construction, the functions $h(r)$ and $k(r)$ are regular at the origin and satisfy
\[
h(0)=a_0,
\qquad
k(0)=b_0.
\]
Moreover, since $H(r)\to1$ and $K(r)\to0$ as $r\to\infty$, it follows that $h(r)\sim r^{-m}$ and $k(r)\sim r^{-n}$ for large $r$.
Replacing the expression for $H(r)$ and $K(r)$ in the BPS equations~(\ref{eq:bps}), we obtain
    \begin{align}
        h'&=
        \left(2k - r^n k^2\right)r^{\gamma}\,h^{\alpha} - \frac{m}{r}h, \nonumber
        \\
        k'&=
        (1-r^n k)\frac{h^{1-\alpha}}{r^{\gamma+2}} - \frac{n}{r}k,
    \end{align}
where $\gamma=(\alpha-1)m+n-\beta-2$. 
Using the relation~(\ref{eq:m}) together with $n=2/m$, one finds after straightforward algebra that $\gamma=-1$
throughout the admissible parameter region.
Therefore, the BPS equations simplify to
    \begin{align}
        rh'&=
        \left(2k - r^nk^2\right)h^{\alpha} - mh,
        \nonumber\\
        rk'&=
        (1-r^n k)h^{1-\alpha} - nk.
        \label{eq:ansatz_system}
    \end{align}
This reformulation has an important practical advantage: the singular behavior of the original BPS system is reduced to a simple pole at $r=0$. In addition, the right-hand sides of Eqs.~(\ref{eq:ansatz_system}) vanish identically at the origin when evaluated on the leading asymptotic data, which makes this form particularly convenient for numerical integration.

To handle large values of $a_0$, especially for the solutions with the negative root $m_-$, we write the auxiliary functions as $h(r)=\e^{\chi(r)}$ and $k(r)=\e^{\kappa(r)}$, and work with the system
    \begin{align}
        r\chi'&=
        (2\mathrm{e}^{\kappa}-r^n\mathrm{e}^{2\kappa})\mathrm{e}^{(\alpha-1)\chi} - m, \nonumber
        \\
        r\kappa'&=
        (\mathrm{e}^{-\kappa}-r^n)\mathrm{e}^{(1-\alpha)\chi} - n,
    \end{align}
imposing the boundary conditions
    \begin{align}
        \chi(r_a)&=\chi_0,\nonumber
        \\
        \kappa(r_a)&=\kappa_0,\nonumber
        \\
        \chi(r_b)&=-m\ln(r_b), \nonumber
        \\
        \kappa(r_b)&=-n\ln(r_b),
    \end{align}
where $\chi_0=\ln(a_0)$ is the free parameter and $\kappa_0=\ln(b_0)=(1-\alpha)\chi_0+\ln(m/2)$.

\begin{figure}[H]
    \centering
    \begin{subfigure}[c]{0.39\textwidth}
        \includegraphics[width=0.9\textwidth]{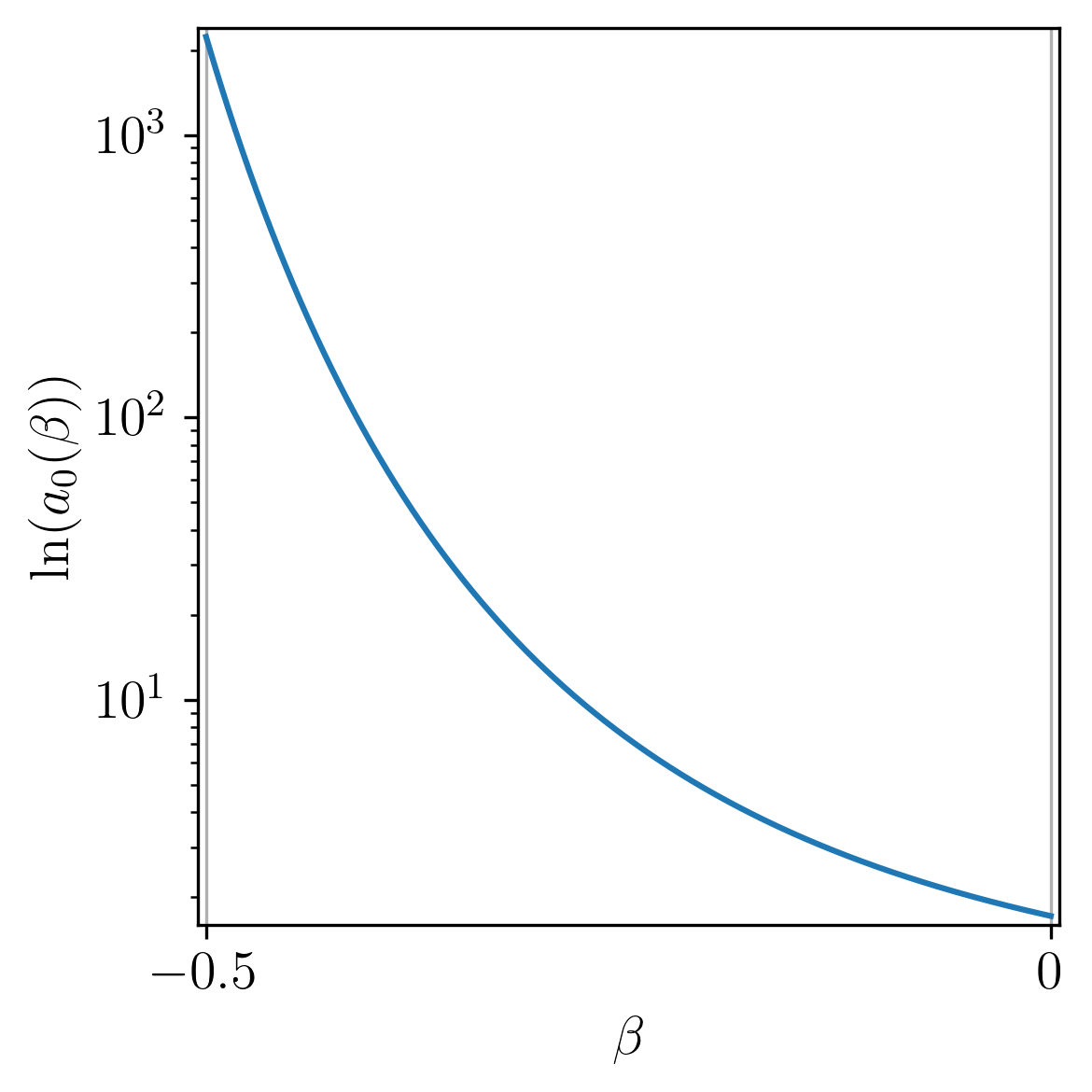}
    \end{subfigure}
    \hspace{4pt}
    \begin{subfigure}[c]{0.49\textwidth}
        \includegraphics[width=0.9\textwidth]{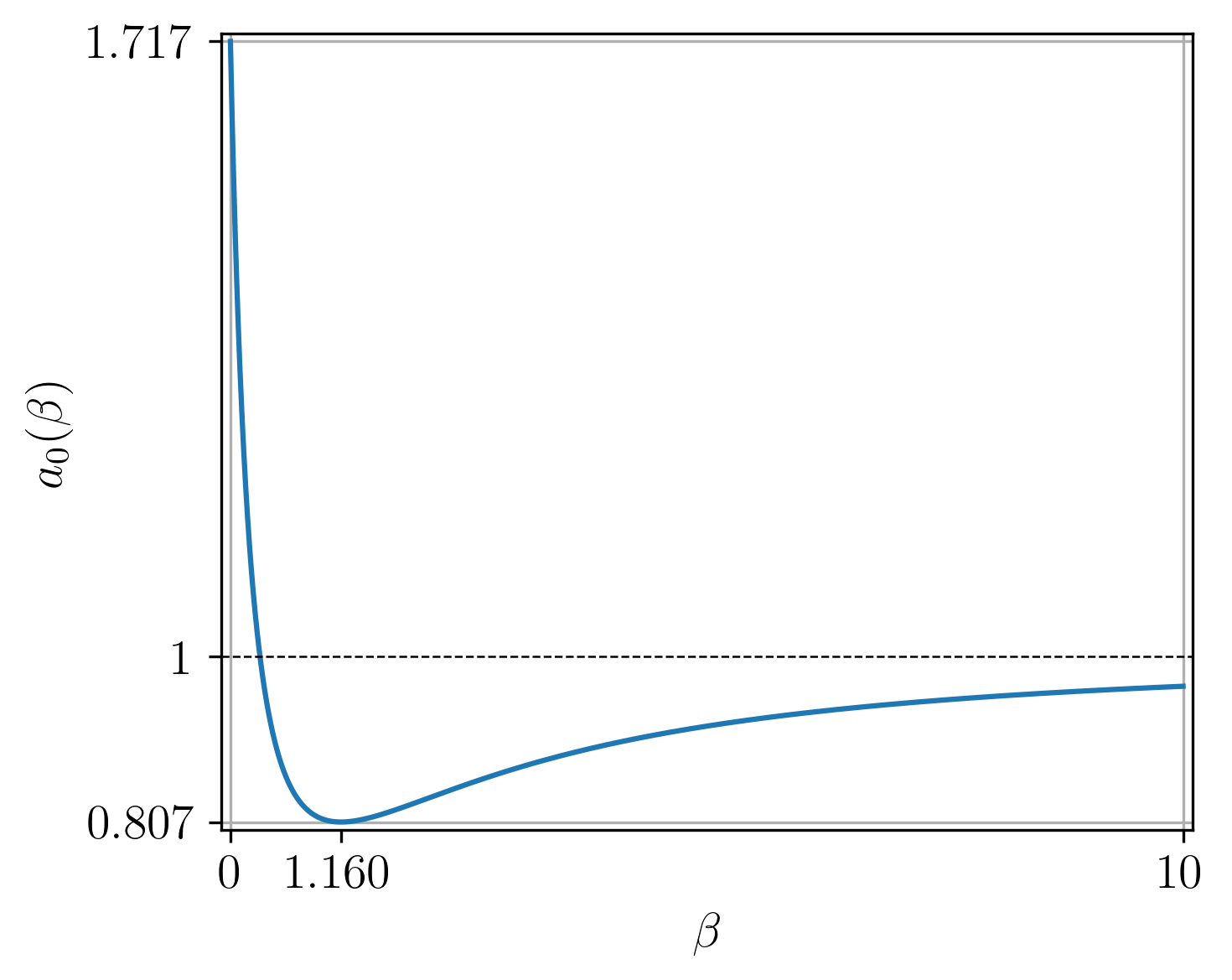}
    \end{subfigure}
    \caption{Analytical behavior of the parameter $a_0(\beta)$ for $\alpha=1$, showing its divergence as $\beta\to-1^{+}$ and its asymptotic approach to unity for large $\beta$.}
    \label{fig:analytical_a0}
\end{figure}

As a consistency check of the numerical implementation, it is useful to consider the analytically solvable case $\alpha=1$. In this case, the exact solution~(\ref{eq:sol1}) allows us to determine the coefficient $a_0$ explicitly. Using the expansion
\begin{equation}
    \Ei(-x)=
    \gamma + \ln(x) - x + \mathcal{O}(x^2),
\end{equation}
where $\gamma$ denotes the Euler-Mascheroni constant, we obtain near $x=0$
    \begin{align}
        H(x) &=
        \mathrm{e}^{C(\gamma-1)}\,x^{C}
        \left(1-\frac{C}{2}x\right)\mathrm{e}^{\mathcal{O}(x^2)}, \nonumber
        \\
        K(x) &=
        1-\frac{x}{2}+\mathcal{O}(x^2).
    \end{align}
Reexpressing these formulas in terms of the radial coordinate $r$, we find
    \begin{align}
        H(r)&\approx
        a_0\,r^{2/(\beta+1)}\left[1-\frac{2r^{\beta+1}}{(\beta+1)^3}\right], \nonumber
        \\
        K(r)&\approx
        1-\frac{r^{\beta+1}}{\beta+1},
    \end{align}
with
\begin{equation}
    a_0(\beta)=
    \exp\!\left[\frac{2(\gamma-1)}{(\beta+1)^2}\right]\left(\frac{2}{\beta+1}\right)^{\!2/(\beta+1)^2}.
\end{equation}
Introducing the variable $u\equiv\beta+1$, and taking the logarithm, we obtain
\begin{equation}
    \ln\big(a_0(u)\big)=
    \frac{2\,[\,(\gamma-1)+\ln 2-\ln u\,]}{u^2}.
\end{equation}
This expression diverges as $u\to0^+$ and tends to zero as $u\to\infty$. Therefore, $a_0(\beta)$ grows without bound as $\beta\to-1^+$ and approaches unity for large $\beta$. The numerically determined values of $a_0$ along the line $\alpha=1$ are expected to reproduce precisely this behavior, as illustrated in Fig.~\ref{fig:analytical_a0}.

\bibliography{references}

@article{Picon-1999-kInflation,
    author = {C. Armendáriz-Picón and T. Damour and V. Mukhanov},
    title = {k-Inflation},
    journal = {Physics Letters B},
    volume = {458},
    number = {2},
    pages = {209-218},
    year = {1999},
    issn = {0370-2693},
    doi = {https://doi.org/10.1016/S0370-2693(99)00603-6},
    url = {https://www.sciencedirect.com/science/article/pii/S0370269399006036}
}

@article{Babichev-2006-Global,
  title = {Global topological $k$-defects},
  author = {Babichev, E.},
  journal = {Phys. Rev. D},
  volume = {74},
  issue = {8},
  pages = {085004},
  numpages = {7},
  year = {2006},
  month = {Oct},
  publisher = {American Physical Society},
  doi = {10.1103/PhysRevD.74.085004},
  url = {https://link.aps.org/doi/10.1103/PhysRevD.74.085004}
}

@article{Adam-2007-kdefects,
    author = {Adam, C and Sánchez-Guillén, J and Wereszczyński, A},
    title = {k-defects as compactons},
    journal = {Journal of Physics A: Mathematical and Theoretical},
    year = {2007},
    month = {oct},
    publisher = {},
    volume = {40},
    number = {45},
    pages = {13625},
    doi = {10.1088/1751-8113/40/45/009},
    url = {https://doi.org/10.1088/1751-8113/40/45/009}
}

@article{Adam-2009-Corrigendum,
    author = {C Adam and J Sanchez-Guillen and A Wereszczynski},
    title = {k-defects as compactons},
    journal = {Journal of Physics A: Mathematical and Theoretical},
    year = {2009},
    month = {feb},
    publisher = {},
    volume = {42},
    number = {8},
    pages = {089801},
    doi = {10.1088/1751-8121/42/8/089801},
    url = {https://doi.org/10.1088/1751-8121/42/8/089801}
}

@article{Hooft-1974-Magnetic,
    author = {'t Hooft, G.},
    title = {Magnetic monopoles in unified gauge theories},
    journal = {Nuclear Physics B},
    volume = {79},
    number = {2},
    pages = {276-284},
    year = {1974},
    issn = {0550-3213},
    doi = {https://doi.org/10.1016/0550-3213(74)90486-6},
    url = {https://www.sciencedirect.com/science/article/pii/0550321374904866}
}

@article{Polyakov-1974-Particle,
    author = {Polyakov, A. M.},
    title = {Particle spectrum in quantum field theory},
    journal = {JETP Lett.},
    volume = {20},
    pages = {194--195},
    year = {1974},
    note = {[Pisma Zh. Eksp. Teor. Fiz. 20, 430 (1974)]}
}

@article{Bogomolny-1975-Stability,
    author = {Bogomolny, E. B.},
    title = {Stability of Classical Solutions},
    journal = {Sov. J. Nucl. Phys.},
    volume = {24},
    pages = {449},
    year = {1976},
    note = {[Yad. Fiz. 24, 861 (1976)]}
}

@article{Prasad-1975-Exact,
  title = {Exact Classical Solution for the 't Hooft Monopole and the Julia-Zee Dyon},
  author = {Prasad, M. K. and Sommerfield, Charles M.},
  journal = {Phys. Rev. Lett.},
  volume = {35},
  issue = {12},
  pages = {760--762},
  numpages = {0},
  year = {1975},
  month = {Sep},
  publisher = {American Physical Society},
  doi = {10.1103/PhysRevLett.35.760},
  url = {https://link.aps.org/doi/10.1103/PhysRevLett.35.760}
}

@article{Casana-2012-Generalized,
  title = {Generalized BPS magnetic monopoles},
  author = {Casana, R. and Ferreira, M. M. and da Hora, E.},
  journal = {Phys. Rev. D},
  volume = {86},
  issue = {8},
  pages = {085034},
  numpages = {5},
  year = {2012},
  month = {Oct},
  publisher = {American Physical Society},
  doi = {10.1103/PhysRevD.86.085034},
  url = {https://link.aps.org/doi/10.1103/PhysRevD.86.085034}
}

@article{Bazeia-2013-Deformed,
    author = {D. Bazeia and R. Casana and M.M. Ferreira and E. {da Hora} and L. Losano},
    title = {Deformed self-dual magnetic monopoles},
    journal = {Physics Letters B},
    volume = {727},
    number = {4},
    pages = {548-553},
    year = {2013},
    issn = {0370-2693},
    doi = {https://doi.org/10.1016/j.physletb.2013.11.001},
    url = {https://www.sciencedirect.com/science/article/pii/S0370269313008939},
}

@article{Bazeia-2018-Magnetic,
  title = {Magnetic monopoles with internal structure},
  author = {Bazeia, D. and Marques, M. A. and Menezes, R.},
  journal = {Phys. Rev. D},
  volume = {97},
  issue = {10},
  pages = {105024},
  numpages = {7},
  year = {2018},
  month = {May},
  publisher = {American Physical Society},
  doi = {10.1103/PhysRevD.97.105024},
  url = {https://link.aps.org/doi/10.1103/PhysRevD.97.105024}
}

@article{Bazeia-2018-Small,
  title = {Small and hollow magnetic monopoles},
  author = {Bazeia, D. and Marques, M. A. and Olmo, Gonzalo J.},
  journal = {Phys. Rev. D},
  volume = {98},
  issue = {2},
  pages = {025017},
  numpages = {8},
  year = {2018},
  month = {Jul},
  publisher = {American Physical Society},
  doi = {10.1103/PhysRevD.98.025017},
  url = {https://link.aps.org/doi/10.1103/PhysRevD.98.025017}
}

@article{Bazeia-2018-Bimagnetic,
  title = {Bimagnetic monopoles},
  author = {Bazeia, D. and Marques, M. A. and Menezes, R.},
  journal = {Phys. Rev. D},
  volume = {98},
  issue = {6},
  pages = {065003},
  numpages = {8},
  year = {2018},
  month = {Sep},
  publisher = {American Physical Society},
  doi = {10.1103/PhysRevD.98.065003},
  url = {https://link.aps.org/doi/10.1103/PhysRevD.98.065003}
}

@article{Bazeia-2021-Multimagnetic,
  title = {Multimagnetic monopoles},
  author = {Bazeia, D. and Liao, M. A. and Marques, M. A.},
  journal = {The European Physical Journal C},
  volume = {81},
  pages = {552},
  year = {2021},
  publisher = {Springer},
  doi = {10.1140/epjc/s10052-021-09352-w},
  url = {https://doi.org/10.1140/epjc/s10052-021-09352-w}
}

@article{Atmaja-2018-BPS,
    author = {Ardian Nata Atmaja and Ilham Prasetyo},
    title = {BPS Equations of Monopole and Dyon in SU(2) Yang-Mills-Higgs Model, Nakamula-Shiraishi Models, and Their Generalized Versions from the BPS Lagrangian Method},
    journal = {Advances in High Energy Physics},
    volume = {2018},
    number = {1},
    pages = {7376534},
    year = {2018},
    doi = {https://doi.org/10.1155/2018/7376534},
    url = {https://doi.org/10.1155/2018/7376534}
}

@article{Atmaja-2022-Are,
    author = {Ardian Nata Atmaja},
    title = {Are there BPS dyons in the generalized SU(2) Yang-Mills-Higgs model?},
    journal = {Eur. Phys. J. C},
    volume = {82},
    pages = {602},
    year = {2022},
    doi = {https://doi.org/10.1140/epjc/s10052-022-10569-6},
    url = {https://doi.org/10.1140/epjc/s10052-022-10569-6}
}

@article{Mulyanto-2026-On,
    author = {Mulyanto and Emir Syahreza Fadhilla and Ardian Nata Atmaja},
    title = {On BPS equations of generalized SU(2) Yang-Mills-Higgs model with scalars-dependent coupling θ-term},
    journal = {Physics Letters B},
    volume = {874},
    pages = {140267},
    year = {2026},
    issn = {0370-2693},
    doi = {https://doi.org/10.1016/j.physletb.2026.140267},
    url = {https://doi.org/10.1016/j.physletb.2026.140267}
}

@article{Benes-2023-Shapes,
  title = {Shapes of magnetic monopoles in effective $SU(2)$ models},
  author = {Bene\ifmmode \check{s}\else \v{s}\fi{}, Petr and Blaschke, Filip},
  journal = {Phys. Rev. D},
  volume = {107},
  issue = {12},
  pages = {125002},
  numpages = {14},
  year = {2023},
  month = {Jun},
  publisher = {American Physical Society},
  doi = {10.1103/PhysRevD.107.125002},
  url = {https://link.aps.org/doi/10.1103/PhysRevD.107.125002}
}

@article{Benes-2026-Magnetic,
    author = {Beneš, Petr and Blaschke, Filip},
    title = {Magnetic Monopoles with an Internal Degree of Freedom},
    journal = {Progress of Theoretical and Experimental Physics},
    volume = {2026},
    number = {2},
    pages = {023B04},
    year = {2026},
    month = {01},
    issn = {2050-3911},
    doi = {10.1093/ptep/ptag004},
    url = {https://doi.org/10.1093/ptep/ptag004}
}

@article{Tong-2014-Vortices,
    author = {Tong, D. and Wong, K.},
    title = {Vortices and Impurities},
    journal = {J. High Energ. Phys.},
    volume = {01},
    pages = {090},
    year = {2014},
    doi = {10.1007/JHEP01(2014)090},
    url = {https://doi.org/10.1007/JHEP01(2014)090}
}

@article{Cockburn-2017-Dynamics,
    author = {Alexander Cockburn and Steffen Krusch and Abera A. Muhamed},
    title = {Dynamics of vortices with magnetic impurities},
    journal = {J. Math. Phys.},
    volume = {58},
    number = {6},
    pages = {063509},
    year = {2017},
    month = {06},
    doi = {10.1063/1.4984980},
    url = {https://doi.org/10.1063/1.4984980}
}

@article{Almeida-2022-Selfdual,
  title = {Self-dual $CP(2)$ vortex-like solitons in the presence of magnetic impurities},
  author = {Almeida, V. and Casana, R. and da Hora, E. and Krusch, S.},
  journal = {Phys. Rev. D},
  volume = {106},
  issue = {1},
  pages = {016010},
  numpages = {14},
  year = {2022},
  month = {Jul},
  publisher = {American Physical Society},
  doi = {10.1103/PhysRevD.106.016010},
  url = {https://link.aps.org/doi/10.1103/PhysRevD.106.016010}
}

@article{Bazeia-2024-Abelian,
    author = {Bazeia, D. and Campos, J. G. F. and Mohammadi, A.},
    title = {Abelian Chern-Simons vortices in the presence of magnetic impurities},
    journal = {J. High Energ. Phys.},
    volume = {12},
    pages = {108},
    year = {2024},
    doi = {10.1007/JHEP12(2024)108},
    url = {https://doi.org/10.1007/JHEP12(2024)108}
}

@article{Gonzalez-2026-Selfdual,
    author = {Gonzalez-Gutierrez, N. H. and Casana, Rodolfo and Santos, Andr\'e C.},
    title = {Self-dual compactons in the gauged restricted baby Skyrme model in the presence of an external magnetic field},
    journal = {J. High Energ. Phys.},
    volume = {01},
    pages = {005},
    year = {2026},
    doi = {10.1007/JHEP01(2026)005},
    url = {https://doi.org/10.1007/JHEP01(2026)005}
}

@article{Adam-2019-BPS,
    author = {Adam, C. and Queiruga, J. M. and Wereszczynski, A.},
    title = {BPS soliton-impurity models and supersymmetry},
    journal = {J. High Energ. Phys.},
    volume = {07},
    pages = {164},
    year = {2019},
    doi = {10.1007/JHEP07(2019)164},
    url = {https://doi.org/10.1007/JHEP07(2019)164}
}

@article{Adam-2019-phi4,
    author = {Adam, C. and Romanczukiewicz, T. and Wereszczynski, A.},
    title = {The $\phi^4$ model with the BPS preserving defect},
    journal = {J. High Energ. Phys.},
    volume = {03},
    pages = {131},
    year = {2019},
    doi = {10.1007/JHEP03(2019)131},
    url = {https://doi.org/10.1007/JHEP03(2019)131}
}

@article{Adam-2019-Solvable,
    author = {Adam, C. and Oles, K. and Queiruga, J. M. and Romanczukiewicz, T. and Wereszczynski, A.},
    title = {Solvable self-dual impurity models},
    journal = {J. High Energ. Phys.},
    volume = {07},
    pages = {150},
    year = {2019},
    doi = {10.1007/JHEP07(2019)150},
    url = {https://doi.org/10.1007/JHEP07(2019)150}
}

@article{Campos-2020-Fermion,
  title = {Fermion transfer in the ${\ensuremath{\phi}}^{4}$ model with a half-BPS preserving impurity},
  author = {Campos, Jo\~ao G. F. and Mohammadi, Azadeh},
  journal = {Phys. Rev. D},
  volume = {102},
  issue = {4},
  pages = {045003},
  numpages = {12},
  year = {2020},
  month = {Aug},
  publisher = {American Physical Society},
  doi = {10.1103/PhysRevD.102.045003},
  url = {https://link.aps.org/doi/10.1103/PhysRevD.102.045003}
}

@article{Slawinska-2025-Kink,
  title = {Kink-kink solutions in Bogomol'nyi-Prasad-Sommerfield impurity models},
  author = {S\l{}awi\ifmmode \acute{n}\else \'{n}\fi{}ska, K.},
  journal = {Phys. Rev. E},
  volume = {111},
  issue = {1},
  pages = {014228},
  numpages = {7},
  year = {2025},
  month = {Jan},
  publisher = {American Physical Society},
  doi = {10.1103/PhysRevE.111.014228},
  url = {https://link.aps.org/doi/10.1103/PhysRevE.111.014228}
}

@article{Bazeia-2025-Generalized,
    author = {Bazeia, D. and Marques, M. A. and Menezes, R.},
    title = {Generalized scalar field models in the presence of impurities},
    journal = {Eur. Phys. J. C},
    volume = {85},
    pages = {836},
    year = {2025},
    doi = {10.1140/epjc/s10052-025-14582-3},
    url = {https://doi.org/10.1140/epjc/s10052-025-14582-3}
}

@article{Bazeia-2025-Magnetic,
    author = {Bazeia, D. and Liao, M. A. and Marques, M. A.},
    title = {Magnetic monopoles in Yang-Mills-Higgs theory with impurities},
    journal = {arXiv preprint},
    eprint = {2505.11215v2},
    archivePrefix = {arXiv},
    primaryClass = {hep-th},
    year = {2025},
    url = {https://arxiv.org/abs/2505.11215v2}
}

\end{document}